\documentclass[aps,twocolumn,showpacs,preprintnumbers,amsmath,amssymb,nofootinbib,superscriptaddress,showkeys]{revtex4}

\usepackage{epsfig}

\begin{document}

\title{Renormalization Group Analysis of Boundary Conditions \\  in
       Potential Scattering}\author{M. Pavon
       Valderrama}\email{mpavon@ugr.es} \affiliation{Departamento de
       F\'{\i}sica At\'omica, Molecular y Nuclear, Universidad de
       Granada, E-18071 Granada, Spain.}\affiliation{The
       H. Niewodnicza\'nski Institute of Nuclear Physics, \\ PL-31342
       Krak\'ow, Poland} \author{E. Ruiz
       Arriola}\email{earriola@ugr.es} \affiliation{Departamento de
       F\'{\i}sica At\'omica, Molecular y Nuclear, Universidad de
       Granada, E-18071 Granada, Spain.}

\date{\today}

\begin{abstract} 
\rule{0ex}{3ex} We analyze how a short distance boundary condition for
the Schr\"odinger equation must change as a function of the boundary
radius by imposing the physical requirement of phase shift
independence on the boundary condition. The resulting equation can be
interpreted as a variable phase equation of a complementary boundary
value problem. We discuss the corresponding infrared fixed points and
the perturbative expansion around them generating a short distance
modified effective range theory.  We also discuss ultraviolet fixed
points, limit cycles and attractors with a given fractality which take
place for singular attractive potentials at the origin. The scaling
behaviour of scattering observables can analytically be determined and
is studied with some emphasis on the low energy nucleon-nucleon
interaction via singular pion exchange potentials. The generalization
to coupled channels is also studied.
\end{abstract}


\maketitle

\newpage 


\section{Introduction}
\label{sec:intro}

One of the most remarkable features of low energy scattering for a
short range spherically symmetric potential is the onset of scale
invariance and rotational invariance. For a short range potential,
$U(r)$, which will be the subject of the present paper, with a typical
size $a$ and in the long wavelength limit $ k a \ll 1 $, the Schr\"odinger's
equation for the reduced wave function in the $l=0$ channel becomes
\begin{eqnarray}
u'' (r) =0  \, . 
\label{eq:sch0} 
\end{eqnarray}
The solutions to this equation are straight lines 
\begin{eqnarray}
u (r) = a + b r \, , 
\end{eqnarray}
where the coefficients $a$ and $b$ (or rather their ratio), are
determined by matching to the solution in the region where the
potential acts. The intersection of the asymptotic solution with the
x-axis determines the scattering length $\alpha=-a/b$.
Eq.~(\ref{eq:sch0}) is obviously invariant under the scale
transformation $ r \to \lambda r $, but the general solution is {\it
not} due to the presence of the potential $U (r)$. There are, however,
two particular situations where the solution {\it also} transforms
well under scaling, namely when either $a=0$ or $b=0$. These two cases
correspond to $\alpha =0 $ (trivial scattering) and $ \alpha = \pm
\infty $ (zero energy bound state) respectively. The presence of a
potential $U(r)$ at $r \sim a$ or finite energy, $ ka \sim 1$, induces
scaling violations which can be computed within perturbation theory,
and obey scaling laws. As a matter of fact, it is interesting to see
what kind of perturbation theory can be constructed around these two
very simple scale invariant cases. To answer this question Wilsonian
renormalization group (RG) methods seem the adequate
tool~\cite{Wilson:1973jj}.

Our interest and focus in the present paper is mainly motivated by
encouraging developments in the last decade in nuclear physics, and
more specifically on the Nucleon-Nucleon (NN) interaction problem in
the framework of effective field theories triggered by Weinberg's
work~\cite{Weinberg:1990rz,Weinberg:1991um} (for reviews see e.g. 
Refs~\cite{vanKolck:1998bw,vanKolck:1999mw,Bedaque:2002mn,Epelbaum:2005pn,Hammer:2006qj}). 
Actually, much of the discussion has been unavoidably linked to the ability 
of designing adequate regularization schemes which in addition to preserve 
the symmetries can be removed beyond perturbation theory, allow to handle
highly singular potentials and provide a one-valued renormalization
group flow of low energy parameters.

In this paper we attack the problem by using the boundary condition
(BC) regularization. It is based on the natural idea that all unknown
information below some scale $R$ can always be parameterized in terms
of a mixed boundary condition at the distance $R$ (see
Sect.~\ref{sec:bcr}).  Strictly speaking, this is exactly true for a
non-relativistic system which can be described through the
Schr\"odinger equation, since it is a second order differential
operator. The long distance physics is assumed to be determined in
terms of a potential $U(r)$ above the boundary radius $R$. The BC
method has been used extensively in the past for the treatment of NN
scattering~\cite{LF67}, and more recently for analyzing the commonly
accepted analysis of phase-shifts~\cite{Stoks:1993tb} in terms of the
chiral expansion~\cite{Rentmeester:1999vw} but always keeping the
short distance cut-off finite. The equivalence between effective field
theory (EFT) and boundary conditions at the origin for short range
potentials has been established in Ref.~\cite{vanKolck:1998bw}. We
assume this equivalence to hold also in the presence of long distance
interactions. Actually, we will see that it is possible to shrink the
boundary to the origin with a smooth limit in the physical
observables. From a Lagrangian viewpoint non-local momentum dependent
terms in the interaction can, after suitable field redefinitions and
using the equations of motion, be expressed as local and
center-of-mass (CM) energy dependent potentials~\cite{Beane:2000fi}.

The boundary condition satisfies a renormalization group equation (see
Sect.~\ref{sec:rg}) which infrared and ultraviolet critical points may
be examined and characterized. Moreover, the boundary condition may be
interpreted in terms of an outer truncated potential problem (see
Sect.~\ref{sec:vf}). The renormalization group equation for the
boundary condition becomes a variable phase
equation~\cite{Calogero:1965}, which has a straightforward physical
interpretation. Besides, one of the virtues of the variable phase
approach is that it always deals with a on-shell problem, i.e., at any
stage of the calculation the variable phase shift exactly corresponds
to a physical phase shift of a certain on-shell problem. The price to
pay is the non-linear character of the equation. Thus, the well known
off-shell ambiguities characteristic of Lippmann-Schwinger equation
(LSE), and which make the discussion on renormalization cumbersome,
particularly when truncations are involved, never appear.

Although some of the aspects presented in this paper have been known
or implicitly assumed for a long time in some way or another, we
believe that the interpretation in terms of a short distance boundary
condition becomes quite transparent and unifying, particularly when
explicit long range, i.e. non-contact, interactions are
considered~\cite{PavonValderrama:2003np,PavonValderrama:2004nb}.  In
the case of contact interactions, a general renormalization group
analysis of elastic one channel potential scattering has been studied
previously in Ref.~\cite{Birse:1998dk} in momentum space within a
Lippmann-Schwinger framework using a sharp momentum cut-off, which
separates between the low and high energy region. The standard low
momentum expansions for both the reaction matrix or the K-matrix (in
fact the effective range expansion) arise as energy perturbations
around the trivial and nontrivial scattering fixed points.  The
situation gets more involved when long distance forces are included
and has been tackled in Ref.~\cite{Barford:2002je} providing the EFT
understanding of the long distance modified effective range
expansion~\cite{Haeringen}. This modified effective range expansion
has been used to eliminate the one pion exchange (OPE) effects in the
$^1S_0$ channel in Ref.~\cite{Steele:1998zc}. The method of
Ref.~\cite{Barford:2002je} has been applied more recently for the
study of peripheral waves~\cite{Birse:2003nz} and is based on delta
shell regularization in coordinate space for the short range part of
the interaction, but cutting off the large momentum components. This
implies, in particular that the long range piece extends down to the
origin. In the case of a singular potential at the origin this
procedure becomes ill defined, because a sharp cut-off in momentum
space does not suppress the short distance components entirely, unless
states of high angular momentum with a short distance suppression
cancelling the singularity are considered~\cite{Birse:2003nz}. The
relation of Wilson like renormalization and power counting has been
treated in Ref.~\cite{Birse:2005um} in the particular case of the
singular tensor component to the OPE potential (see also
Ref.~\cite{PavonValderrama:2005gu,Nogga:2005hy,Epelbaum:2006pt} and
\cite{Valderrama:2005wv,PavonValderrama:2005uj} for the two pion
exchange (TPE) extension.). A momentum space treatment of Wilsonian
renormalization ideas has also been proposed in
Ref.~\cite{Bogner:2001jn} allowing for a determination of model
independent low momentum potentials out of several realistic NN
potentials~\cite{Bogner:2001gq,Holt:2003rj,Bogner:2003wn}. The role of
redundant operators in the absence of long range potentials has been
discussed in Ref.~\cite{Harada:2005tw}.

A particular advantage of the BC has to do with the treatment of
non-perturbative renormalization when long distance potentials are
included. Although the momentum space treatment accommodates more
general situations such as non-local potentials than those described
here, we feel that for the most frequent case of long distance physics
with local potentials the analysis in coordinate space becomes more
transparent~\footnote{Actually particle exchange implies in the
non-relativistic limit a local and perhaps energy dependent
potential}. Another important reason to prefer coordinate space in our
analysis is that the Schr\"odinger equation defines a second order
boundary value problem, and hence a sharp separation between long and
short range physics becomes very natural. In particular, the short
distance unknown physics may be handled in the spirit of old and
modern works as a general boundary condition on the wave function at
the origin. The question of how the origin should be approached is
delicate, and depends on the postulated long range potential. We will
discuss this issue along this paper in detail.

The paper is organized as follows. In Sect.~(\ref{sec:bcr}) we
elaborate on the boundary condition regularization, and its advantages
as compared to other coordinate and momentum space regularizations. In
Sect.~(\ref{sec:rg}) we make a comprehensive renormalization group
analysis of boundary conditions. The corresponding infrared fixed
points are determined and identified, and the renormalization group
flow due to energy perturbations and potential perturbations is
discussed. For power-like singular potentials at the origin we also
establish the fixed points, limit cycles and attractors. In
Sect.~(\ref{sec:vf}) we establish the relation between the
renormalization group flow of boundary conditions and the well known
variable phase approach to potential scattering. This provides a nice
interpretation of the boundary condition regularization, which
suggests several working schemes. 
The generalization of higher partial waves and coupled inelastic channels
is rather straightforward and is presented in a sketchy manner in
Sect.~(\ref{sec:higher_pw}). 
Finally, in Sect.~(\ref{sec:conclusions}) we draw our conclusions.

Since the present paper had its main motivation in the study of the 
NN interaction problem in effective field theories, most of the examples
considered along this work are taken from NN scattering in the $^1S_0$
singlet channel. According to Weinberg's power counting, the long range
reduced potential between the nucleons in this channel can be written as 
a low energy expansion which takes 
the form~\cite{Kaiser:1997mw,Rentmeester:1999vw}
\begin{equation}
M_N\,V_{\rm NN} = U_{\rm NN} = U_{\rm LO} + U_{\rm NLO} + U_{\rm NNLO} + \dots 
\end{equation}
where ${\rm LO}$ refers to leading order, ${\rm NLO}$ to next to leading 
order, ${\rm NNLO}$ to next to next to leading order, and so on;
$U_{\rm LO}$ is the well-known one pion exchange (OPE) potential, 
while $U_{\rm NLO}$ and $U_{\rm NNLO}$ come mainly from 
two pion exchange (TPE), although they contain some minor 
contributions to the OPE potential. 
In this work, for convenience, we will mainly use the terms OPE and TPE 
for the potentials, which refer to
\begin{equation}
U_{\rm OPE} = U_{\rm LO} \quad U_{\rm TPE} = U_{\rm LO} + U_{\rm NLO}
+ U_{\rm NNLO} \, .
\end{equation}
It should be noted that in the $^1S_0$ channel the OPE piece is just the
well known Yukawa potential
\begin{equation}
U_{\rm OPE}(R) = - \frac{1}{a R}\,e^{-m R} \, ,
\end{equation}
where $a = 16 \pi f^2 / (M_N\,m^2\,g^2) = 0.7\,{\rm fm}$ and 
$m = 0.699\,{\rm fm}^{-1}$ is the pion mass ($f = 0.468\,{\rm fm}^{-1}$ 
is the pion weak decay constant, $g = 1.29$ is the pion axial coupling
and $M_N = 4.758\,{\rm fm}^{-1}$ is the nucleon mass).
For the TPE potential, it is enough to know that for distances below the 
pion Compton wave length, $m R \ll 1$, it behaves as
\begin{equation}
U_{\rm TPE}(R) \to  - \frac{a^4}{R^6} \, ,
\end{equation}
where $a \simeq 1.64\,{\rm fm}$ in the singlet channel, although it depends 
on the set of parameters one uses 
(for further details on the conventions used along this work for
the TPE potential, see Refs.~\cite{Valderrama:2005wv,PavonValderrama:2005uj}).

\section{Boundary condition regularization} 
\label{sec:bcr}

For simplicity, let us consider the Schr\"odinger equation for
$S$-wave scattering with reduced potential $U(r)= 2\mu V(r)$ and
reduced wave function $u(r)$,
\begin{equation}
-u_k'' (r) + U(r) u_k(r) = k^2 u_k  \, , 
\label{eq:sch}
\end{equation}
with the asymptotic behaviour at infinity   
\begin{equation}
u_k (r) \to \sin ( k r + \delta(k)) \, , 
\label{eq:asymp}
\end{equation}
and subject to the mixed boundary condition
\begin{equation}
u_k' (R) - L_k (R) u_k (R) =0   \, . 
\label{eq:bc} 
\end{equation}
The coefficient $L_k(R) $ encodes the physics below the scale $ r \le
R$ down to the origin at the momentum $k$~\footnote{This is the most
general boundary condition which makes the Hamiltonian self-adjoint in
the interval $ R \le r < \infty $. The hard core boundary condition,
$u(R)=0$ corresponds to {\it formally} take the limit $ L \to \infty $
needs a separate discussion, since there is no continuous dependence
on the parameter $L$ at $L=\infty$.}. For a given value of the inner
boundary radius $R$, we get a solution $u(r,R)$ which depends both on
the distance $r$ and on the inner boundary radius $R$. Obviously, the
phase shift $\delta(k)$ inherits this $R$ dependence. The boundary
condition represents our lack of explicit knowledge at some low
scales, $ r \le R$, while we assume complete information on the
potential, actually a local one, $U(r)$ for $r > R$. For a given $R$
we do not expect the scattering phase shift to depend strongly on this
lack of explicit knowledge for wavelengths larger than $R$, i.e. $ kR
\ll 1$. We will show below how this statement can be made more
quantitative.

Finally, as in any method we want ultimately to remove the regulator,
i.e. to take the limit of the boundary radius to zero, $R \to 0$. This
implicitly requires to extend the potential to short distances. In
practice, we expect that for a radius much smaller than any other
length scale in the problem a smooth limit should be obtained. As we
will see in practical cases of interest in NN scattering some cautions
must be taken, because although the limit in the bare parameters is
not necessarily smooth, the physical results turn out to be indeed
well behaved under certain circumstances.

The BC regularization method has several advantages over other
methods. For clarity, we list them here, although their complete
meaning will become obvious along the paper. 

\begin{itemize}
\item It does not break any symmetry. In the present context this is a
trivial statement, because it is applied to potential scattering once
the CM motion and angular dependence are separated. 

\item It can incorporate any higher order derivative interactions. 
This is easily done by making an energy expansion of the
boundary condition. This excludes the subtraction method 
of Ref.~\cite{Frederico:1999ps}.

\item It is a non-perturbative regularization. This is particularly
interesting if one wants to discuss perturbative approximations, or
power-counting schemes. A good test to those perturbative treatments
is to see whether higher order corrections are indeed small by
comparing with the full non-perturbative solution. In dimensional
regularization (DR) there is no way to make such calculations, unless
the regulators are analytically
removed~\cite{Kaplan:1996nv,Nieves:2003uu} or the potential has a
particularly simple separable structure~\cite{Phillips:1999bf}.
 
\item It can be applied to attractive or repulsive singular potentials
at the origin~\footnote{By singular potential we understand a
potential fulfilling the condition $\lim_{r \to 0} r^2 |U(r)| = \infty
$ or $\lim_{r \to 0} r^2 U(r) < -1/4 $.}. A crucial feature for these
potentials is that they are non-perturbatively renormalizable but
become perturbatively non-renormalizable~\cite{Case:1950}. Again,
dimensional regularization both in the minimal subtraction (MS) or
power divergence subtraction (PDS) scheme is unable to handle this
problem, and to date there is no calculation dealing with singular
potentials in DR. Also the delta shell regularization in coordinate
space at short distances is excluded, because it assumes the singular
potential to act at distances below the short distance
regulator~\cite{Barford:2002je}.
 
\item It generates a one valued renormalization group flow because it
involves {\it one} distance scale only. In contrast, regularization by
a potential depends on at least {\it two} distance scales:
the {\it range} of the potential as well as the {\it strength}
($1/\sqrt{U}$ has dimensions of length) of the potential.  The
square-well short distance regulators advocated in
Ref.~\cite{Beane:2000wh,Beane:2001bc} provide a multiple branched RG
flow structure, which does influence the phase-shifts. It is not clear
from that work which branch should one take, {\it a priori}. In fact,
these infinite branches reflect the well known non-uniqueness of the
inverse scattering problem, rather than the short distance singularity
of the long range potential.

\item It allows a numerical elimination of the cut-off. In the
calculations we will show below, we will always make sure that
physical results are insensitive to the actual value of the short
distance regulator. This improves e.g. on Ref.~\cite{Epelbaum:1999dj}
where a finite cut-off was imposed.

\item It provides a one-to-one mapping between the expansion of the BC
and the physical scattering amplitude or inverse amplitude, both for
the case of natural and unnatural scattering length. In other words,
if the short distance wave function is truncated to a finite given
order in a low energy expansion, the amplitude is also truncated at
the same order. This property is shared with dimensional
regularization in the MS scheme in the case of small scattering
length.

\item It is a uniquely defined regularization in the sense that the
logarithmic derivative of the wave function computed from the
asymptotic one can be uniquely determined from experiment by just
integrating Schr\"odinger equation from infinity downwards to the
origin.
\end{itemize} 

One of the problems which this regularization manages to deal with
rather transparently is the disentanglement between short and long
distance physics.  Anticipating some of the results to come, we
propose to solve~\footnote{For the time being we will assume a
completely regular potential. Singular potentials will be discussed
separately below.}
\begin{equation}
-u_k '' (r) + U(r) u_k (r) = k^2 u_k (r)\, , 
\label{eq:sch_k} 
\end{equation}
subject to the boundary condition at the origin and normalization at
infinity
\begin{eqnarray}
\frac{u_k' (0^+) }{u_k(0^+)} &=& k \cot \delta_S (k) \, ,\\ 
u_k (r) &\to& \frac{\sin (k r + \delta(k))}{\sin \delta(k)} \, .
\label{eq:bc_sch} 
\end{eqnarray}
We use the notation $u_k (0^+) = \lim_{R_S \to 0^+ } u_k( R_S )$,
since we will see that in general {\it a limit must be taken}.  In the
absence of long range potential $U(r)=0$ the phase shift is given by
$\delta_S(k)$. On the other hand, if we take $\delta_S(k)=0$ we get a
standard problem with a regular boundary condition at the origin,
$u_k(0)=0$. The actual problem is that $\delta_S(k) $ is unknown.  At
low energies both the {\it full} phase-shift $\delta(k) $ and the {\it
short distance} phase-shift $\delta_S(k)$ can be described by some low
energy approximation, like e.g., an effective range expansion,
\begin{eqnarray}
k \cot \delta_S (k) &=& - \frac{1}{\alpha_{0,S}}+ \frac12 r_{0,S} k^2 +
\dots
\label{eq:ere_short} \\ 
k \cot \delta (k) &=& - \frac{1}{\alpha_{0}}+ \frac12 r_{0} k^2 +
\dots
\label{eq:ere_full}
\end{eqnarray} 
where $ \alpha_{0,S}$ is the zero range scattering length, $ r_{0,S}$
the zero range effective range, and $\alpha_0$ and $r_0$ the full
ones.  If we also make an expansion at low energies of the reduced
wave function
\begin{eqnarray}
u_k (r)= u_0 (r) + k^2 u_2 (r) + \dots   
\end{eqnarray} 
we get a recurrent hierarchy of equations, namely  
\begin{eqnarray} 
-u_0 '' (r) + U(r) u_0 (r) &=& 0  \, , \qquad \label{eq:u0} \\ 
\alpha_S u_0' (0^+) +  u_0\,(0^+) &=& 0 \, , \nonumber \\  
u_0 (r) &\to&   1- \frac{r}{\alpha}  \, , \nonumber
\end{eqnarray} 
at zeroth order and 
\begin{eqnarray} 
-u_2 '' (r) + U(r) u_2 (r) &=& u_0 (r) \, , \label{eq:u2} \\ 
\alpha_S u_2' (0^+) + u_2 (0^+) &=& \frac12 r_S \alpha_S u_0 (0^+) \, , 
\nonumber \\ 
u_2 (r) &\to& \frac{r}{6 \alpha}\left(r^2 -3 \alpha r + 3 \alpha r_0 \right) 
\, , \nonumber
\end{eqnarray} 
at second order and so on. These equations suggest a scheme to proceed
in practice. If the short distance physics could be deduced entirely
from the potential we would set $ \alpha_{0,S}=0$, $r_{0,S}=0$, and so
on. Then, the full phase shift $ \delta(k)$ and hence the full low
energy threshold parameters would be determined entirely from the
solutions of the regular problem at the origin~\footnote{We remind
that we are assuming a regular potential at the origin.}. On a leading
order (LO) approximation, one can improve on that by treating $\alpha$
and $U(r)$ as independent variables and predict $\delta$ and the
remaining parameters of the effective range expansion , i.e. $r_0$,
$v_2$, and so on.  In the next-to-leading order approximation (NLO)
$\alpha_0$, $r_0$ and $U(r)$ are regarded as independent variables.

The standard way to proceed would be to integrate the equations,
Eq.~(\ref{eq:u0}), (\ref{eq:u2}) and so on, from the origin (or a
sufficiently small radius $R_S$) and then to adjust the short distance
parameters to get the proper threshold parameters. Instead, one can
simply integrate from infinity downwards, with a known value of
$\alpha_0 $, using Eq.~(\ref{eq:u0}) to obtain $\alpha_{0,S}$ and then
one can use Eq.(\ref{eq:sch_k}) together with Eq.~(\ref{eq:bc_sch})
and Eq.~(\ref{eq:ere_short}) to compute $ \delta(k) $ for any energy
with a given truncated boundary condition. This procedure provides by
definition the low energy parameters we started with and takes into
account that the long range potential determines the form of the wave
function at long distances. The only parameter in the procedure is the
short distance radius $R_S$, which we expect to produce a smooth limit
for the phase shift when we remove it by taking the limit $R_S \to 0$
(in practice $R_S $ should be smaller than any other length scale in
the problem).

Unfortunately, numerical downwards integration of the Schr\"odinger
equation is a rather unstable and delicate procedure because at short
distances the irregular solution starts dominating and high precision
may be required to determine the low energy parameters at short
distances, in a way as to recover the long distance wave
function. This causes a sort of practical and spurious irreversibility
triggered by the irregular solution; downwards and upwards integration
may not necessarily be faithfully represented as inverse operations of
each other at the numerical level (see
e.g. \cite{PavonValderrama:2004nb} for further details).

\section{Renormalization Group Analysis of Boundary Conditions} 
\label{sec:rg}

\subsection{RG equation} 
\label{sub:RG_eq}

We want to determine the evolution of the boundary condition, $L_k (R)
$, on the boundary radius, $R$, by imposing the physical requirement
of independence of phase shifts. This is very much in the spirit of
the derivation of the Callan-Symanzik equation as applied to the
renormalization of Green's functions in Quantum Field Theory. The
purpose is to establish contact with the methods of
Ref.~\cite{Birse:1998dk} where the analysis is carried out entirely in
momentum space, and to show that the whole discussion can quite
naturally be carried out in coordinate space. In
Appendix~\ref{sec:potential_rg} we show how the method can also be
applied to the case where one uses a square well
potential~\cite{Beane:2000wh} to regulate the short distance physics
yielding a multivalued evolution.  (See also the discussion in
Ref.~\cite{Eiras:2001hu}.).

In order to proceed further, we make the infinitesimal change of the
boundary radius $ R \to R + \Delta R$ and take into account the total
derivative
\begin{equation}
\frac{\partial u(r,R)}{\partial R}  = u_R (r,R) \, .  
\end{equation}
Then, the derivative of the boundary condition with respect to the
boundary radius is given by
\begin{eqnarray}
u'' (R,R) &+& u_R ' (R,R) -
L' _k (R) u(R,R) \\ &-& L_k (R) \left( u' (R,R)+ u_R (R,R) \right)  =0 \, .
\nonumber 
\label{eq:bc_def} 
\end{eqnarray}
Deriving also Schr\"odinger's equation with respect to the inner
boundary radius $R$ we get
\begin{equation}
-u_R '' (r,R) + U(r) u_R (r,R) = k^2 u_R (r,R) \, , 
\label{eq:sch_def}
\end{equation}
and the asymptotic wave function 
\begin{equation}
u (r,R) \to \sin (k r + \delta (k)) \, , 
\end{equation}
we get 
\begin{eqnarray}
u_R (r,R) &\to& \cos (k r + \delta (k))\,\delta_R (k) \, ,\nonumber \\ u'
(r,R) & \to & \cos (k r + \delta (k)) \, , \\ 
u_R' (r,R) & \to & -\sin (k r + \delta (k))\,\delta_R (k) \nonumber \, .
\label{eq:bc_asy}
\end{eqnarray}
Thus, using Lagrange's identity we get 
\begin{eqnarray}
0= - u_R  u'' - u_R'' u = \left(- u_R u' + u_R' u \right)' \, .   
\end{eqnarray} 
Integrating between $R$ and $\infty$ and using the boundary condition,
Eq.~(\ref{eq:bc}) and Eq.~(\ref{eq:bc_asy}), we finally get
\begin{equation}
- k \frac{ d \delta }{dR} = \left[ k^2 - U(R) + L_k' (R) + L_k (R)^2
  \right] u(R,R)^2 \, .
\label{eq:del1}
\end{equation}
This equation tells us how the phase shift changes as the inner radius
is changed. If we require the phase shift {\it not} to be dependent on
the particular choice of $R$ (renormalization group invariance) we
get~\footnote{This defines RG invariance by a continuous change in the
cut-off parameter. Note that there is another possible solution to the
equation, namely the zeros of the wave function, $u_k (R_n (k),R) =0$,
at some discrete set of short distance cut-offs $R_n (k) $ which
obviously depend on energy. This is equivalent a hard core potential,
as opposed to the soft boundary condition (\ref{eq:bc}). We analyze
this possibility later on.}
\begin{equation}
- L_k' (R) = k^2 - U(R) + L_k (R)^2 \, .
\label{eq:renorm} 
\end{equation}
This equation governs the evolution of the boundary condition, i.e.,
the logarithmic derivative of the wave function at the boundary, which
shows that in order to guarantee independence of the phase shifts with
respect to $R$ at {\it all } energies the boundary condition must
also depend on energy, a not surprising result.

It is instructive to see that Eq.~(\ref{eq:renorm}) can be almost
trivially deduced, by applying Schr\"odinger's equation to the
boundary condition at the point $R^+$, but the relation to renormalization 
and the approach of Ref.~\cite{Birse:1998dk} is less obvious.
 
Eq.~(\ref{eq:del1}) has interesting consequences as regards the low
energy limit of the boundary condition and provides in addition an
error estimate of the phase shift for $k \to 0$. If we go to the zero
energy limit, the phase shift behaves as $\delta \sim - \alpha k $,
with $\alpha$ the s-wave scattering length and Eq.~(\ref{eq:del1})
becomes,
\begin{eqnarray}
- \frac{ d \alpha }{dR} &=& \left[ - U(R) + L_0 ' (R) + L_0 (R)^2
  \right] \times \, \nonumber\\ 
&& \times \,\lim_{k \to 0}\left[\frac{u(R,R)^2}{k^2} \right] \, .
\label{eq:al1}
\end{eqnarray}
The limit in Eq.~(\ref{eq:al1}) is finite since the normalization at
infinity of the wave function in the limit $ k\to 0 $ is given by $ u
\sim k (r- \alpha) $. The independence of the scattering length with
respect to the BC implies the energy independent evolution equation
\begin{equation}
- L_0' (R) = - U(R)  + L_0 (R)^2 \, .
\end{equation}
If we assume this equation we get 
\begin{equation}
-\frac1k \frac{ d \delta }{dR} =  k^2  \frac{u(R,R)^2}{k^2} = 
{\cal O} (k^2) \, ,
\label{eq:del2}
\end{equation}
where we have used that in our normalization $ u = {\cal O} (k) $ for
small $k$.  This means that by making the physical scattering length
BC independent the phase shifts are $R$ independent only to order
$k^2$, if $ R \sim 1/k $. This argument can be extended to higher
orders in $ kR $; if we solve the evolution equation to order $k^n $
the error in the phase shift is $ {\cal O} (k^{2n+2}) $.

The previous estimate has, in addition a direct application in the
renormalization of singular potentials (both attractive and repulsive)
at the origin, of the form
\begin{equation} 
U(R) \to \pm \frac{1}{a^2} \left( \frac{a}{R} \right)^n \qquad (R
\to 0) \, ,
\end{equation} 
for $n \ge 2 $. If we approach the origin $R \to 0 $, we may neglect
the energy term in Eq.~(\ref{eq:del1}). This means that the condition
for the phase shift to be independent on the boundary condition
becomes the condition of independence of the scattering length. This,
in turn means that if the theory is renormalizable at zero energy it
is renormalizable at any energy. We will see below an alternative and
more appealing formulation of this fact. It has been known for a
long time by a detailed study of the wave functions close to the
origin using the WKB approximation~\cite{Case:1950}, and it is very
rewarding to provide such a simple derivation of this result within
the present framework.
  
The BC parameter $L_k (R)$ has dimensions of inverse length, so it is
natural to measure it in units of the boundary radius $R$,
\begin{equation} 
L_k (R) = \frac{\xi_k (R)}R \, . 
\end{equation} 
The equation satisfied by $\xi_k(R)$ is 
\begin{equation} 
R \frac{ d \xi_k }{ d R} = \xi_k ( 1- \xi_k ) + \left[ U(R)-k^2
\right] R^2 \, . 
\label{eq:bc_pert} 
\end{equation} 
This is a Ricatti type equation. By using the superposition principle
for the wave function 
\begin{eqnarray}
u_k (R) = u_{k,c} (R)  +  k \cot \delta (k) \,  u_{k,s} (R) \, , 
\end{eqnarray} 
with $ u_{k,c} \to \cos (k R) $ and $ u_{k,s} \to \sin (k R) /k $ we have 
that 
\begin{eqnarray}
\xi_k (R) = R \frac{u_k' (R)}{u_k(R)} = R \frac{ u_{k,c}' (R) + k \cot
\delta (k) \, u_{k,s}' (R)}{ u_{k,c} (R) + k \cot \delta (k) \,
u_{k,s} (R) }
\end{eqnarray} 
whence the phase shift can be explicitly determined.  Using the
independence of the phase shift on the cut-off radius it can be easily
shown that two solutions $\xi_k(R)$ and $\xi_k (R_0) $ are related by
a Moebius bilinear transformation (see Ref.~\cite{Ince:1926}),
\begin{eqnarray}
\xi_k (R) = \frac{A(R,R_0) \xi_k (R_0)+ B (R, R_0) }{C (R, R_0)
\xi_k(R_0)+ D (R, R_0)} \, ,
\label{eq_moebius} 
\end{eqnarray}   
where $A$, $B$, $C$ and $D$ depend on the potential $U$ {\it only}.
The former expression disentangles the boundary condition
parameterizing the unknown short distance information from the known
long range potential. Actually, the matrix
\begin{eqnarray}
{\bf M} (R,R_0) =
\begin{pmatrix}
 A(R,R_0) &  B(R,R_0) \\ 
C(R,R_0) &  D(R,R_0)  
\end{pmatrix} 
\end{eqnarray}   
satisfies the group properties
\begin{eqnarray}
 {\bf M} (R,R') {\bf M} (R',R'') = {\bf M} (R,R'') \, ,
\label{eq:cut-off-group}
\end{eqnarray} 
which faithfully represents the dilatation group for the short
distance cut-off $ R\to \lambda R$. We will study below the stability
structure of this group corresponding to the the infrared limit $ R
\to \infty $ and the ultraviolet limit $ R \to 0$. For a study of the
periodic case in $\log (R) $ see e.g. Ref.~\cite{Campos:1997}.

\subsection{Long distance Fixed Points at zero energy} 

For a potential of typical size $a$ and at low energies $ k a \ll 1 $
we may look at the region $ a \ll R \ll 1/k$ where both the potential
and the energy can be neglected, yielding the equation
\begin{equation} 
R \frac{ d \xi_0 }{ d R} = \xi_0 ( 1- \xi_0) \qquad , \qquad a \ll R \ll
1/k \, . 
\label{eq:ren}
\end{equation} 
Note that the equation is scale invariant under the change $ R
\to \lambda R$ within the interval $a \ll R \ll 1/k $.  Obviously,
having a finite interval breaks the scale invariance, since the
interval boundaries also change. Thus, both the potential and the
energy are scaling violating perturbations. 

Fixed points in the dimensionless boundary condition $\xi$ are defined
as those fulfilling the condition $\xi_0' =0$.  Hence,
Eq.~(\ref{eq:ren}) has two fixed points at $\xi_0=0 $ and
$\xi_0=1$. The point $\xi_0=0$ is unstable since any small
perturbation of its value at say $r \sim a $ results in increasingly 
large deviations from the fixed point, as can be seen directly by analyzing
the differential equation, Eq.~(\ref{eq:ren}). On the contrary, the
point $\xi_0=1$ is stable. Since both fixed points are associated 
with long distance behaviour they correspond to infrared (IR) fixed points. 
The physical interpretation of fixed points in the
present context is clear; by varying the BC in the range $a \ll R \ll 1/k$
according to Eq.~(\ref{eq:ren}), we guarantee physics independence
at low energy. The fixed points represent very special situations
where the BC itself does not change with the boundary radius, i.e. it
is scale invariant. For such a situation, we can characterize the
boundary condition by a {\it number} instead of a function of the
boundary radius. In a real case we do not expect to have exact fixed
points, but only some approximation to them. For this case it will be
interesting to choose the scale $a \ll R \ll 1/k$ to describe the
boundary condition function because we expect a weak dependence on the
scale.

For the unstable point, $\xi_0=0$, the situation is such that an
extreme fine tuning of the BC is required to have this scale
invariance. On the contrary, the stable fixed point, $\xi_0=1$, does
not require this accurate fine tuning.

To analyze the physical situation corresponding to these fixed points
let us now take into account that $ L_0(R) = u'_0(R) / u_0(R) $. In the
region $a \ll R \ll 1/k$ we have the asymptotic wave function (we use
the normalization condition $u_0(0) = 1$),
\begin{equation} 
u_0(R) = 1-\frac{R}{\alpha} \, ,
\end{equation} 
yielding
\begin{equation} 
L_0(R) = \frac{\xi_0(R)}R = \frac1{R-\alpha} \, .  
\end{equation} 
Thus, the unstable fixed point $\xi_0=0$ corresponds to $\alpha \to
\infty$ (zero energy bound state) and the stable fixed point $\xi=1$
to $\alpha=0$ (trivial scattering). These conclusions are in full
agreement with the momentum space analysis of Ref.~\cite{Birse:1998dk}. 

\subsection{Positive Energy Perturbation} 
\label{sec:positive}

In the region where the potential does not act $r \gg a$, the equation
satisfied by $\xi_k (R)$ is
\begin{equation} 
R \frac{ d \xi_k  }{ d R} = \xi_k ( 1- \xi_k) -k^2 R^2  \, , 
\end{equation} 
which solution is 
\begin{equation} 
\xi_k (R)  = k R \cot \left[ k R+ \delta (k) \right] \, . 
\label{eq:del_bc}
\end{equation}
We have fixed the arbitrary constant by imposing that the
extrapolation of the logarithmic derivative to the origin can be
related to the scattering phase shift, $\delta(k)$. Then we have 
\begin{equation}
k \cot \delta (k) = k\, \frac{ k R  + \xi_k (R) \cot(kR) }{ kR \cot(kR) -
\xi_k (R) } \, . 
\end{equation}  
Thus, an expansion of the scaled boundary condition in the form
\begin{eqnarray} 
\xi_k (R) &=& \xi_0 (R) + (k R)^2 \xi_2 (R) + (k R)^4 \xi_4 (R) + \dots
\nonumber \\
\label{eq:bc_low}
\end{eqnarray} 
valid in the region $a \ll R \ll 1/k$ yields an expansion for the phase
shift as given by Eq.~(\ref{eq:del_bc}).
\begin{eqnarray}
k \cot \delta (k) =  \frac1{R} \frac{ \xi_0 }{1-\xi_0}+ \frac{\xi_0^2-
3 \xi_0 + 3\xi_2 + 3 }{3 ( \xi_0-1)^2} R k^2 + \dots \,\, . \nonumber \\
\end{eqnarray} 
Whereas written in this way the expansion for $k \cot \delta $ around
the trivial fixed point $\xi_0 = 1$ induces increasingly large
contributions for increasing orders in $k$, spoiling the convergence
of the low energy expansion, in the case of the nontrivial fixed
point $\xi_0 = 0$ this yields a perfectly well defined expansion. 
A similar situation occurs for the expansion in $ \tan \delta /k$,
although with opposite fixed points corresponding to the divergent 
and convergent case. Thus the perturbation theory around the nontrivial 
fixed point (large scattering length) corresponds to an 
effective range expansion of the form
\begin{eqnarray}
k \cot \delta (k) = - \frac1{\alpha} + \frac12 r_0 k^2 + v_2 k^4 +
\dots 
\label{eq:kcot} 
\end{eqnarray} 
whereas for the trivial fixed point (small scattering length) the low
energy expansion reads,  
\begin{eqnarray}
\frac{\tan \delta (k)}k = - \alpha - \beta k^2 - \gamma k^4 + \dots \,\, .
\label{eq:tank} 
\end{eqnarray}
Matching both expansions, Eq.~(\ref{eq:kcot}) and Eq.~(\ref{eq:tank})
we get the identifications
\begin{eqnarray}
\beta &=& \frac12 r_0 \alpha^2 \, ,  \\ 
\gamma &=& \frac14 \alpha \left( \alpha r_0^2 + v_2 \right) \, . 
\end{eqnarray} 
The convergence radius of these low energy expansions has to do with
the longest distance singularities of the potential. For Yukawa like 
behaviour, $U \sim e^{-m R} / R$, one has $|k| < m / 2$, due to the
branch cut at $k = \pm i m /2$.

\begin{figure}[]
\begin{center}
\epsfig{figure=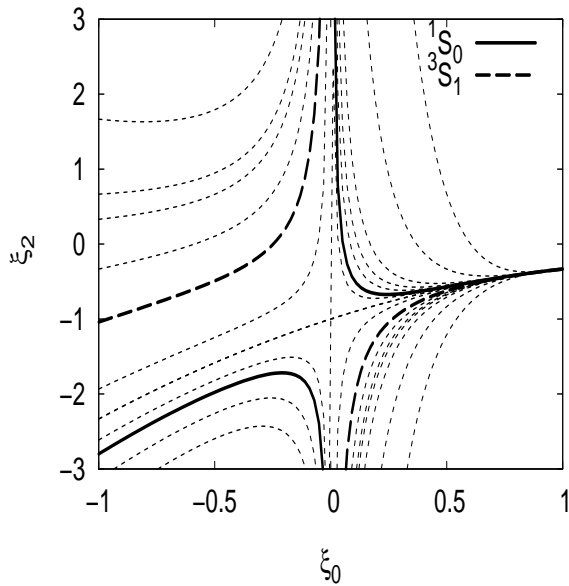,height=8cm,width=8cm}
\end{center}
\caption{The relation between the s-wave parameters $\xi_2$ and
$\xi_0 $ for the pure short range theory defined as the energy
dimensionless coefficients of the logarithmic derivative of the
asymptotic wave function $R\,u_k'(R)/u_k(R) = \xi_0 (R) + (kR)^2 \xi_2 (R)
+ \dots$ depending on particular choices of the ratio $r_0 / \alpha$
between the effective range and the scattering length. We also plot
the flow for NN-scattering the case of singlet $^1S_0$ and triplet
$^3S_1$ states.}
\label{fig:renorm}
\end{figure}

Comparing the expansions (\ref{eq:bc_low}) and (\ref{eq:kcot}), we have
\begin{widetext}
\begin{eqnarray} 
\xi_0 (R) &=& \frac{R}{R-\alpha} \, , \label{eq:xi_0} \\ \xi_2 (R) &=&
\frac{6\,\alpha\,R^2 - 2\,R^3 + \alpha^2\,\left( -6\,R + 3 r_0 \right)
} {6\,{\left( \alpha - R \right) }^2\,R} \, , \label{eq:xi_2} \\ \xi_4
(R) &=& \frac{-\left( -24\,\alpha\,R^5 + 4\,R^6 - 15\,\alpha^3\,
\left( 4\,R^3 - 6\,R^2 r_0 + 3\,R r_0^2 - 12 v_2 \right) +
30\,\alpha^2\,\left( 2\,R^4 - R^3 r_0 - 6\,R v_2 \right) \right)
}{180\,R^3\,{\left( -\alpha + R \right) }^3} \, . \label{eq:xi_4}
\end{eqnarray} 
\end{widetext}
If we express the energy correction to the scaled boundary condition
$\xi_2 $ in terms of $\xi_0$ by eliminating the distance $R$ we get
\begin{eqnarray} 
\xi_2 &=& \frac{\xi_0 (\xi_0^2 -3 \xi_0 +3 )(3 r_0 /\alpha - 2)- 3 r_0
/\alpha }{6 \xi_0}  \, .  
\end{eqnarray} 
The renormalization group flow corresponding to the dependence of
$\xi_2 $ in terms of $\xi_0 $ is depicted in Fig.~(\ref{fig:renorm})
for several values of the ratio between the effective range $r_0$ and
the scattering length $\alpha$. Similar pictures could be obtained for
any other pair of $\xi_n $ variables. For illustration purposes we
also show the flow for the case of singlet $^1S_0$ and triplet $^3S_1$
states in the NN interaction where $\alpha_0 = -23.74 \, {\rm fm} $,  $r_0
= 2.77 \, {\rm fm} $ and $\alpha_0 = 5.42 \, {\rm fm} $ , $r_0 = 1.75 \, {\rm
fm} $ respectively. 

\subsection{Zero energy Short distance critical points}

\subsubsection{Fixed points and Cycles.} 

We analyze now the short distance behaviour of Eq.(\ref{eq:bc_pert})
corresponding to the ultraviolet regime. Generally speaking, if the
solution to the Schr\"odinger equation is written as a linear
combination of a regular and irregular solution, 
$u_0(r) = c_0\,u_{\rm irreg}(r) + c_1\,u _{\rm reg} (r)$, then
\begin{eqnarray}
\xi_0 (R) = R\,\frac{u_0'(R)}{u_0(R)} = R 
\frac{c_0 u_{\rm irreg}'(R) + c_1 u _{\rm reg}'(R)}
{c_0 u_{\rm irreg}(R) + c_1 u _{\rm reg} (R)} \, ,
\end{eqnarray} 
where the ratio $c_0 / c_1$ can be fixed by choosing $\xi_0 (R_0)$ with
$R_0$ some reference scale. Obviously, regular solutions are unstable
fixed points whereas irregular solutions correspond to stable fixed
points since $u_{\rm irreg}$ always takes over when $R \to 0$.  

We will first consider (Sect.~\ref{subsub:UV-regular}) the case of
regular potentials, for which the following condition is fulfilled
\begin{eqnarray}
\lim_{R \to 0} \,R^2\,U(R) = 0
\end{eqnarray}
and afterwards the case of singular potentials which behave as $1/R^n$
at short distances.  In this latter case we are going to make a
further distinction between those power-law potentials with $n > 2$
(Sect.~\ref{subsub:UV-singular}) and the inverse square potential,
$U(R) = g / R^2$ (Sect.~\ref{subsub:UV-ir2}). For a short review of
the short-distance solutions of singular power-law potentials see
Appendix~\ref{app:singular}.

\subsubsection{Regular potentials.}
\label{subsub:UV-regular}

For a regular potential the wave function behaves linearly at short
distances, $u(R) = c_0 + c_1\,R$, yielding
\begin{eqnarray}
\xi_0(R) = \frac{c_1\,R}{c_0 + c_1\,R} \, , 
\end{eqnarray}
which has the fixed points $\xi_0 = 0, 1$. The first one, $\xi_0 = 0$ is stable
and corresponds to irregular solution of the Schr\"odinger equation.
The convergence towards this fixed point is linear in $R$
\begin{eqnarray}
\xi_0(R) \to \frac{c_1}{c_0}\,R + {\cal O}(R^2) \, .
\end{eqnarray}
The other fixed point, $\xi_0 = 1$ is unstable and corresponds to the regular
solution $u(R) \sim R$.

A special case is that of the Yukawa potential
\begin{eqnarray}
U(R) = - \frac{1}{a\,R}\,e^{-m\,R} \, ,
\label{eq:yukawa}
\end{eqnarray}
which has a $1/R$ singularity for short distances. For this potential 
the behaviour of the wave function at short distances is given by
\begin{eqnarray}
u(R) = 
c_0\,\left[ 1 + m\,R - \frac{3 R}{2 a} - \frac{R}{a} \log{(\frac{R}{a})}\right]
+ c_1\,R \, ,
\end{eqnarray}
for which the same fixed points than discussed before for a common
regular potential are reproduced (namely $\xi_0=0,1$) , but with logarithmic convergence
towards the stable fixed point $\xi_0 = 0$
\begin{eqnarray}
\xi_0(R) \to \frac{c_1}{c_0}\,R + m\,R - \frac{5 R}{2 a} - 
\frac{R}{a} \log{(\frac{R}{a})} + {\cal O} (R^2) \, ,
\end{eqnarray}
from which an $R\,\log{R}$ convergence trend is deduced (which, incidentally  
is independent of the actual $c_1/c_0$ ratio).

\subsubsection{Power-law singular potentials.}
\label{subsub:UV-singular}

For a power-law potential which behaves as
\begin{eqnarray}
U (R) = \pm \frac{1}{a^2}\left(\frac{a}{R}\right)^n 
\qquad\, ,\quad n > 2 \, ,
\end{eqnarray} 
as $R \to 0 $ we have one scale only, so we can define the variable
$R= a\,x$, in such a way that the coupling constant becomes one.  The
solution to the Schr\"odinger equation for the repulsive case reads
\begin{eqnarray} 
u (x) &=& c_0 \sqrt{x} K_{\frac1{n-2}} \left( \frac{x^{1-n/2}}{n/2-1}
\right) + \nonumber \\ && 
c_1 \sqrt{x} I_{\frac1{n-2}} \left( \frac{x^{1-n/2}}{n/2-1} \right) \, ,
\end{eqnarray}
where $ c_{0,1}$ are integration constants and $K_\nu (z) $ and $
I_\nu (z) $ are regular (exponentially increasing) and irregular
(exponentially decreasing) modified Bessel functions respectively. 
The fixed points are $\xi_0 = \pm \infty $ corresponding to take $c_0 = 0$
and $c_1 = 0$ respectively; $\xi_0 = + \infty$ is unstable, while
$\xi_0 = - \infty$ is stable.
 
For the attractive $n> 2$ case, the solution to the Schr\"odinger
equation for the attractive case reads
\begin{eqnarray} 
u (x) &=& c_0 \sqrt{x} J_{\frac1{n-2}} \left( \frac{x^{1-n/2}}{n/2-1}
\right) + \nonumber \\ && 
c_1 \sqrt{x} J_{-\frac1{n-2}} \left( \frac{x^{1-n/2}}{n/2-1} \right) \, ,
\end{eqnarray}
where $J_\nu$ are oscillating spherical Bessel functions. From here
one can obtain the dimensionless logarithmic derivative $\xi_0 (R)$,
while the ratio $c_0/c_1$ may be determined by choosing $\xi_0(R_0)$. 
Close to the origin we can use the asymptotic expansions
\begin{eqnarray} 
J_\nu (z) &\to& \sqrt{2/\pi z} \cos \left( ( \nu/2 + 1/4) \pi -z \right)
\end{eqnarray} 
and hence we have
\begin{eqnarray} 
\xi_0  (R) = \frac{\xi_0 (R_0 ) \cot \Phi -1 }{\xi_0 (R_0) + \cot \Phi} \, , 
\end{eqnarray}
with 
\begin{eqnarray} 
\Phi= \frac1{n-2}\left[\left(\frac{R}{R_0}\right)^{1-n/2} -1 \right] \, ,
\end{eqnarray}
where $R_0$ is some short distance reference scale. 
So we see that for $n\ge 2$ we have an attractor $(R,\xi_0(R))$ 
with a fractal asymptotic ($R \to 0$) dimension $d=2-2/n$.
The fractal dimension can be deduced from the scaling properties of the zeros
of $\xi_0(R)$, $R_N^{n/2-1} \sim 1 / N$, being $R_N$ the N-th zero.

\begin{figure*}[]
\begin{center}
\epsfig{figure=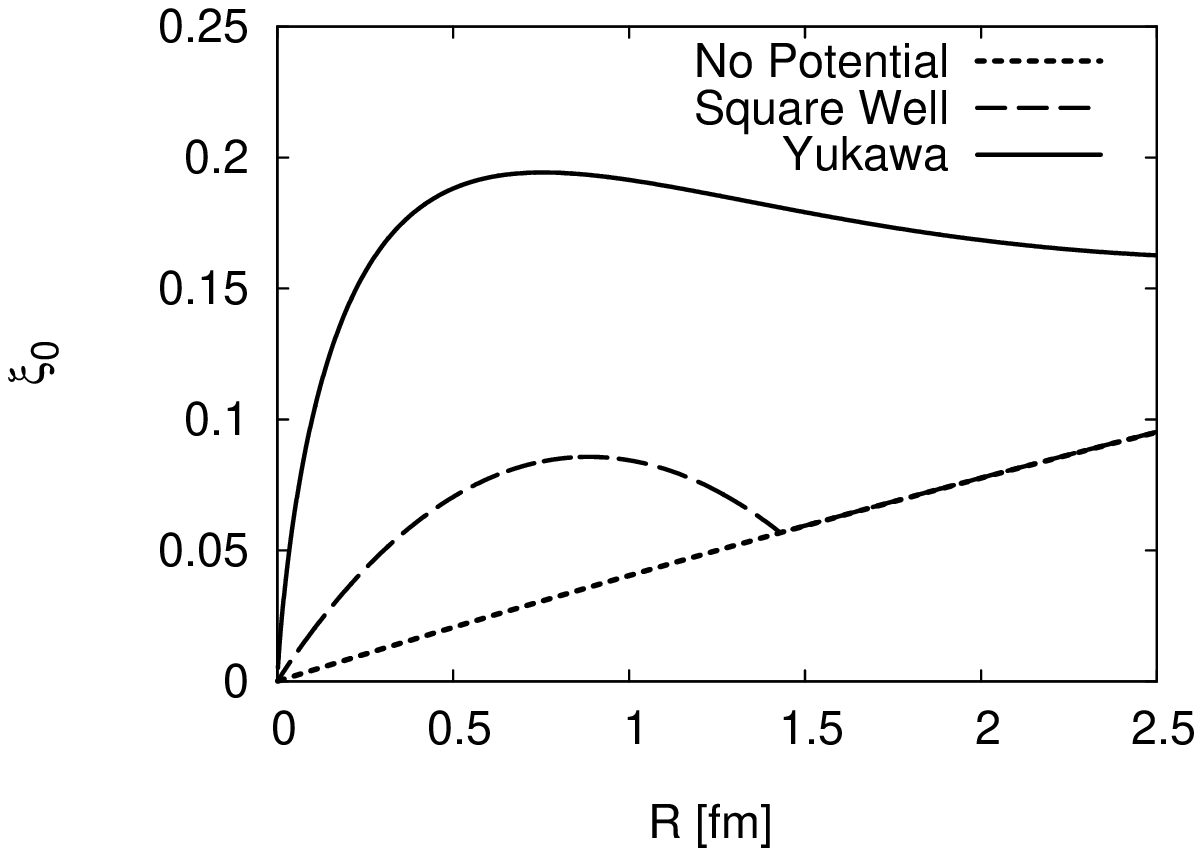,height=5.5cm,width=5.5cm}
\epsfig{figure=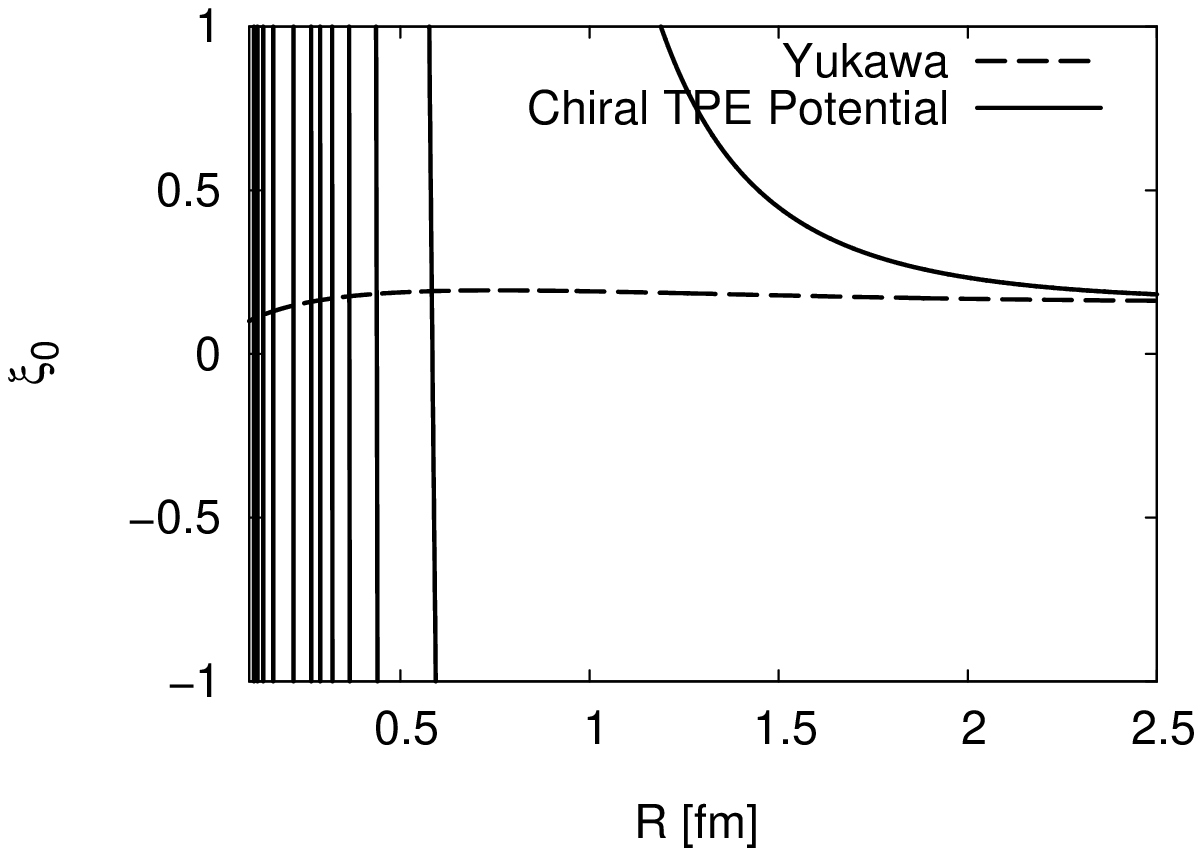,height=5.5cm,width=5.5cm}
\end{center}
\caption{
Running of $\xi_0(R)$ for regular and singular potentials.
In the left panel we show $\xi_0(R)$ for the case with no potential in which
$\xi_0(R) = R / (R - \alpha_0)$, for a square well potential and for a
Yukawa potential (the One Pion Exchange potential in the $^1S_0$ channel).
In the right channel we show $\xi_0(R)$ for the case of the Chiral Two Pion
Exchange potential which for short distances behaves as $- a^4 / R^6$; the
lines representing the renormalization flow of $\xi_0(R)$ becomes more dense
in its way to the origin, giving rise to a nontrivial fractal dimension
$d = 2 - 2/n = 5/3$ in the $R \to 0$ limit.
These examples are taken from neutron-proton scattering in the singlet s-wave
$^1S_0$ channel, in which the scattering length is 
$\alpha_0 = -23.74\,{\rm fm}$, and $R$ is expressed in fm.
}
\label{fig:xi0-running}
\end{figure*}

\begin{figure*}[]
\begin{center}
\epsfig{figure=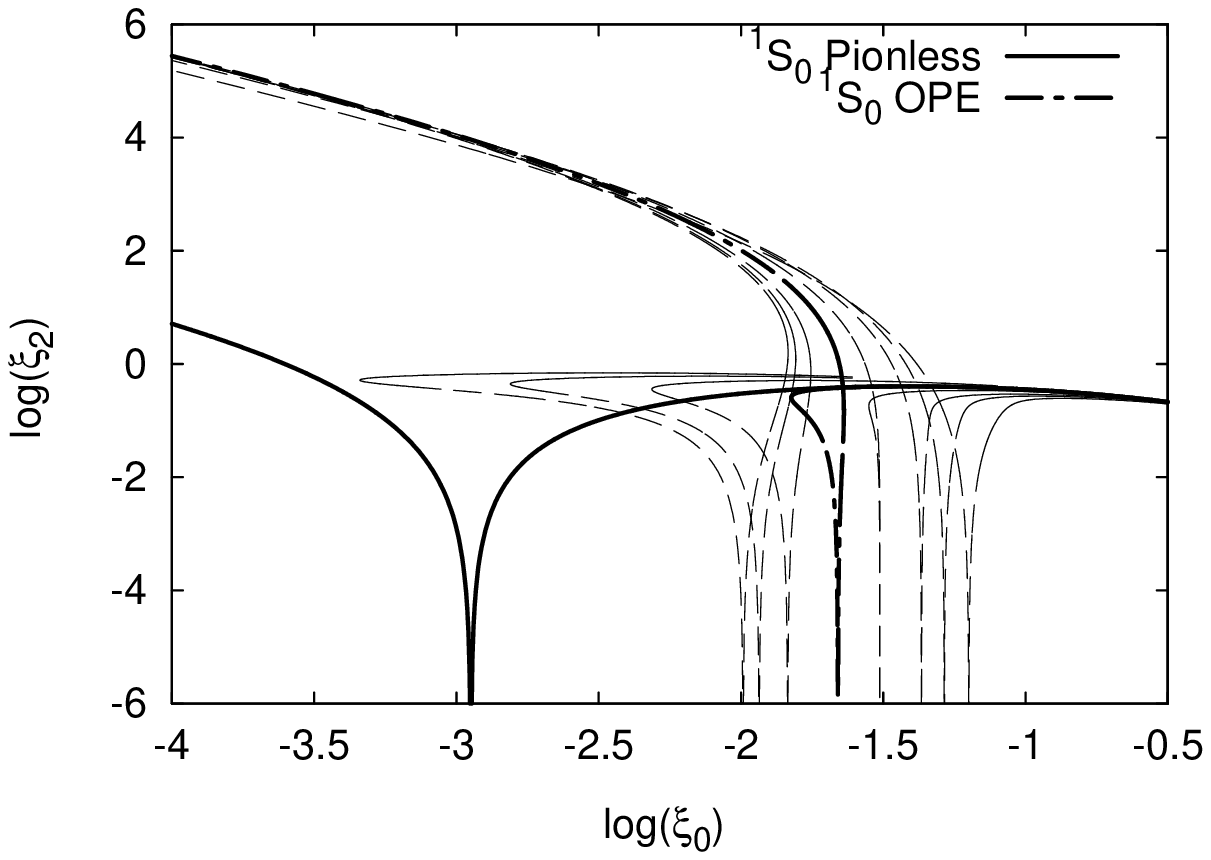,height=5.5cm,width=5.5cm}
\epsfig{figure=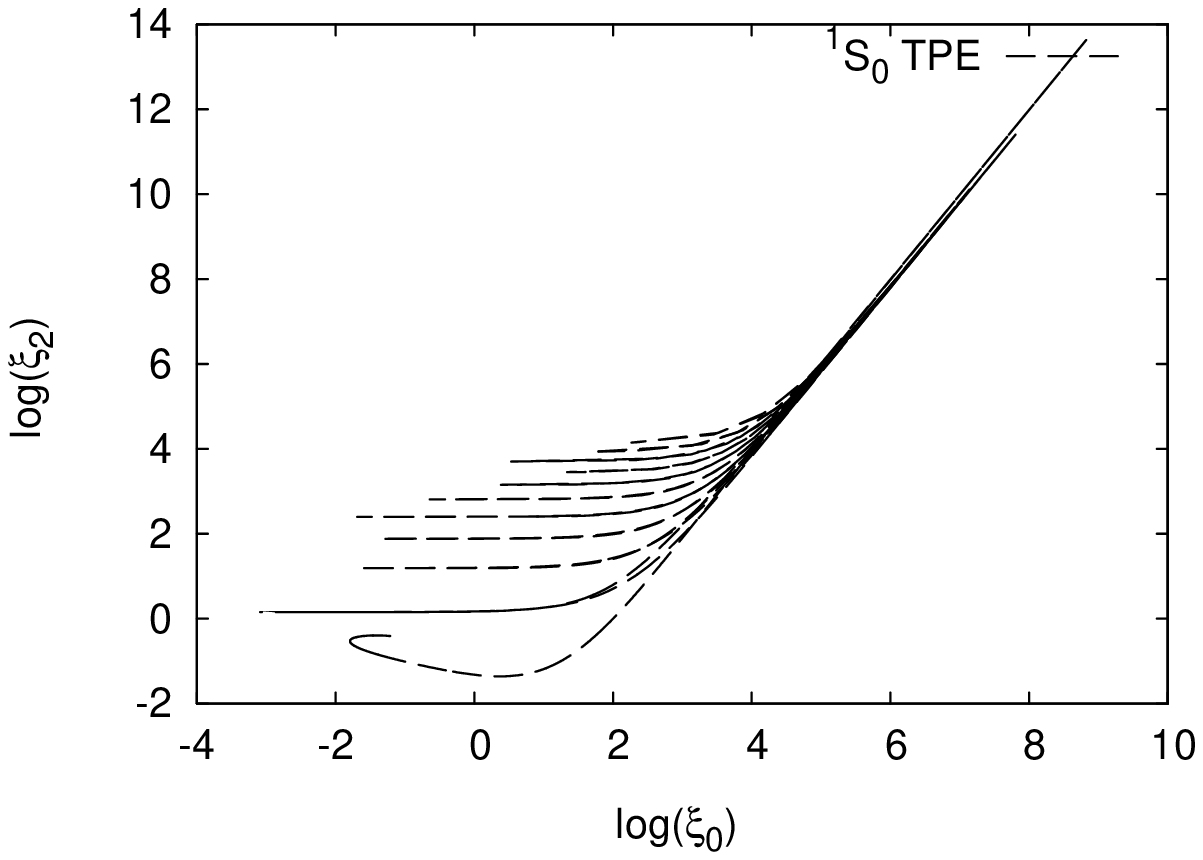,height=5.5cm,width=5.5cm}
\epsfig{figure=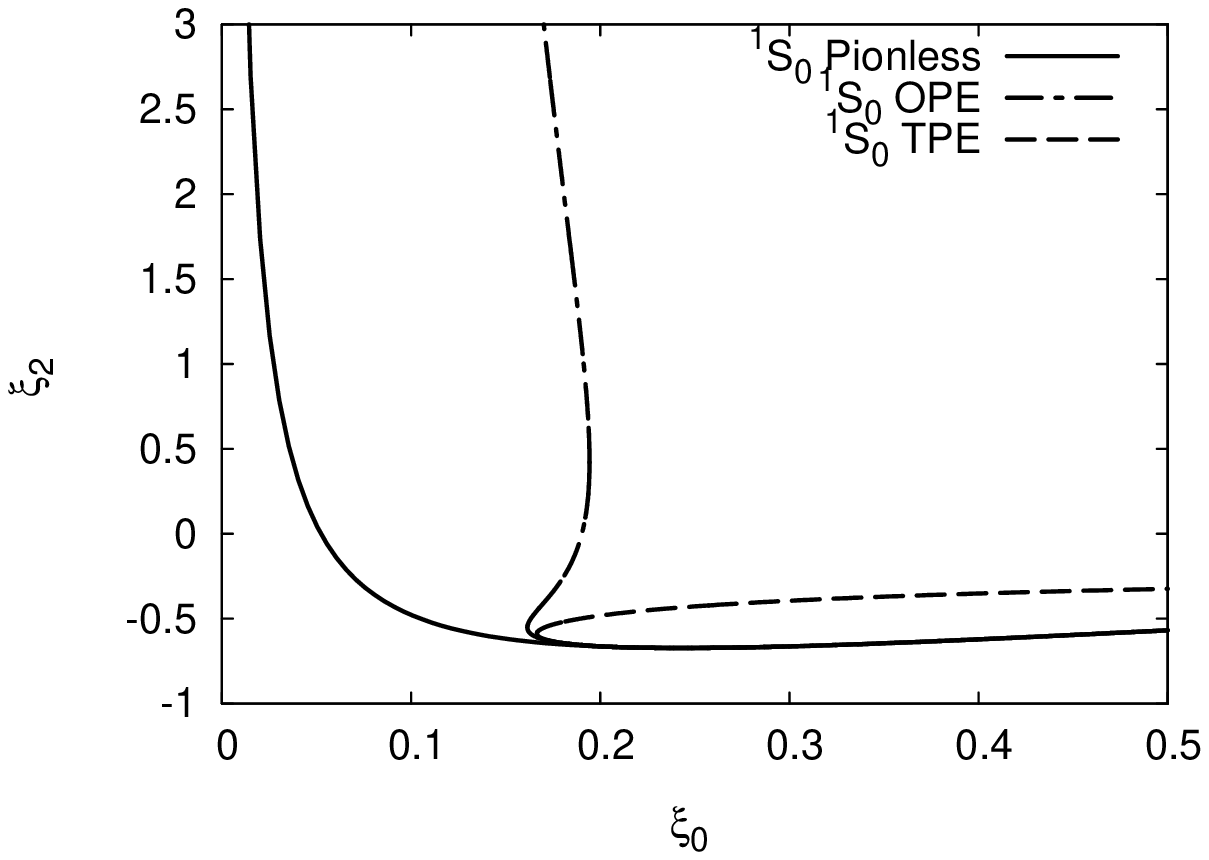,height=5.5cm,width=5.5cm}
\end{center}
\caption{ Running of $\xi_0$ vs $\xi_2$ for regular and singular
potentials (note the log scale).  In the left panel we show the case
of a Yukawa potential (the One Pion Exchange potential in the $^1S_0$
channel).  In the middle channel we show the case of the Chiral Two
Pion Exchange potential. In the left panel we compare the case with
no potential to that of a OPE and TPE potential. These examples are
taken from neutron-proton scattering in the singlet s-wave $^1S_0$
channel, in which the scattering length is $\alpha_0 = -23.74\,{\rm
fm}$, and $r_0=2.77 {\rm fm}$. These plots should be compared with
Fig.~(\ref{fig:renorm}).} 
\label{fig:xi0-xi2-pot}
\end{figure*}

\subsubsection{Inverse square potential}
\label{subsub:UV-ir2}

The inverse square potential, $U=g/R^2$, requires a separate
study. Note that in that case the RG equation, Eq.(\ref{eq:bc_pert}),
is formally invariant under the continuous scaling transformation, 
$R \to \mu R$. This symmetry is, however, explicitly broken by an initial
condition $ \xi_0 (R_0) $.  Using Eq.~(\ref{eq:bc_pert}) we get the
zero energy fixed points at
\begin{eqnarray}
\xi_0 = \pm \sqrt{1+ 4 g}  
\end{eqnarray} 
for $g > -1/4$. The positive and negative roots correspond to stable
and unstable fixed points respectively.  For $g < -1/4 $ we have
purely imaginary solutions.  This is the signal for a limit
cycle. Actually, the solutions are given by
\begin{eqnarray} 
u_0 (R) &=& c_0 R^{\lambda_-} + c_1 R^{\lambda_+} \nonumber \\ 
&& \lambda_{\pm} = (1 \pm \sqrt{1+ 4 g})/2 \quad {\rm for} \quad g > -1/4 \, ,
\nonumber\\
\\ 
\nonumber\\  
u_0 (R) &=& 
c_0 \sqrt{R} \cos\left( \lambda \log R \right) +
c_1 \sqrt{R} \sin\left(\lambda \log R \right) 
\nonumber\\ 
&& \lambda = \sqrt{-1- 4 g} \quad {\rm for} \quad g < -1/4 \, .
\end{eqnarray}
So, in the first case we obtain 
\begin{eqnarray}
\xi_0(R) = \frac{c_0 \lambda_- R^{\lambda_-} + c_1 \lambda_+ R^{\lambda_+}}
{c_0 R^{\lambda_-} + c_1 R^{\lambda_+}}
\quad {\rm for} \quad g > -1/4 \, ,
\end{eqnarray}
which has an attractive fixed point at $\xi_0 = \lambda_-$, 
and a repulsive one at $\xi_0 = \lambda_+$. 
In the latter case, $g < -1/4$, we obtain
\begin{eqnarray} 
\xi_0(R) &=& \lambda 
\cot\left[ \tan^{-1} \frac{2\lambda}{2 \xi_0 (R_0)-1} + 
\lambda \log \frac{R}{R_0} \right]+ \frac{1}{2} \, .
\nonumber\\
\end{eqnarray} 
In this case the discrete scaling property $\xi_0 ( R e^{N
\pi/\lambda}) = \xi_0 (R) $ with $N=0,\pm 1,\pm 2, \dots $ typical of
the limit cycles~\cite{Braaten:2004rn} is reproduced. For instance, if
we consider $\xi_0(R_N)=0 $, then the sequence of the zeros of $\xi_0$
is given by $R_{N+1} = e^{\pi/\lambda} R_N$. This corresponds to the
Russian doll renormalization (see e.g. Ref.~\cite{Leclair:2003xj}).
Coordinate and momentum space analyses have been treated in
\cite{Braaten:2004pg} and \cite{Hammer:2005sa} respectively. 

\subsubsection{Overview of the ultraviolet limit}

On the light of the previous discussion we obtain for the short
distance dimensionless boundary condition, $\xi_0(R)$, the following
behaviour

\begin{itemize}
\item For a regular potential we have two fixed points, an attractive one,
corresponding to the irregular solution of the Schr\"{o}dinger equation, 
and a repulsive one, corresponding to the regular solution.
This means that all solutions go to the irregular solution at the origin.

\item For a singular potential with $ n = 2 $ and coupling $g > -1/4$ 
we have two fixed points. For $g < -1/4$ there are limit cycles.

\item For a repulsive singular potential with $n > 2$ we have two fixed 
points. The attractive one corresponds to the irregular solution.

\item For an attractive singular potential with $n > 2$ we have an
attractor with asymptotic fractal dimension $d=2-2/n $.
\end{itemize} 

In Fig.~(\ref{fig:xi0-running}) we present three of the possible
situations. The examples are taken from neutron-proton scattering in
the singlet s-wave $^1S_0$ channel, for which the scattering length is
$\alpha_0 = -23.74\,{\rm fm}$. In the regular potentials example we
use for the discussion the One Pion Exchange (OPE) potential, which,
for this channel, takes the form of an usual Yukawa potential as the
one given in Eq.~(\ref{eq:yukawa}), for which $m = 0.699\,{\rm
fm}^{-1}\,(= 138\,{\rm MeV})$ and $a = 0.7\,{\rm fm}$, and a Square
Well potential with range $R_0 = 1/m = 1.43\,{\rm fm}$ and depth $U_0
= -0.1\,{\rm fm}^2$.  In the singular potential example we use the
chiral Two Pion Exchange (TPE) potential, which behaves as $-a^4 /
R^6$ at distances below the pion Compton wave length. For completeness
we draw also in Fig.~(\ref{fig:xi0-xi2-pot}) the behaviour of $\xi_0$
vs. $\xi_2$ very much in spirit of our study of
Sect.~\ref{sec:positive} on positive energy perturbations (see
Fig.~(\ref{fig:renorm}) ).

\subsection{Low energy expansion of the BC with a potential}

If we relax the condition $R\gg a$, we can undertake a low energy
expansion to get the set of differential equations for the low energy
scaled BC coefficients,
\begin{eqnarray}
R \frac{ d \xi_0 }{ d R} &=& \xi_0 ( 1- \xi_0) + U(R) R^2 \, , 
\label{eq:xi_0'} \\   
R \frac{ d \xi_2 }{ d R} &=& - \xi_2 (1+ 2 \xi_0) - 1  \, , 
\label{eq:xi_2'} \\   
R \frac{ d \xi_4 }{ d R} &=& - \xi_4 (3+ 2 \xi_0 )  -\xi_2^2 \, ,
\label{eq:xi_4'} 
\end{eqnarray} 
valid for any $R$. The r.h.s. are the corresponding beta functions of
the renormalization group flow. Note that unlike the standard RG
equations a manifest scale dependence shows up due to the potential
$U(R)$ which describes the long range physics. The fulfillment of
these equations guarantees cut-off independence of low energy
parameters, $\alpha$, $r_0$, $v_2$ and so on for any value of the
cut-off $R_S$. Moreover, these equations exhibit a natural hierarchy;
the solution of a given coefficient $ \xi_{2n} $ depends only on the
previous ones $\xi_{2n-2} , \dots , \xi_0$~\footnote{This is in
contrast to momentum space parameterizations of the short distance
potential via momentum dependent perturbations $ V_s (k', k) = C_0 +
C_2 (k^2 + k'^2) + \dots $ where operator mixing occurs. In practice
this means that if at zeroth order one keeps only the counterterm
$C_0$ its running is fixed by the regularization method and a
renormalization condition, most naturally fixing the scattering length
$\alpha$. When the new term $C_2$ is considered, the running of the
lower order counterterm $C_0 $ is modified and both $C_0$ and $C_2$
are intertwined, their values at a given cut-off being fixed by
$\alpha$ and the effective range. A way out is to consider energy
dependent perturbations instead of momentum dependent
ones~\cite{Birse:1998dk,Barford:2002je} and the hierarchy that we find
is recovered.}.  Furthermore, they are non-perturbative in the
potential and do not require any off-shell information, as in the
momentum space treatments~\cite{Barford:2002je}. Actually, the
solution for the first equation in the absence of a potential reads,
\begin{eqnarray}
\xi_0( R) = \frac{R}{R-\alpha}   \, ,     \qquad R \gg    a \, , 
\end{eqnarray} 
where $\alpha$ is the integration constant which can be identified
with the physical scattering length. In the presence of the potential
$U(R)$, this suggests a solution of the form
\begin{eqnarray}
\xi_0( R) = \frac{R}{R-\alpha_0(R)} \, ,
\end{eqnarray} 
where $\alpha(R)$ is an undetermined coefficient, satisfying
Eq.~(\ref{eq:valpha_0}) of Sect.~\ref{sec:vf}. We will show in the
next section that $\alpha_0 (R)$ is the scattering length
corresponding to the truncated potential $U(r)$ for $r < R$ and zero
otherwise. Actually, making use of the analogy we can solve the
equations, (\ref{eq:xi_0'}), (\ref{eq:xi_2'}) and (\ref{eq:xi_4'}) in
a more efficient manner, see Eqs.~(\ref{eq:valpha}), (\ref{eq:vr0})
and (\ref{eq:vv2}).

The set of equations (\ref{eq:xi_0'}), (\ref{eq:xi_2'}) and
(\ref{eq:xi_4'}) have to be solved with some initial conditions. For
asymptotically large distances we must have $\xi_0' =0$ , and hence
\begin{eqnarray}
\xi_0(\infty) &=& 1 \qquad ( \alpha \neq \infty ) \, , \\ 
\xi_0(\infty) &=& 0 \qquad ( \alpha = \infty ) \, .
\end{eqnarray} 
Thus, unless $\alpha=\infty $, all solutions go asymptotically to the
stable fixed point, according to the general theory. If $\alpha \gg a
$ then there is a region for $ r \sim a $ where $ \xi_0 \ll 1 $ and
$\xi_0 $ remains almost constant. Likewise for $\alpha \ll a $ we have
$ \xi_0 \sim 1 $ for $ r \ge a$. Finally, if $\alpha \sim a $ we have
$ \xi_0 \gg 1 $ only in the region $ r \sim a$.

If we solve around the fixed point $\xi_0 =1  $ we get 
\begin{eqnarray}
\xi_0(R) &=& 1 + R \int_R^\infty U(r) dr  \, .
\end{eqnarray} 
Going to $R \to \infty $ since already at $ R \sim a$ we may have 
\begin{eqnarray}
\xi_0(a) &=& \frac{a}{a-\alpha} \, .
\end{eqnarray} 
Thus, if $ a \gg \alpha $ we have $ \xi_0 (a) \sim 1$ and if $ a \ll
\alpha $ then $ \xi_0 (a) \sim 0 $, even if $\xi_0 (\infty) = 1$
(unless $\alpha=\infty$, in which case we do have $\xi_0(\infty) = 0$). 
Thus, taking the scale $R \sim a$ seems like a
good choice to be as close as possible to the fixed point situation
$\xi' \sim 0 $.

\subsection{Error estimates}
\label{sub:error_estimates}

\subsubsection{General Considerations}

Exact renormalization group invariance requires the knowledge of the complete 
phase shifts at all energies for the fulfillment of Eq.~(\ref{eq:renorm}),
but practical computations demand the use of a limited amount of physical 
information as input in order to have predictive power.
One example is a theory in which we know the value of the scattering length, 
$\alpha_0$, and consequently we want to fix its renormalization group flow, 
which can be obtained from the low energy limit of Eq.~(\ref{eq:del1}), 
yielding
\begin{eqnarray}
\frac{d\,\alpha}{d R} &=& \lim_{k \to 0}\,{\left(\frac{u_k(R,R)}{k}\right)}^2 
\nonumber\\ && \times \left[ L_0'(R) + {L_0(R)}^2 - U(R) \right] = 0 \, .
\end{eqnarray}
Thus, $\alpha_0(R) = \alpha_0(R_0)$ implies the fulfillment of 
Eq.~(\ref{eq:xi_0'}), but not of Eq.~(\ref{eq:xi_2'}) or higher order ones.
This theory can be shown equivalent to truncating the boundary condition 
(see Appendix~\ref{app:truncation}) in the $R \to 0$ limit
\begin{equation}
\lim_{R \to 0}\,L_k(R) = \lim_{R \to 0}\,L_0(R) \, ,
\end{equation}
with $L_0(R)$ fulfilling the RG equation
\begin{equation}
L_0'(R) + {L_0(R)}^2 = U(R)
\end{equation}
subjected to the asymptotic boundary condition 
$L_0(R) \to 1 / (R - \alpha_0)$ at $R \to \infty$.

We can try to improve the description of the phase shifts by making RG 
independent not only the scattering length $\alpha_0$, but also the 
effective range $r_0$, or even higher order parameters of the 
effective range expansion. In case we fix $\alpha_0$, $r_0$, $v_2$, $\dots$, 
$v_n$, we can write the boundary condition as
\begin{eqnarray}
L_k(R) &=& L_0 + k^2\,L_2 + \dots + k^{2 n}\,L_{2 n} + \dots
\end{eqnarray}
where $L_0$, $L_2$, $\dots$, $L_{2 n}$ obey their respective renormalization
group equations~\footnote{
The relation between the truncation of the boundary condition and the fixation
of low energy scattering observables, i.e. the different terms of the effective
range expansion, can be made clear by rewriting Eq.~(\ref{eq:del1}) as
\begin{eqnarray}
\frac{d}{d R}\,k\,\cot{\delta} &=& 
\left[ L_k'(R) + L_k^2(R) + k^2 - U(R) \right] \nonumber\\
&& \times\,u(R,R)^2 \, , \nonumber
\end{eqnarray}
which is obtained in just the same way as Eq.~(\ref{eq:del1}), but changing the
asymptotic normalization of $u(r,R)$ to
\begin{eqnarray}
u(r,R) \to \cos{k\,r} + k\,\cot{\delta(k,R)}\,\frac{\sin{k\,r}}{k} \, , 
\nonumber
\end{eqnarray}
from this it is obvious that fixing $\alpha_0$ is equivalent to the fulfillment
of Eq.~(\ref{eq:xi_0'}), fixing $\alpha_0$ and $r_0$ to fulfilling 
Eq.~(\ref{eq:xi_0'}) and Eq.~(\ref{eq:xi_2'}), and so on. 
}, while the higher order terms, represented by the dots, do not.

\subsubsection{Zeroth Order Truncation}

If we truncate the BC to zeroth order in the energy, i.e. Eq.~(\ref{eq:xi_0'})
is fulfilled while Eq.~(\ref{eq:xi_2'}) and higher order ones are not, then 
the scattering length is independent on the short distance cut-off $R$.
The truncation is made in order to fix the scattering length
of the system, while the effective range, the shape parameter and higher 
order terms of the expansion of $k\,\cot{\delta}$ are determined by both
the scattering length and the long range potential.

This situation is equivalent to take $L_k(R) = L_0(R)$, 
from which the cut-off dependence of the phase shift  can be deduced by 
making the substitution $L_k(R) \to L_0(R)$ in Eq.~(\ref{eq:del1}), 
yielding
\begin{equation}
\label{eq:LO-truncation}
\frac{d\,\delta}{d R} = - k^3\,{\left( \frac{u_k(R,R)}{k} \right)}^2 \, .
\end{equation}
For a regular potential, for which we have the behaviour $u_k(R,R)/k
\sim c_0 + c_1\,R$ for small cut-offs, the phase shift shows a linear
dependence in the cut-off radius which slope increases with the
momentum $k$, so for small cut-off radii we have
\begin{equation}
\frac{d\,\delta}{d R} \sim - k^3\,c_0^2
\end{equation}
and consequently we find a linear dependence of the phase shift with respect
to the cut-off radius, $\Delta\,\delta \sim R$.

For an attractive singular potential, which behaves as $U(R) \sim
-1/R^n$, the situation changes, since the wave function is very much
suppressed at short distances, $u_k(R,R)/k \sim R^{n/4}$ as compared
to the regular potential case.  Then we obtain, for small cut-offs,
the behaviour
\begin{equation}
\frac{d\,\delta}{d R} \sim - k^3\,R^{n/2} \, ,
\end{equation}
from which one can deduce a cut-off error $\Delta\,\delta \sim R^{n/2 + 1}$,
and we recover renormalizability, i.e. cut-off independence, 
in the small cut-off limit
\begin{equation}
\lim_{R \to 0}\,\frac{d\,\delta}{d R} = 0 \, .
\end{equation}

\begin{figure*}[]
\begin{center}
\epsfig{figure=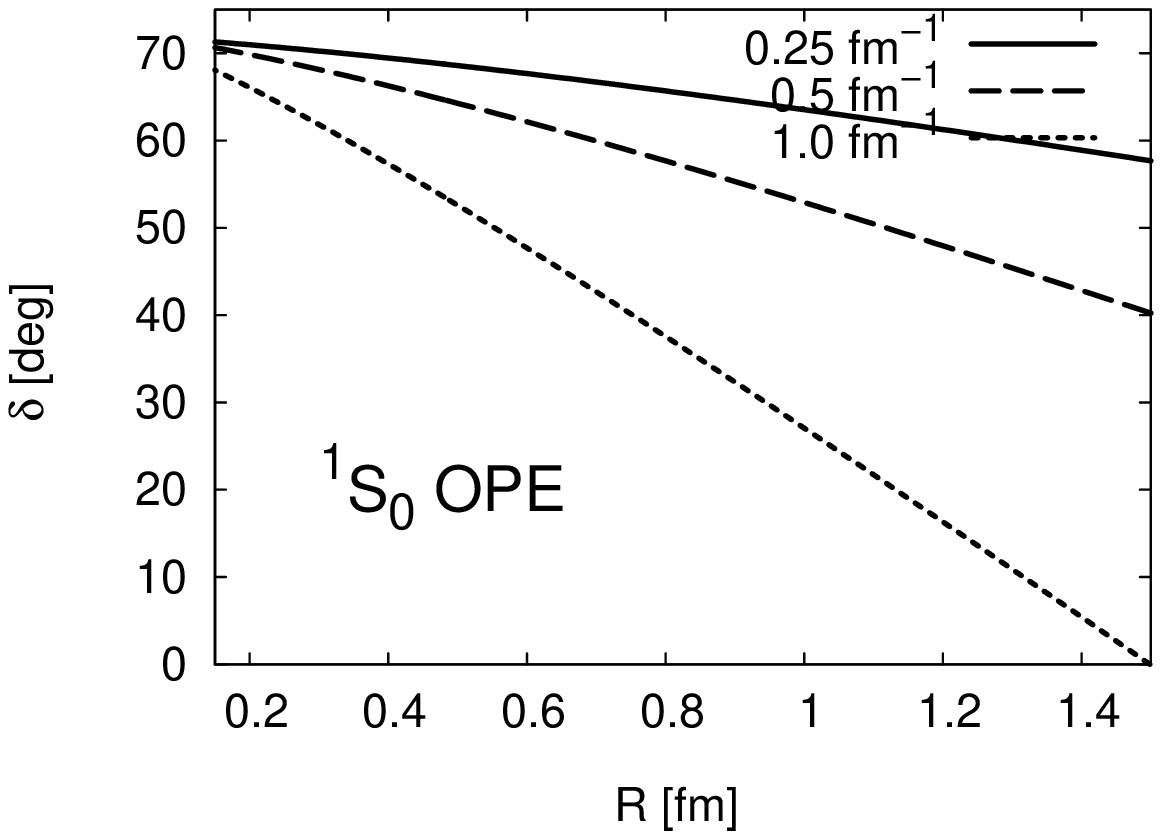,height=5.5cm,width=5.5cm}
\epsfig{figure=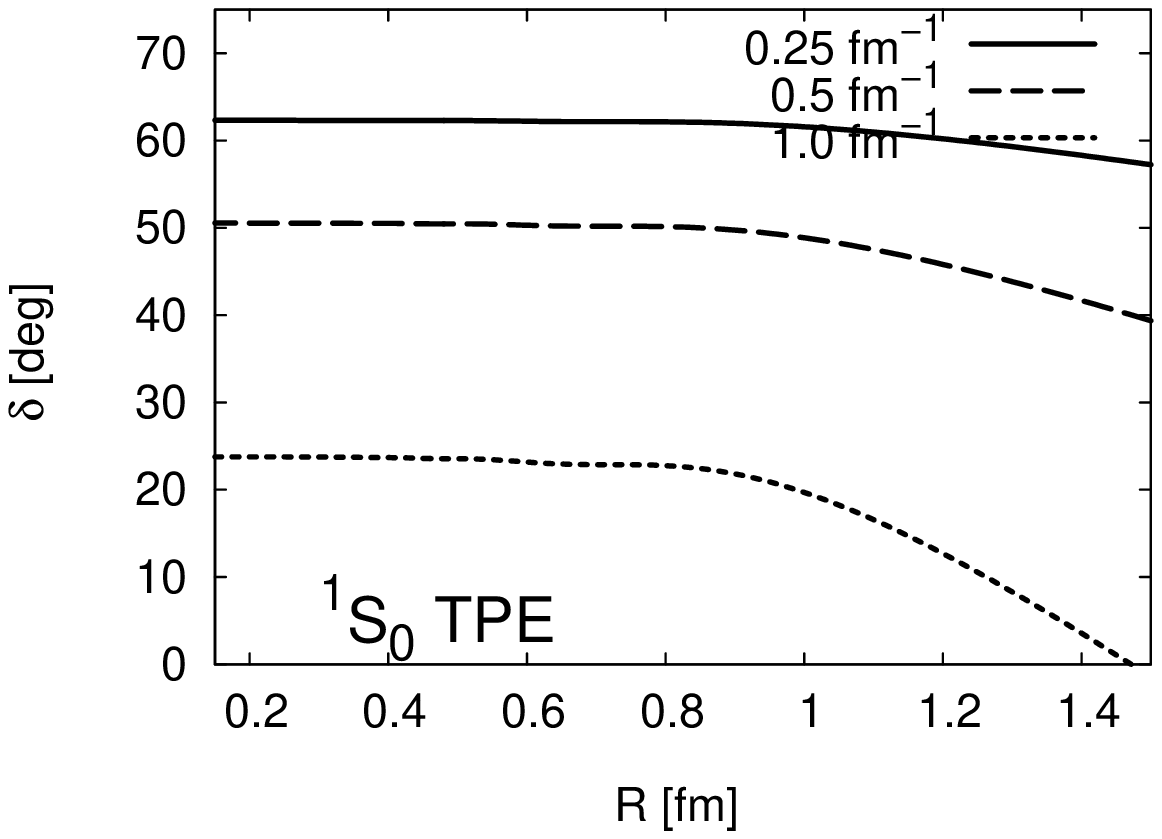,height=5.5cm,width=5.5cm}
\end{center}
\caption{
Cut-off dependence of the phase shifts for NN scattering in the $^1S_0$ channel
with the OPE and TPE potentials, which behave as $1/R$ and $1/R^6$ 
respectively. For computing the phase shift we use a energy independent 
boundary condition fulfilling the equation $L_0'(R) + L_0^2 (R) - U(R) = 0$,
which in turns mean that the scattering length is independent on the cut-off
(we take $\alpha_0 = -23.74\,{\rm fm}$). We show the phase shifts for
center of mass momenta of $k_{cm} = 0.25, 0.5, 1.0\,{\rm fm}^{-1}$.
}
\label{fig:delta_Rs}
\end{figure*}

This cut-off dependence is depicted in Fig.~(\ref{fig:delta_Rs}), in
which we show the nucleon-nucleon phase shift in the $^1S_0$ channel
as a function of the cut-off $R$ for various center-of-mass
momenta. The phase shift $\delta\,(k,R)$ is computed using an energy
independent boundary condition which turns out to be equivalent to
fixing the scattering length to some given value, as for example the
experimental one $\alpha_0 = -23.74\,{\rm fm}$, for both the one and
two pion exchange potentials.  The OPE potential is just an usual
Yukawa potential, while the TPE potential behaves as $ - a^4 / R^6$
for distances smaller than the pion Compton wavelength. As we can see
in the figures, in the OPE case the phase shift show a linear
dependence in $R$ near the origin, while in the TPE case the phase
shift becomes insensitive to the cut-off much earlier.

Finally, for a repulsive singular potential behaving as $U(R) \sim
1/R^n$, the phase shifts shows an exponential behaviour as the cut-off
is removed.  This can be understood from the behaviour of the wave
function at short distances (see Appendix~\ref{app:singular})
\begin{eqnarray}
\frac{u_k(R,R)}{k} \sim R^{n/2}\,e^{+ (a / R)^{n/2-1} / (n/2-1)} \, ,
\end{eqnarray}
where we have taken the irregular solution since for an arbitrary
scattering length this is the UV fixed point and the regular solution
may be discarded. Then, for small cut-off
\begin{equation}
\label{eq:LO-repulsive}
\frac{d\,\delta}{d R} \sim - k^3\,R^{n/2}\,e^{+ 2 (a / R)^{n/2-1} / (n/2-1)} 
\, ,
\end{equation}
from which one can see that the phase shift develops a very strong
cut-off dependence when one tries to fix the scattering length to an
arbitrary value, unlike the previous cases.

\subsubsection{Relation to Orthogonality Constraints}

The truncation of the boundary condition at zeroth order is equivalent to
the fulfillment of orthogonality relations between different energy solutions,
which was used in Refs.~\cite{PavonValderrama:2005gu,Valderrama:2005wv,PavonValderrama:2005uj} 
to renormalize and obtain model-independent predictions for the $^1S_0$ 
singlet and $^3S_1-{}^3D_1$ triplet (deuteron) channels.
Orthogonality constraints imply that different energy wave functions
fulfill the integral relation
\begin{equation}
\int_{0}^{\infty}\,u_k(r) u_{k'}(r)\,dr = \delta\,(k-k') \, .
\end{equation}
If we use the corresponding Lagrange identity between $u_k$ and $u_k'$, we can
rewrite the previous orthogonality constraint as
\begin{equation}
(k^2 - k'^2)\,\int_{0}^{\infty}\,u_k(r) u_{k'}(r)\,dr = 
{u'}_k\,u_{k'} - u_k {u'}_{k'} \Big|_0 \, ,
\end{equation}
so for $k' \neq k$, and including a short distance cut-off $R$, we obtain
\begin{equation}
\frac{{u'}_k(R)}{u_k(R)} = \frac{{u'}_{k'}(R)}{u_{k'}(R)} \, .
\end{equation}
Finally, taking $k' = 0$ as a reference state
\begin{equation}
\frac{{u'}_k(R)}{u_k(R)} = \frac{{u'}_{0}(R)}{u_{0}(R)} \, ,
\end{equation}
or, equivalently, $L_k(R) = L_0(R)$, i.e., imposing orthogonality between
different energy solutions is equivalent to truncate the boundary condition at
zeroth order.

\subsubsection{Higher Order Truncations and Inconsistencies}

Naively one can think that a truncated energy expansion of the
boundary condition would lead to a systematically more accurate
description of the data. In this section we will analyze this problem
and show cases where this naive expectation is not fulfilled. 

To begin with we can try to truncate the boundary condition, for
example, to second order in the energy 
\begin{eqnarray}
L_k(R) = L_0(R) + k^2\,L_2(R) \, ,
\end{eqnarray}
so Eqs.~(\ref{eq:xi_0'}) and ~(\ref{eq:xi_2'}) are fulfilled while 
Eq.~(\ref{eq:xi_4'}) and higher order ones are not.
In such a case both the scattering length and the effective range are cut-off
independent, while the phase shift is not. In a similar fashion to the previous
case, the cut-off dependence of the phase shift can be estimated, yielding
\begin{equation}
\frac{d\,\delta}{d R} 
= - k^5\,L_2(R)^2\,{\left( \frac{u_k(R,R)}{k} \right)}^2 \, ,
\label{eq:NLO-truncation}
\end{equation}
which can be easily computed for regular potentials.  For such potentials  
we can take into account that (see Sect.~\ref{sec:bcr})
\begin{eqnarray}
\lim_{R \to 0}\,L_k(R) &=& k\,\cot{\delta_S(k)} \nonumber\\
&=& - \frac{1}{\alpha_S} + \frac{1}{2}\,r_S\,k^2 + \dots 
\end{eqnarray}
being $\alpha_S$ and $r_S$ the short-range scattering length and
effective range. Inserting the previous expression on
Eq.~(\ref{eq:NLO-truncation}), we arrive to the following result
\begin{eqnarray}
\lim_{R \to 0}\,\frac{d\,\delta}{d R} 
= - k^5\,{\left( \frac{r_S}{2} \right)}^2\,
{\left( \frac{u_k(0,0)}{k} \right)}^2 \neq 0 \, .
\end{eqnarray}
This means that by adding more terms in the energy expansion of the
boundary condition, i.e. fixing higher order parameters in the
effective range expansion, we will find the same linear dependence as
in the zeroth order case of the phase shift on the cut-off $R$ in the
ultraviolet limit.  With each new order in the energy expansion, the
slope of $\delta\,(k,R)$ in the $R \to 0$ limit is progressively
suppressed in energy with respect to the zeroth order case, e.g. by a
factor of $(k\,r_S)^2$ at second order, $k^4\,r_S\,v_S$ at fourth
order, etc.

\begin{figure*}[]
\begin{center}
\epsfig{figure=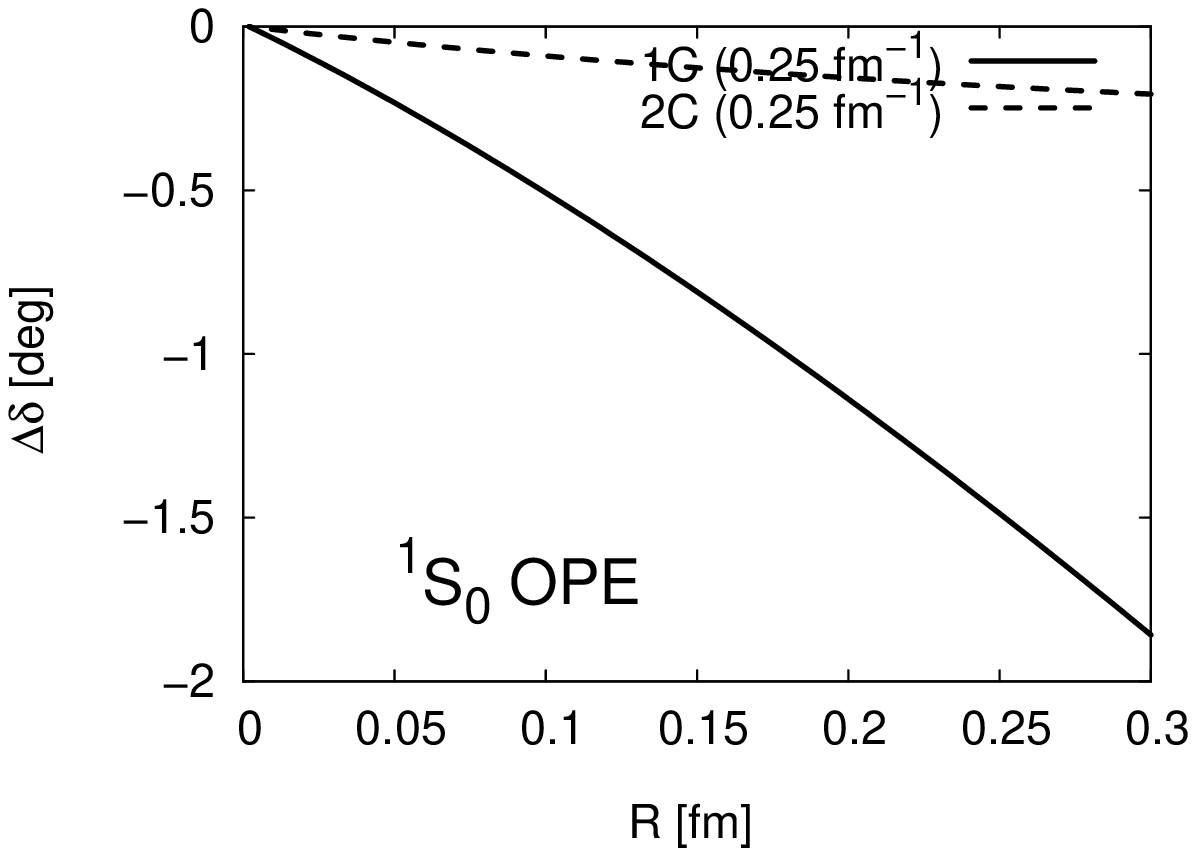,height=5.5cm,width=5.5cm}
\epsfig{figure=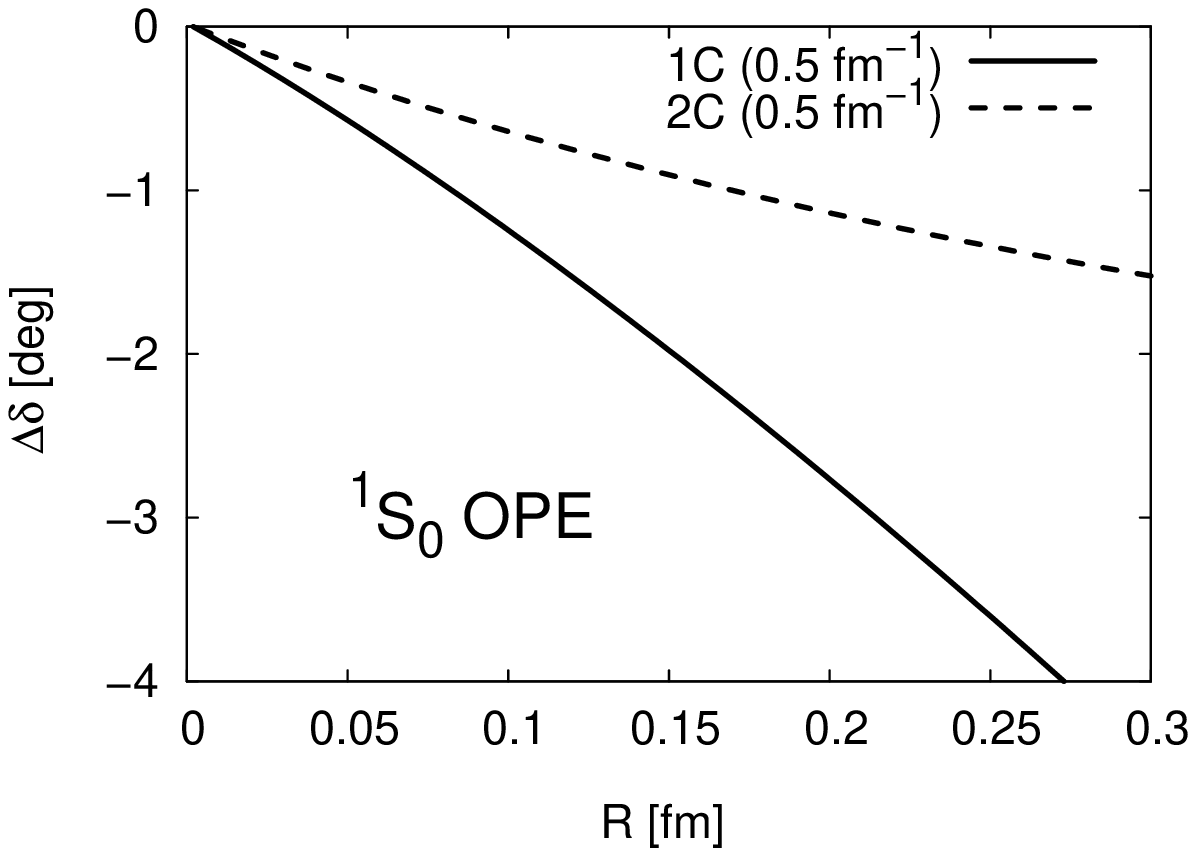,height=5.5cm,width=5.5cm}
\epsfig{figure=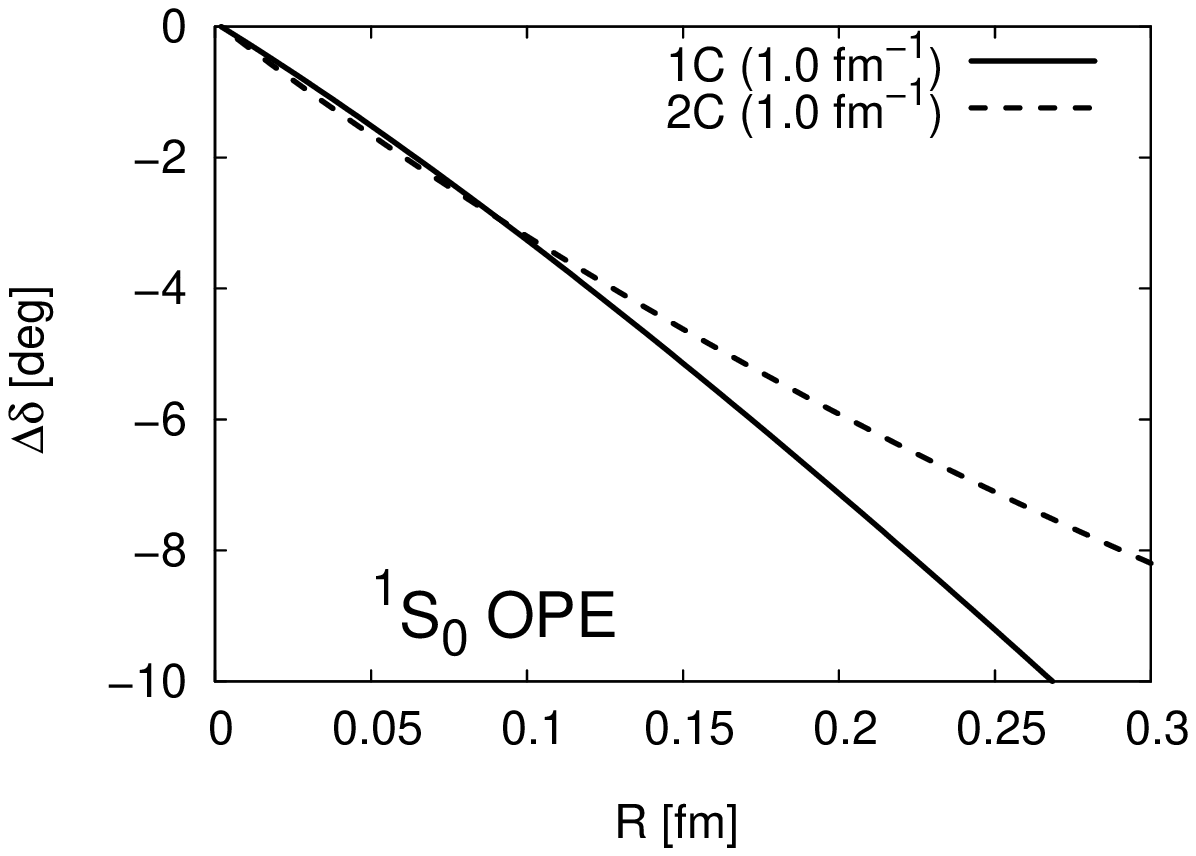,height=5.5cm,width=5.5cm}
\end{center}
\caption{
Cut-off dependence of the phase shifts for NN scattering in the $^1S_0$ channel
with the OPE potential when we fix one (1C curve) or two (2C curve) parameters,
namely the scattering length and the effective range (we take 
$\alpha_0 = -23.74\,{\rm fm}$ and $r_0 = 2.77\,{\rm fm}$). 
We show the convergence of the phase shifts for center of mass momenta of 
$k_{cm} = 0.25$, $0.5$, $1.0\,{\rm fm}^{-1}$ (left, central and right panel
respectively). We can see that the 2p curve is more convergent than the 1p
curve for $0.25$ and $0.5\,{\rm fm}^{-1}$, while only slightly worse for 
$1.0\,{\rm fm}^{-1}$.
}
\label{fig:delta_Rs_cm}
\end{figure*}

One example of this energy suppression is given by the OPE potential
in the $^1S_0$ channel, for which $r_S = 4.46\,{\rm fm}$ (see
Sect.~\ref{subsec:vf-tp}).  In this case, the cut-off dependence is in
principle smoothened for $k < 0.45\,{\rm fm}^{-1}$ ($\sim 88\,{\rm
MeV}$), indeed a bit more due to the higher order terms $k^7$, $k^9$,
... which have been ignored here.  The cut-off dependence for
different energies is depicted in Fig.~(\ref{fig:delta_Rs_cm}), both for
the case where $\alpha$ and both $\alpha$ and $r_0$ parameters are
fixed.  As we can see, the effective cut-off dependence is still a bit
smoother for two parameters at $k = 0.5\,{\rm fm}^{-1}$, and at $k =
1.0\,{\rm fm}^{-1}$ is only slightly worse than with one parameter.

In the case of attractive singular potentials the renormalization group 
analysis becomes more complex. The first step is the evaluation of $L_2$,
which is computed in Appendix~\ref{app:truncation}, yielding for $R \to 0$
\begin{eqnarray}
L_2 (R) \to
{\left( \frac{\alpha_0}{u_0 (R,R)}\right)}^2 \frac{\Delta\,r_0}{2} \, ,
\end{eqnarray}
where $\Delta\,r_0$ is the short distance contribution to the effective 
range~\footnote{Thus, the total effective range of the system is 
$$r_0 = \Delta\,r_0 + r_0(\alpha_0)$$
where $r_0(\alpha_0)$ is the effective range due to the zero energy boundary
condition used to fix the scattering length of the system.
It can be computed with the well known integral formula
$$r_0(\alpha_0) = 2\,\int_{0}^{\infty}\,dr\,\left[\hat{v}_0^2(r) - 
\hat{u}_0^2(r)\right]\, ,$$ where $\hat{v}_0(r) = 1 - r/\alpha_0$ and 
$\hat{u}_0(r)$ is the solution of the zero energy reduced Schr\"odinger 
equation with the asymptotic normalization $\hat{u}_0(r) \to \hat{v}_0(r)$ 
for large distances.}, and the zero energy wave function is defined by
\begin{eqnarray}
u_0(R,R) = \lim_{k \to 0}\,\frac{u_k(R,R)}{k} \, .
\end{eqnarray}
The similarity with the $L_2(R) \to r_S / 2$ result for regular potentials 
is striking. In fact the formula above can be used to obtain $r_S$ for a 
regular potential (for which $u_0(R,R)$ goes to a constant value at 
short distances).

\begin{figure*}[]
\begin{center}
\epsfig{figure=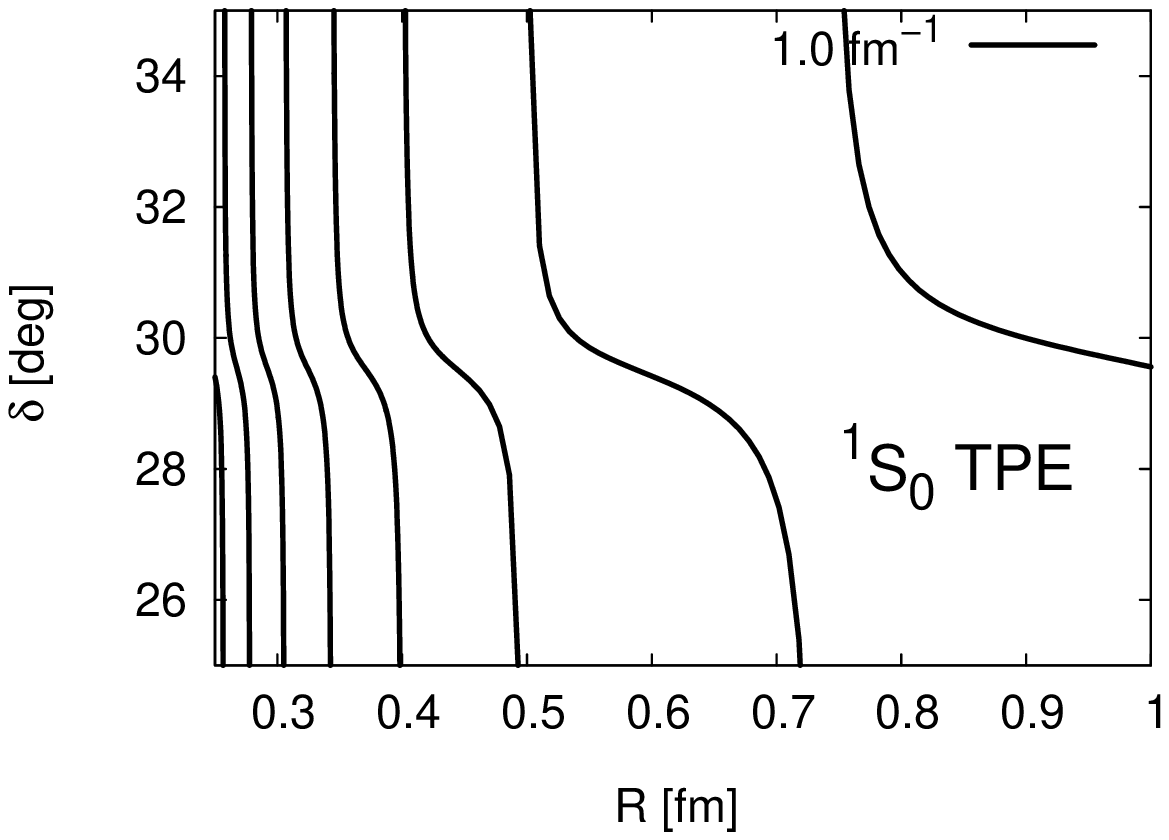,height=5.5cm,width=5.5cm}
\epsfig{figure=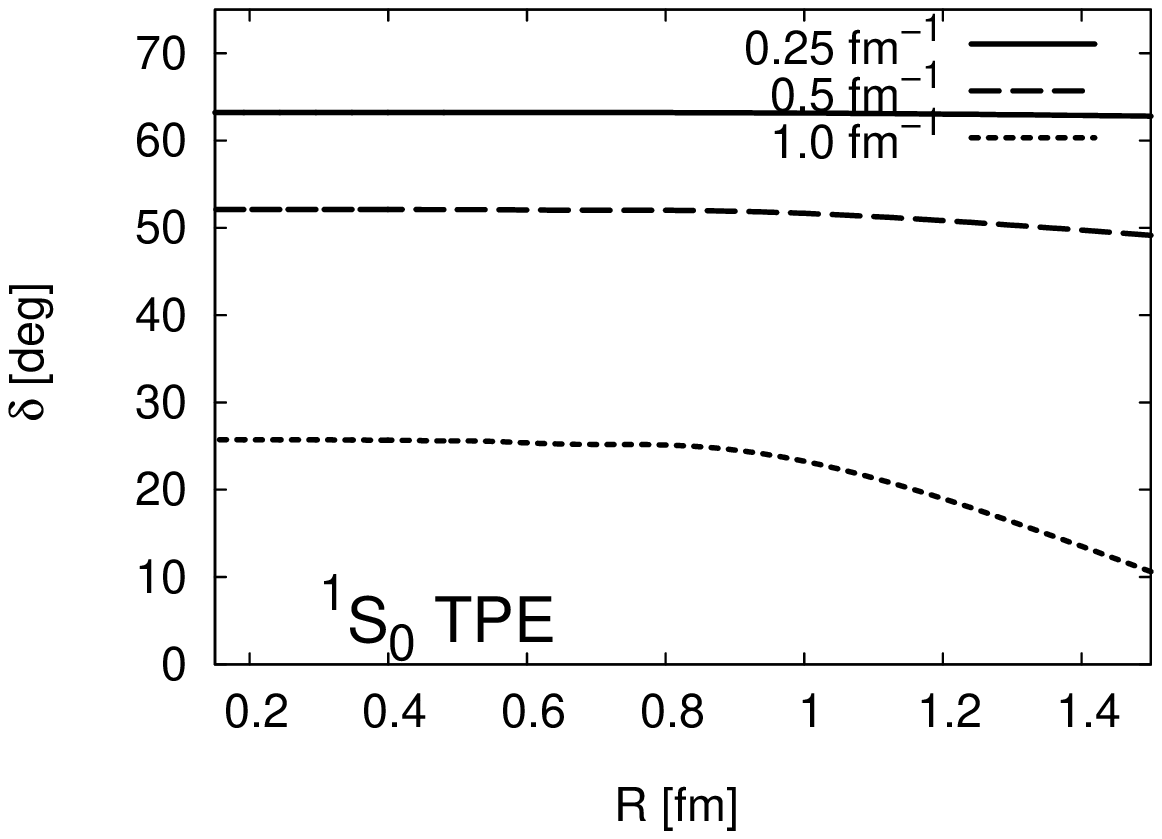,height=5.5cm,width=5.5cm}
\end{center}
\caption{
Cut-off dependence of the phase shifts for NN scattering in the $^1S_0$ channel
with TPE potential (an attractive singular potential) when fixing the 
scattering length and effective range of the system.
In the left panel, the boundary condition has been truncated at order $k^2$
in the momentum expansion, $L_k = L_0 + k^2\,L_2$, thus giving rise to cycles
due to the incorrect renormalization procedure used.
In the right panel, the boundary condition is expressed as a Pad\'e 
approximant, such that we recover renormalizability in the $R \to 0$ limit
(we take $\alpha_0 = -23.74\,{\rm fm}$ and $r_0 = 2.77\,{\rm fm}$). 
For the convergent case, we show the phase shifts for center of mass 
momenta of $k_{cm} = 0.25, 0.5, 1.0\,{\rm fm}^{-1}$.
}
\label{fig:delta_2s_Rs}
\end{figure*}

Now we can evaluate the convergence of the phase shift with respect to 
the cut-off for $L_k = L_0 + k^2\,L_2$, which is given by 
\begin{eqnarray}
\lim_{R \to 0}\,\frac{d\,\delta}{d R} &=& 
- k^5\,{\left( \frac{\Delta r_0}{2} \right)}^2\,
{\left( \frac{\alpha_0^2}{u_0(R,R)} \right)}^2 \neq 0 \, .
\nonumber\\
\end{eqnarray}
This behaviour is depicted in Fig.~(\ref{fig:delta_2s_Rs}) and at
first sight looks weird. With every zero of the wave function there is
a $\pi$ jump in the phase shifts, meaning that the system is
explicitly sensitive to the appearance of deeply bound states,
something that should not have practical consequences within the
domain of applicability of a correctly formulated effective
theory. There is no obvious and unique $R \to 0$ limit.

The previous result seems to suggest that one cannot fix both
$\alpha_0$ and $r_0$ and obtain renormalized phase shifts. However, a
further analysis below will show that it is indeed possible and the
results are well defined and unique.~\footnote{In fact, we can consider for
example the shape parameter $v_2$, which has the following integral
definition
\begin{eqnarray}
v_2 = \int_{0}^{\infty}\,dr\,
\left[ \hat{v}_0\,\hat{v}_2 - \hat{u}_0\,\hat{u}_2 \right]\, , \nonumber 
\end{eqnarray}
where $\hat{v}_0(r) = 1 - r / \alpha_0$ and 
$\hat{v}_2(r) = (r^2 - 3\,\alpha_0 r + 3\,\alpha_0\,r_0)\,r / 6 \alpha_0$,
while $\hat{u}_0$ and $\hat{u}_2$ are solutions of Eqs.(\ref{eq:u0}) and 
(\ref{eq:u2}) respectively, subjected to the asymptotic boundary conditions
$\hat{u}_0 \to \hat{v}_0$ and $\hat{u}_2 \to \hat{v}_2$.
Since the solutions for attractive singular potentials behave at short
distances as $r^{n/4}$ times some trigonometric function, $v_2$ is convergent.
A similar argument can be applied to $v_3$, $v_4$, and so on, and even to the 
phase shifts, thus giving a well defined limit when the cut-off is removed.}

Instead of a naive truncation of the boundary condition as we have
done, we suggest to reorder the expansion as follows
\begin{eqnarray}
L_k(R) = L_0(R) + k^2\,L_2(R) + k^4\,\Delta_4(R) + \dots
\end{eqnarray}
where $L_0$ and $L_2$ fulfill exactly the RG equations, but higher
order terms do not (and hence the notation $\Delta_4$, $\Delta_6$,
etc, for these terms).  In such a case we can see that the scale
dependence of the phase shift is given by
\begin{eqnarray}
\frac{d\,\delta}{d R} &=& 
- k^5\,\left[ L_2(R)^2 + 2\,L_0\,\Delta_4 + \Delta_4'(R) \right]
\nonumber\\
&& \times\,{\left( \frac{u_k(R,R)}{k} \right)}^2 \, .
\label{eq:NLO-correct-truncation}
\end{eqnarray}
Then, analyzing the behaviour of $L_0$, $L_2$ and $\Delta_4$
(see Appendix~\ref{app:truncation}), we find that
\begin{eqnarray}
\label{eq:NLO-correct-dependence}
\frac{d\,\delta}{d R} 
&=& - k^5\,u_0(R,R)\,u_2(R,R) + {\cal O}(k^7) \nonumber\\
&\sim& - k^5\,R^{n/2} \, ,
\end{eqnarray}
which means that the convergence pattern that emerges when we fix both
$\alpha$ and $r_0$ is similar to that found when we fixed $\alpha_0$
only, i.e. there is no improvement on the short distance scaling
suppression. The same arguments, but considering $\Delta_6$,
$\Delta_8$, etc, can be applied to higher orders in the momentum
expansion of the phase shifts yielding identical short distance
scaling suppression. 

The nontrivial fact is that fixing both the scattering length and the
effective range is not exactly equivalent to any truncation of the
boundary condition $L_k(R)$.  All the terms in the energy expansion
are equally singular, and therefore equally relevant. The correct
parameterization of the truncated boundary condition corresponds to a
unique Pad\'{e} approximant looking representation of the boundary
condition
\begin{eqnarray}
\label{eq:NLO-Pade}
L_k(R) = \frac{u'_0 + k^2\,u'_2}{u_0 + k^2\,u_2} + k^4\,\Delta(R) \, ,
\end{eqnarray}
where $\Delta(R)$ is a remainder. Indeed, for short distances, the
Pad\'{e} behaves as $1/R^{n/2}$, while the remainder $\Delta(R) \sim
R$, and hence can be safely ignored when the cut-off is removed. This
expression for $L_k(R)$ gives an $R^{n/2 + 1}$ UV scaling for the
phase shift, regardless on how many low energy parameters $\alpha$,
$r_0$, $v_2$ and so on are fixed. The form of this Pad\'{e}, which
could be visualized as a renormalization group improvement is driven
by the Moebius bilinear transformation, Eq.~(\ref{eq_moebius}) which
embodies the dilatation group properties of the short distance cut-off
given in Eq.~(\ref{eq:cut-off-group}). A different Pad\'{e} approximant
will lead to spurious cut-off dependences, which will jeopardize the
$R \to 0$ scaling behaviour. 

\subsubsection{Remarks on Regular Solutions}

The RG behaviour of regular solutions represents a very special case. 
In it we fix the scattering observables to the values corresponding
to the regular solution at the origin. By doing this the convergence
with respect to the cut-off is improved noticeably.

The first case we are going to consider is the trivial one, that of a
regular potential in which we do not fix anything, but rather enforce
the regular solution $u(R,R) = c_0 R$ at a small cut-off radius
$R$. For such a case we have the cut-off dependence 
\begin{eqnarray}
\frac{d\,\delta}{d R} = k\,[U(R) - k^2]\,
{\left( \frac{u_k(R,R)}{k} \right)}^2\, ,
\end{eqnarray}
since $L_k(R) = 1 / R$. In it $R$ can be interpreted as the starting 
integration point in any usual integration procedure for the 
Schr\"{o}dinger equation.
For a potential which goes to a constant value at the origin, $U(R) \to U_0$,
we have $\Delta \delta \sim R^3$, while for a Coulomb-like potential at short
distances, like the Yukawa potential, we have $\Delta \delta \sim R^2$.

When one fixes the scattering length to the regular value, the
convergence rate is given by Eq.~(\ref{eq:LO-truncation}). In such a
case, there would be a small improvement for a Yukawa potential, which
will now converge as $\Delta \delta \sim R^3$.  These results should
be compared with the convergence when fixing the scattering length to
an arbitrary value, which is $\Delta \delta \sim R$.  When one
additionally fixes the effective range, Eq.~(\ref{eq:NLO-truncation})
describes the convergence rate. In case the effective range is taken
to be the regular value, we have that the short distance effective
range is zero, $r_S = 0$; $L_2(R)$ can be estimated, for example, by
the Green function methods used in Appendix~\ref{app:singular}, giving
$L_2(R) \sim R$.  Then we have $\Delta \delta \sim R^5$, in accordance
with naive expectations that increasing the number of counterterms
will smoothen the cut-off dependence. Further suppressions will take
place at higher orders.

The second case is when the potential is repulsive singular. As we saw
previously, setting the scattering length generates an exponentially
divergent cut-off dependence in the ultraviolet limit, given by
Eq.~(\ref{eq:LO-repulsive}).  The only way in which one can obtain
renormalization invariance with a repulsive singular potential is by
taking the scattering length corresponding to the regular solution at
the origin, for which we obtain a rapidly convergent result for the
phase shift
\begin{equation}
\frac{d\,\delta}{d R} \sim - k^3\,R^{n/2}\,e^{-2 (a / R)^{n/2-1} / (n/2-1)}\, ,
\end{equation}
in which the sign of the exponential has changed with respect to 
Eq.~(\ref{eq:LO-repulsive}), since we are taking the regular 
solution of a $1/R^n$ potential.
When we fix the regular effective range via a truncation, we get the cut-off
dependence found in Eq.~(\ref{eq:NLO-truncation}). Since $L_2(R) \sim R$
for regular solutions, we get an extra $R^2$ suppression over the original
one with one counterterm. Instead, if we use a Pad\'e approximant,
Eq.~(\ref{eq:NLO-Pade}), we will get the convergence pattern of 
Eq.~(\ref{eq:NLO-correct-dependence}), thus giving an extra 
$R^{n/2+1}$ suppression.

The last case to consider is the one of an attractive singular
potential.  For it there is not a unique regular solution; the proper
combination may be fixed by fixing the scattering length. Then the
situation looks similar to the repulsive singular potentials. If we
fix the effective range to the value obtained when fixing the
scattering length, we will obtain an extra $R^{n/2+1}$ or $R^2$
suppression depending on the usage or not of the Pad\'{e} described in
Eq.(\ref{eq:NLO-Pade}).

\subsubsection{Overview of the Error Estimates and 
its Relation to Power Counting}

We can summarize the previous discussion about the truncation of the boundary
condition in the following points

\begin{itemize}
\item For an attractive singular potential it is necessary to fix at least
one scattering observable in order to obtain a finite phase shift in the 
$R \to 0$ limit.

\item For a repulsive singular potential one cannot fix any scattering 
observable and at the same time obtain a finite phase shift in 
the $R \to 0$ limit. This means that one is led to an alternative: 
either doing a finite cut-off computation or choosing the regular
solution and remove the cut-off.

\item For a regular potential one can choose either to fix any scattering
observable, and hence take the irregular solution, or not.

\item Fixing more scattering observables do not lead to a further (qualitative)
short distance suppression, but a more singular potential (if attractive) 
improves the convergence.
\end{itemize}

At this point we should stress the close connection between the short
distance suppression when one fixes a certain number of scattering
observables and the issue of power counting in effective theories.
For posing the discussion in a more standard language, we will refer
to the number of physical observables we are fixing as the number of
independent counterterms in the theory. It should be noted that fixing
physical observables is not exactly the same as setting counterterms.
Counterterms are not observables, but they represent the unknown short
distance potential as a low energy expansion in terms of the delta
function and its derivatives. On the other hand, setting physical
observables is the visible effect for the need of
counterterms.~\footnote{Actually, in momentum space one may find
redundant counterterms appearing in the short distance potential at
fourth order in momentum $V_s (k', k) = C_0 + C_2 (k^2 + k'^2) + C_4
(k'^4 + k^4 ) + D_4 k^2 k'^2 + \dots $. Obviously, $C_4$ and $D_4 $
are redundant or else do not correspond to the same order.  This
actually shows that there are may appear more counterterms than
renormalization conditions.}

Although there are no clear rules in the literature, we will take the
point of view that (i) the cut-off should be removed, (ii) it is the
cut-off dependence what drives the construction of an acceptable power
counting, and (iii) the long range potential is going to be treated
non-perturbatively at any order (for a singular potential this is
absolutely necessary).  These assumptions have been adopted in our
previous
works~\cite{PavonValderrama:2005gu,Valderrama:2005wv,PavonValderrama:2005uj}
but it is fair to mention that they are not universally agreed
upon. (i) is for example accepted in Ref.~\cite{Nogga:2005hy}, while
rejected in
Refs.~\cite{Epelbaum:2003gr,Epelbaum:2003xx,Epelbaum:2004fk}, (ii) is
explicitly used in the RG analysis of
Birse~\cite{Birse:1998dk,Barford:2002je,Birse:2005um}, and in
Refs.~\cite{Beane:2001bc,Nogga:2005hy} for promoting certain
counterterms, and (iii) is accepted in
Refs.~\cite{Epelbaum:2003gr,Epelbaum:2003xx,Epelbaum:2004fk} (the same
which rejected (i)), but not in Ref.~\cite{Nogga:2005hy}, in which it
is advocated the perturbative treatment of the potential beyond LO,
while LO remains non-perturbative (although no actual computation is
done for NLO or NNLO). The amazing aspect of all these disagreements
is that there is no operational definition of what would be a valid
criterion and discrepancies are utterly based on the favourite
prejudices of different authors (including ourselves). Here we are not
going to discuss the validity or convenience of these assumptions
(this has been partially done in our previous
work~\cite{PavonValderrama:2005gu,Valderrama:2005wv,PavonValderrama:2005uj})
but looking instead for their consequences.

The first consequence of our previous analysis is that the long range
potential and the counterterms (or short range potential) are not
independent: the singularity structure of the potential determines
whether counterterms should be included or not.  Only in the very
special case of long range regular potentials are the counterterms
independent (see also
\cite{PavonValderrama:2005gu,Valderrama:2005wv,PavonValderrama:2005uj}).
This disagrees with power counting schemes based on naive dimensional
analysis, such as Weinberg's, in which all the counterterms with
appropriate dimensions are included in the computation, regardless on
the structure of the long range potential.  But if the cut-off is to
be removed, one must always include a counterterm if the long range
potential is singular attractive (in agreement with the previous
findings of Ref.~\cite{Beane:2000wh} and the results of
Ref.~\cite{Nogga:2005hy}) and one cannot include any counterterm at
all in case it is singular repulsive

From a naive viewpoint, it might seem counterintuitive that for a
repulsive singular potential no counterterm can be included, but as we
discuss now it is indeed quite natural.  If we consider a singular
repulsive potential with a characteristic long distance scale $a$,
then the physics associated with a short distance scale $a_S$, such
that $a_S \ll a$, would not affect at all the long distance physics,
since the long distance potential itself would act as a potential
barrier which destroys any effect coming from scales $a_S$ smaller
than $a$. Then, there must be a strong short distance insensitivity
which manifests itself as the dominance of the regular solution and
thus a lack of counterterms when the cut-off is removed. 

The issue of repulsive singular potentials raises a unexpected
consequence for the power counting of the long distance potential: if
the potential is singular attractive at a given order, it should remain
singular attractive at higher orders. If the long distance potential
between two particles is known to be attractive, the effective theory
should reproduce this feature to all orders in the expansion of the
potential (if this is going to be used in any non-perturbative
computation). Of course the full interaction between two particles
cannot truly be singular attractive at all distances; if this were to
be the case the system would collapse. However, in the effective
theory it is the unknown short distance physics which are repulsive,
and these interactions are not explicitly modelled in effective
theories, but implicitly via counterterms.

Another important issue which arises from the previous RG analysis
regards the number of counterterms and/or renormalization conditions
one should include when one has an attractive singular potential.  As
we have seen, the qualitatively power law cut-off scaling behaviour
does not depend at all on the number of counterterms included in the
computation, but only on the power law divergence of the potential
near the origin. Although this seems to suggest that there is no
reason for adding counterterms beyond the first one, when we look at
the quantitative cut-off dependence we can see that the situation may
change. One clear example is given by the TPE potential at NNLO in the
$^1S_0$ singlet channel. If we look at how much the phase shifts
change between the cut-offs $R = 0.15\,{\rm fm}$ and $R = 1.5\,{\rm
fm}$ at a center of mass momentum $k = 1.0\,{\rm fm}^{-1}$, we see
\begin{eqnarray}
{|\Delta\,\delta(k)|}_{1C} &\simeq& {25.21}^{o} \\ 
{|\Delta\,\delta(k)|}_{2C} &\simeq& {15.11}^{o} 
\end{eqnarray} 
where the subscripts $1C$ and $2C$ refer to the number of counterterms used.
As one can appreciate, there is a noticeable improvement of the convergence
when a second counterterm is added, and which can justify its 
inclusion~\footnote{As a matter of fact, it is interesting to notice that
a third counterterm (to fix $v_2 = -0.48\,{\rm fm}$) does not improve at
all the convergence, but worsens it 
(${|\Delta\,\delta(k)|}_{3C} \simeq {56.50}^{o}$).}.
It should be noted that this improvement is not solely related with the 
second counterterm, but also with the fact that the effective range 
prediction when one fixes the scattering length, 
which is $r_0(\alpha_0) = 2.86\,{\rm fm}$, 
is very close to the experimental value, $r_0 = 2.77\,{\rm fm}$, 
to which we fix the second counterterm.
Taking into account that the convergence when fixing two counterterms depends 
on the $\Delta\,r_0$ needed to correctly fix the effective range
(if it were zero we would have an $R^7$ convergence pattern instead $R^4$)
it is not a surprise that there is dramatic improvement.

\begin{figure*}[]
\begin{center}
\epsfig{figure=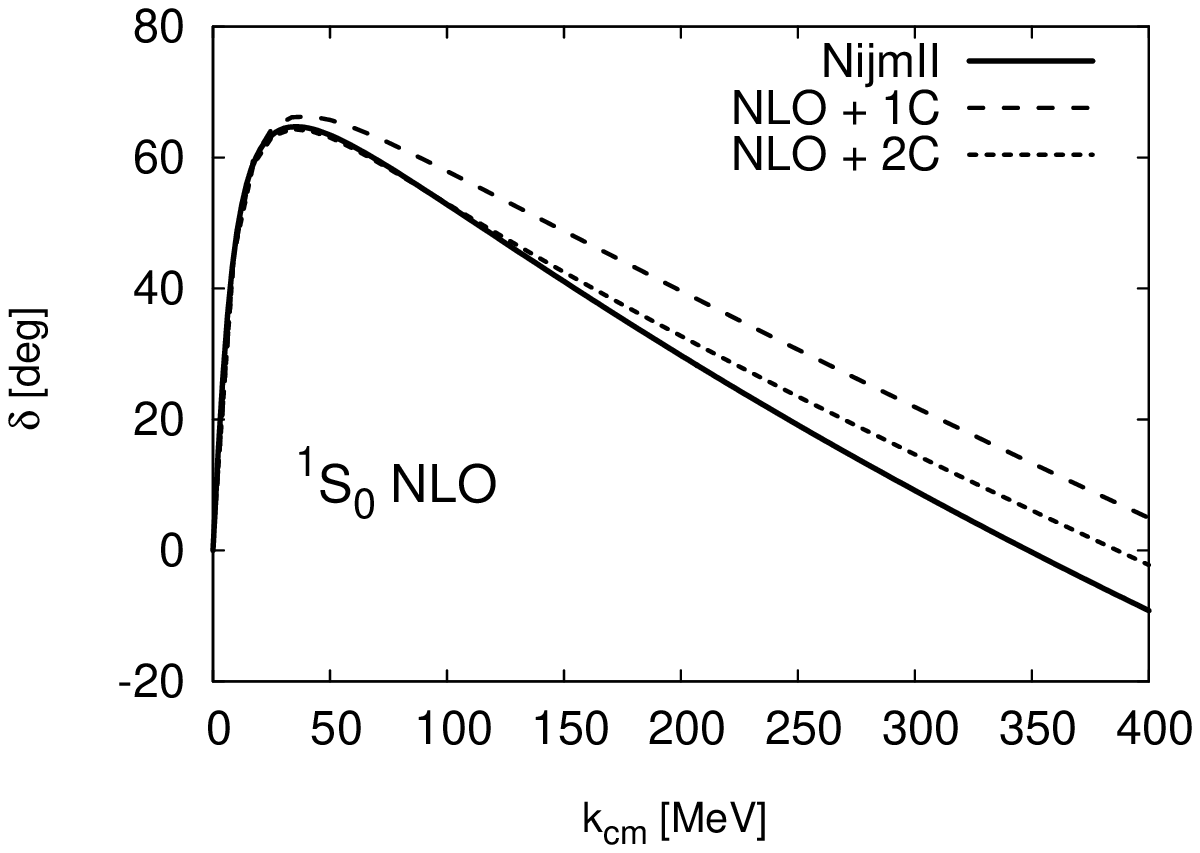,height=5.5cm,width=5.5cm}
\epsfig{figure=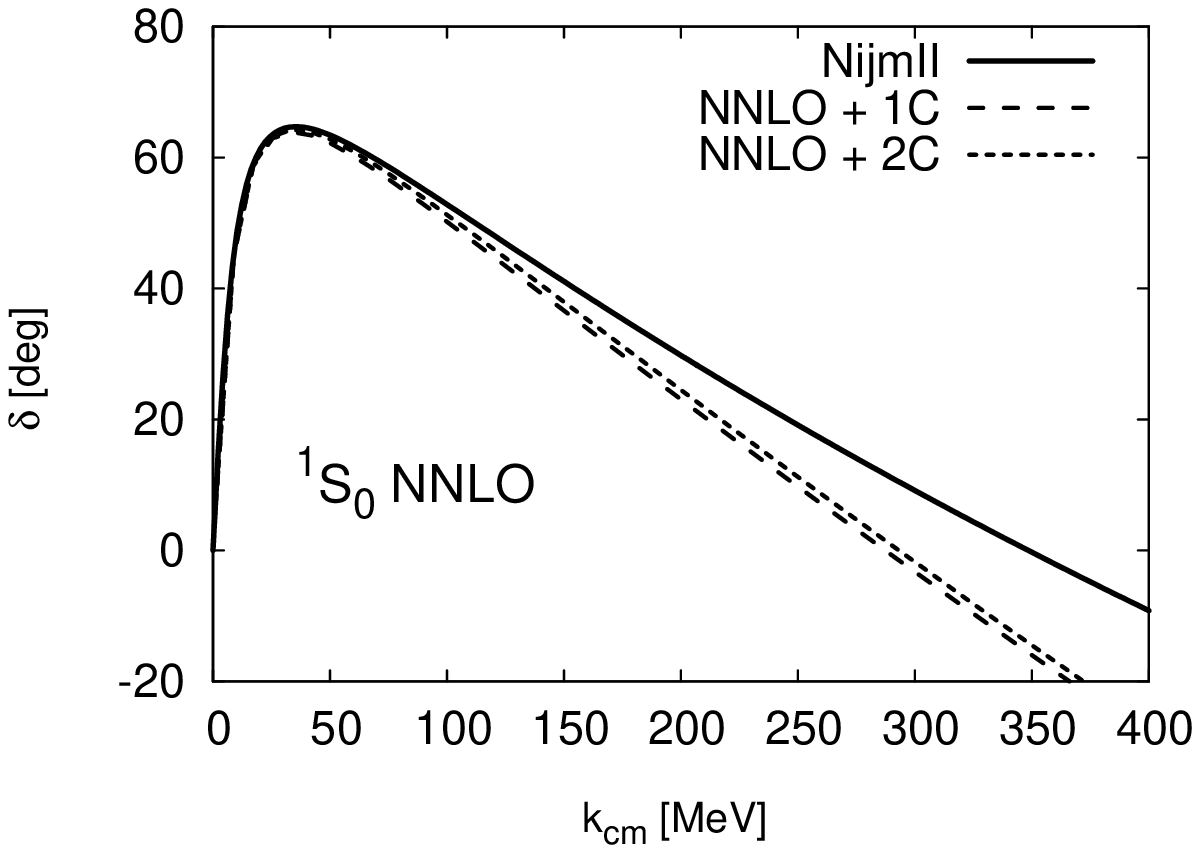,height=5.5cm,width=5.5cm}
\end{center}
\caption{
Phase shifts for the $^1S_0$ single channel for the NLO TPE potential
(left panel) and the NNLO TPE potential (right panel). 
We show the renormalized phase shifts computed when fixing one (1C curve)
or two (2C curve) parameters, i.e. the scattering length and the 
effective range ($\alpha_0 = -23.74\,{\rm fm}$ and $r_0 = 2.77\,{\rm fm}$).
}
\label{fig:ps-TPE-counterterms}
\end{figure*}

On the other hand, there are also good reasons not to include these
extra counterterms.  As argued in
Refs.~\cite{PavonValderrama:2005gu,Valderrama:2005wv,PavonValderrama:2005uj},
the inclusion of more counterterms than the minimum required for
finiteness break orthogonality constraints between different energy
solutions, and in some cases, as the singlet at NLO~\footnote{The NLO
prediction for the effective range is $2.29\,{\rm
fm}$~\cite{Valderrama:2005wv}, smaller than the experimental one.},
also the Wigner causality bound~\cite{Phillips:1996ae}, although the
unphysical consequences of breaking this bound have not been studied.
Another good reason is the lack of an obvious improvement in the NNLO
results in the singlet channel, as can be seen in
Fig.~(\ref{fig:ps-TPE-counterterms}).  While NLO phases noticeably
improve when fixing the effective range, the improvement for NNLO
phases is very small. Thus it may be a better option to preserve
orthogonality. Actually, Fig.~(\ref{fig:ps-TPE-counterterms}) suggests
that it would be more instructive and perhaps profitable to improve on
the long distance potential than adding more and more
counterterms~\footnote{This remark applies equally when going from a
pure effective range expansion, which describes the low energy data
with arbitrary accuracy but randomly agrees to the intermediate energy
region sensitive to the explicit pion exchange potential.}. Work along
these lines is on the way~\cite{EPR07}. 

It is also important to notice that the renormalizability of singular
potentials depends on a very specific representation of the short
range physics, a Pad\'{e} approximant if we write them in terms of an
energy dependent boundary condition.  This casts doubts on the
renormalizability of momentum space treatments in which the contact
interactions are parameterized as an {\it ad-hoc} expansion of deltas
and their derivatives, since only a very precise short distance
interaction will be able to renormalize the Lippmann-Schwinger
equation.

\section{Boundary condition and Variable phase equation}
\label{sec:vf}

\subsection{Equivalence between BC and Variable Phase} 

The boundary condition, Eq.~(\ref{eq:bc}), corresponds to the inner
radius $R$ of a boundary value problem defined in the region $R \le
r < \infty $. We can give a physically appealing and computationally
convenient interpretation of this BC in terms of a complementary outer
boundary value problem in the region $0 < r \le R$. If we consider the
family of potentials $U (r,R) = U(r) \theta(R-r)$, which corresponds
to a set of truncated potentials, $U(r)$, at distances below a certain
radius, $r < R$, acting only from the origin to the boundary radius
$R$ we would have at the boundary, $r=R$, the asymptotic wave function
\begin{eqnarray}
u(r) = \sin (kr + \delta(k,R))  \qquad r > R  \, ,
\end{eqnarray}
where we keep explicitly the dependence on $R$ of the phase shift.
The logarithmic derivative at the boundary $r=R$ from the left is
therefore
\begin{eqnarray}
L_k (R)  = k \cot (kR + \delta(k,R))  \, . 
\label{eq:bc-vf}
\end{eqnarray}
If we identify this expression with that of Eq.~(\ref{eq:bc}) we get
the following equation for $\delta(k,R)$
\begin{equation}
\frac{d \delta (k,R) }{dR} = -\frac1k U(R) \sin^2 (k R+ \delta(k,R))
\, .
\label{eq:vfe}
\end{equation}
This is a variable phase equation of the type analyzed in
Ref.~\cite{Calogero:1965}, describing the evolution of the
phase-shift, $\delta(k,R)$, corresponding to the truncated potential,
$ U(r,R)= U(r) \theta(R-r) $. The standard derivation for a regular
wave function at the origin, $u(0)=0$, is well known. For the case of
general solutions including also energy dependence see
e.g.~\cite{PavonValderrama:2003np,PavonValderrama:2004nb}. 

\begin{figure}[]
\begin{center}
\epsfig{figure=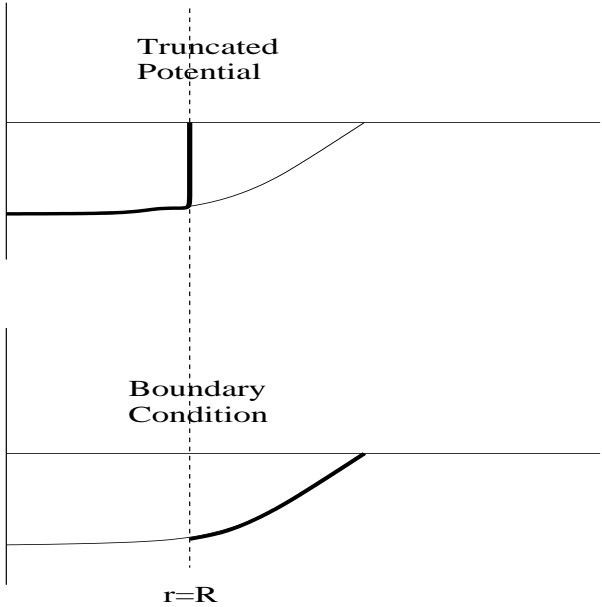,height=8cm,width=8cm}
\end{center}
\caption{The relation between the boundary condition problem defining
the {\it outer} problem and the variable phase equation defining the
{\it inner} problem. The logarithmic derivative at the boundary
provides the identification $L_k(R)= k\,\cot(kR+\delta(k,R))$. $\delta
(k,R) $ is the phase-shift produced by the potential$U(r)$ truncated
at $r > R$. The physical phase-shift is given by
$\delta(k)=\delta(k,\infty)$. The boundary condition at the origin is
$ L_k (0^+) = k \cot \delta_S (k) $ where $ \delta_S(k)$ corresponds to
the short distance phase shift, i.e., corresponding to $U(r)=0$
everywhere for $r> 0$ (see main text).}
\label{fig:bcvf}
\end{figure}

Thus, the renormalization group equation for the boundary condition of
an inner truncated potential $U(r,R) = U(r)\,\theta(r-R)$ is solved by
the logarithmic derivative of an outer cut-off potential $\bar
U(r,R)= U(r)\,\theta(R-r) $ through the variable phase equation
Eq.~(\ref{eq:bc-vf}). This result builds a one to one relation between
the evolution of effective boundary conditions of the outer problem
and that of a variable phases of the inner problem, 
which is illustrated in fig.~(\ref{fig:bcvf}). Obviously, the
physical phase shift can be obtained from the variable phase as an
asymptotic limit
\begin{eqnarray}
\delta(k) = \delta(k,\infty )  \, . 
\end{eqnarray} 
On the other hand, the boundary condition extrapolated to the origin
is given by 
\begin{eqnarray}
L_k (0^+) = k \, \cot \delta(k,0^+) \, . 
\end{eqnarray} 
In the standard variable phase approach~\cite{Calogero:1965} one
assumes $\delta (k,0)=0 $ corresponding to the absence of zero range
interaction, and hence $L_k (0)= \infty$, i.e. a regularity condition
at the origin, $u(0)=0$. In the present context it makes sense to
define the short range phase-shifts as the variable phase extrapolated
to the origin, once the potential has been completely switched off
\begin{eqnarray}
\delta_S (k) = \delta(k,0^+) \equiv \lim_{R \to 0^+} \delta( k, R) \, .    
\end{eqnarray}  
In fact, it turns out that the way the former limit must be taken is a
bit subtle, particularly in the case of singular potentials (see also
below). In the absence of a potential one gets a constant variable
phase, and hence we would simply get $ \delta (k) = \delta_S (k) $. 

On these grounds, in the presence of a long range potential, the total phase
shift $\delta(k)$ can be understood as a long distance distortion of the 
short range phase shift $\delta_S (k)$
\begin{eqnarray}
\delta_S(k) \longrightarrow \mbox{(Long range distortion)}
\longrightarrow \delta(k)
\end{eqnarray}
In this interpretation the boundary condition regularization can be used
to disentangle the short and long range physics from $\delta(k)$ to obtain
$\delta_S(k)$
\begin{eqnarray}
\delta (k) \longrightarrow \mbox{(Remove the distortion)}
\longrightarrow \delta_S(k)
\end{eqnarray}

Another aspect of the variable-phase approach is the fact that we
always deal with a given on-shell problem, that corresponding to the
truncated potential. So, in the whole process there is no need to
invoke any smooth off-shell behaviour, although we are changing the
Hilbert space when the boundary radius is moved. 

For later purposes it is convenient to introduce the variable
effective range $\hat{M}$-matrix and its inverse the reaction
$\hat{V}$-matrix,
\begin{eqnarray}
\hat{M}(k,R) &=& k \cot \delta (k,R) \, , \\ 
\hat{V}(k,R) &=& \frac{\tan \delta (k,R)}k \, ,   
\end{eqnarray} 
where direct insertion in Eq.~(\ref{eq:vfe}) yields  
\begin{eqnarray}
\frac{ d \hat{M}(k,R)}{dR} = U(R) 
\left[ \hat{M}(k,R) \frac{\sin k R}k + \cos k R \right]^2
\label{eq:vK}
\end{eqnarray}
and
\begin{eqnarray}
\frac{ d \hat{V} (k,R)}{dR} = -U(R) \left[ \frac{\sin k R}k + 
\hat{V}(k,R) \cos k R \right]^2
\label{eq:vR}
\end{eqnarray}
respectively.

\subsection{Renormalization of Low energy Threshold Parameters} 
\label{subsec:vf-tp}

Using the equivalence discussed above between the variable phase and
boundary condition problems, it is instructive to do a low energy
expansion. At zero energy we have,
\begin{eqnarray}
\alpha_0(R) \equiv -\lim_{k\to 0} \frac{\delta(k)}k  \, ,
\end{eqnarray}  
so one can obtain 
\begin{eqnarray}
\frac{d \alpha_0}{dR} &=& U(R) \left( R - \alpha_0 \right)^2 \, .
\label{eq:valpha_0}
\end{eqnarray} 
For increasing $R$ this equation describes how the scattering length
evolves as the potential $U(r)$ is switched on. The physical
scattering length is given by the value of $\alpha_0(R)$ at $R = \infty$, 
that is, $\alpha_0 = \alpha_0 (\infty) $, provided we specify an initial 
condition at any point, say the origin. In the standard approach, one takes 
$\alpha(0)=0$. However, this is not the only possibility. We may use
$R= \infty$ as the initial condition, and compute $\alpha(0)$ from
there by integrating Eq.~(\ref{eq:valpha_0}) for decreasing $R$. This
can be interpreted as the evolution of the scattering length as the
potential is switched off for $r > R$. If we use an arbitrary value
for $\alpha$ at $R = \infty $ we obtain in general $\alpha (0) \equiv
\alpha_S \neq 0$. Thus, we may interpret the value of $\alpha_S $ as
the scattering length corresponding to switching off the potential
entirely.

Going beyond the zero energy limit $k=0$ is not uniquely defined,
because there are many equivalent ways of parameterizing the phase
shifts by a low energy expansion. The coefficients of the expansion
are, however, well defined. If for definiteness we use the effective
range expansion for the running $\hat{M}$-matrix
\begin{eqnarray}
\hat{M}(k,R) &=& k\,\cot \delta (k,R) \nonumber\\
&=& -\frac1{\alpha_0 (R) } + \frac12 r_0 (R)  k^2 + 
v_2 (R) k^4 + \cdots \nonumber\\  
\end{eqnarray}
one has the set of
equations~\cite{PavonValderrama:2003np,PavonValderrama:2004nb}
\begin{eqnarray}
\frac{d \alpha_0}{dR} &=& U(R) \left( R - \alpha_0 \right)^2 \, ,
\label{eq:valpha} \\ \frac{ d r_0}{dR} &=& 2 U(R) R^2 \left( 1- \frac
{R}{ \alpha_0 } \right) \left( \frac{r_0}R + \frac{R}{3 \alpha_0 } -1
\right) \, , \label{eq:vr0}\\ \frac{d v_2 }{d R} &=& R^4 U(R) \left\{
\frac14 \left( \frac{r_0}R + \frac{R}{3 \alpha_0 } -1 \right)^2 \right.
\label{eq:vv2}\\ &+& \left.
2\left( 1- \frac{R}{ \alpha_0} \right) \left(-\frac1{12} \frac{r_0}R
+ \frac{v_2}{R^3} - \frac1{120} \frac{R}{ \alpha_0} +
\frac1{24}\right) \right\} \, . \nonumber 
\end{eqnarray}
These equations have to be supplemented with the initial
conditions~\cite{PavonValderrama:2003np,PavonValderrama:2004nb}
\begin{eqnarray}
\alpha (0^+) &=& \alpha_S \qquad \, \qquad \alpha(\infty) = \alpha \, , \\ 
r_0(0^+ )    &=& r_{0,S} \qquad \, \qquad r_0 (\infty)   = r_0 \, ,   \\ 
v_2(0^+ )    &=& v_{0,S} \qquad \, \qquad v_2 (\infty)   = v_2 \, .  \\ 
\end{eqnarray}
It should be noted here that $\alpha_S$, $r_{0,S}$ and $v_{2,S}$ can 
only be defined for regular potentials~\footnote{For attractive singular
potentials $\alpha_0(R)$, $r_0(R)$ and $v_2(R)$ behave as highly oscillating
functions for $R \to 0$, so no definite limit can be obtained for $R=0$.}.
The set of Eqs.~(\ref{eq:valpha}), (\ref{eq:vr0}) and (\ref{eq:vv2})
for the running low energy threshold parameters solve the set of
equations for the parameters $\xi_0 (R) $, $\xi_2 (R) $ and $\xi_4 (R)$, 
Eqs.~(\ref{eq:xi_0'}), (\ref{eq:xi_2'}) and (\ref{eq:xi_4'}) if in
the relations (\ref{eq:xi_0}), (\ref{eq:xi_2}) and (\ref{eq:xi_4}) one
substitutes the asymptotic low energy threshold parameters for the
running ones.

The previous reasoning may also be applied to a low
energy expansion of the variable reaction $\hat{V}$-matrix
\begin{eqnarray}
\hat{V}(k,R) &=& \frac{\tan \delta (k,R)}k \nonumber\\
&=& -\alpha_0 (R) - \frac12 \beta_0 (R) k^2 + \cdots
\end{eqnarray}
which generates the following equations for $\alpha_0$ and $\beta_0$ 
\begin{eqnarray}
\frac{d \alpha_0}{dR} &=& U(R) \left( \alpha_0 -R \right)^2 \, ,
 \label{eq:valpha0} \\ \frac{ d \beta_0}{dR} &=& -U(R) \left(R -
 \alpha_0 \right) \left( R^3 - 3 \alpha_0 R^2 + 3 \beta_0 
 \right) \, . \label{eq:vb0}
\end{eqnarray}
Obviously, we have the relation
\begin{eqnarray}
\beta_0 (R) = r_0 (R) \alpha_0 (R)^2  \, .
\end{eqnarray} 
It is straightforward to check that the set of Eqs.~(\ref{eq:valpha}),
and (\ref{eq:vr0}) and Eqs.~(\ref{eq:valpha0}) and (\ref{eq:vb0})
are mutually compatible, as it should be.

All low energy expansions share a common hierarchy for the low energy
parameters; the evolution of a given low energy parameter contributing
to a given order depends only on the evolution of lower order low
energy parameters. The set of equations express the evolution of the
low energy parameters at zero range when the long distance
contribution is switched on. Conversely, they offer a possibility to
determine the zero range low energy parameters from the total ones by
downwards evolution in the cut-off variable $R$.

\subsubsection{Application to NN Scattering}

The previous equations, Eqs.~(\ref{eq:valpha}), (\ref{eq:vr0}) and 
(\ref{eq:vv2}), can be used to study the renormalization behaviour of the 
low energy parameters for neutron-proton scattering in the $^1S_0$ channel
with the OPE potential. Although the OPE potential is regular for the $^1S_0$
channel, it behaves as $1/r$ for distances below the pion Compton wave length.
The consequence of this mild singularity is that the short distance scattering
length, $\alpha_S = \alpha_{0}(0^+) = 0$, is zero, regardless the fact that
we are not taking the regular solution of the OPE potential. 
This can be understood by studying the behaviour of $\alpha_0(R)$ for the
ultraviolet limit, $R \to 0$ (see Ref.~\cite{PavonValderrama:2003np}). 
We can solve Eq.~(\ref{eq:valpha}) for short distances and large scattering 
lengths with the Yukawa potential, given by Eq.~(\ref{eq:yukawa}), yielding
\begin{equation}
\frac{1}{\alpha_0(R)} \simeq \frac{1}{a} \log{R} + C \, ,
\end{equation}
where $C$ is an integration constant, and which is valid for small
distances ($m R \ll 1$) and large scattering lengths ($R \ll
\alpha_0(R)$). This solution of Eq.~(\ref{eq:valpha}) is equivalent to
take the irregular solution of the Schr\"odinger equation or the
stable ultraviolet fixed point studied in
Sect.~\ref{subsub:UV-regular}.  On the other hand, if we assume the
scattering length to be small compared to the cut-off scale,
$\alpha_0(R) \ll R$, we obtain
\begin{equation}
{\alpha_0(R)} \simeq - \frac{R^2}{2\,a} \, ,
\end{equation}
which turns out to be equivalent to take the regular solution of the
Schr\"odinger equation or the unstable ultraviolet fixed point for the
Yukawa potential. So the difference between the regular and irregular
solution lies in the ultraviolet behaviour of $\alpha_0(R)$: although
they both approach the same ultraviolet limit, $\alpha_0(0^+) = 0$,
the trend is dissimilar.

We can integrate Eqs.~(\ref{eq:valpha}), (\ref{eq:vr0}) and (\ref{eq:vv2})
for the other low energy parameters of NN scattering in the $^1S_0$ channel 
with OPE, by taking as initial conditions for $R = \infty$ the experimental 
scattering length and effective range, while for the shape parameter $v_2$
we use the value obtained from the Nijmegen II potential, yielding
\begin{eqnarray}
\alpha (0^+) = 0\phantom{{\rm fermi}} 
&\quad& \alpha(\infty) = -23.74\,{\rm fm} \, , \\ 
r_0(0^+ )  = 4.46\,{\rm fm} &\quad& r_0 (\infty) = 2.77\,{\rm fm} \, , \\ 
v_2(0^+ ) = 1.24\,{\rm fm}^3 &\quad& v_2 (\infty) = -0.48\,{\rm fm}^3 \, .
\end{eqnarray}
In Ref.~\cite{PavonValderrama:2003np} we can find a more detailed discussion
although different results are obtained due to the low infrared cut used 
($R_{\infty} = 10\,{\rm fm}$)~\footnote{Here $R_{\infty} = 20\,{\rm fm}$ is 
used instead.}. In Ref.~\cite{Richardson:1999hj} the result
$r_{0,S} = 4.0\,{\rm fm}$ is obtained by a renormalization analysis 
in momentum space, and in Ref.~\cite{Steele:1998zc} they obtain 
$r_{0,S} = 3.10\,{\rm fm}$ although an ultraviolet cut-off of half the rho 
mass $\Lambda = m_{\rho} / 2$ is employed to perform the calculation; 
if we use $R = \pi / 2\Lambda = \pi/m_{\rho} \sim 0.8\,{\rm fm}$, 
equivalent to the previous cut-off, we obtain $r_{0,S} = 2.86\,{\rm fm}$.

\subsection{Low energy expansion of the short distance interaction} 

So far, all we have done has to do with relating long and short range
physics along the trajectory defined by the long distance
potential. If it was for that nothing would be achieved. We propose to
make a low energy expansion of the short range physics. According to
our previous discussion for a truncated potential it makes sense to
expand either the $\hat{M}$-matrix or the $\hat{V}$-matrix in powers of $k$,
\begin{eqnarray}
\hat{M}_S &=& k \cot \delta_S = \frac{u_k' (0^+)}{u_k(0^+)} \nonumber\\ 
&=& -\frac1{\alpha_S} + \frac12 r_S k^2 + v_S k^4 + \dots \label{eq:KS} 
\\ 
\hat{V}_S &=&  \frac{\tan \delta_S}k = \frac{u_k (0^+)}{u_k' (0^+)} \nonumber\\
&=& -\alpha_S - \frac12 \beta_S k^2 - \gamma_S k^4 + \dots 
\label{eq:VS} 
\end{eqnarray} 
respectively.  To achieve consistency with the low energy expansions,
Eq.~(\ref{eq:kcot}) and (\ref{eq:tank}), up to some order of the full
$\hat{M}$-matrix and $\hat{V}$-matrix we have to compute the short distance 
low energy parameters by integrating downwards the set of
Eqs.~(\ref{eq:valpha}), (\ref{eq:vr0}) and (\ref{eq:vv2}) or
Eqs.~(\ref{eq:valpha}) and (\ref{eq:vb0}). This way we exactly
reproduce the low energy expansion up to a desired order, and generate
all higher orders in energy according to the long distance part of the
potential. In particular, we also generate at any level of truncation
the remaining higher order parameters. Thus, in the LO approximation
we consider $ \alpha $ and $ U(r)$ as independent parameters,
corresponding to keep one term in Eq.~(\ref{eq:KS}), and hence setting
$ r_0( 0^+) = v_2 (0^+) = \dots =0$. One then obtains $r_0 $, $v_2$, etc., 
from $\alpha$ and $U(r)$, and as a consequence the phase shift $\delta(k)$.  
In the NLO approximation one considers $\alpha$ and $r_0$ as independent 
variables and predicts $v_2$,$v_3$ etc., and hence the phase shift 
$\delta(k) $ from the knowledge of $ \alpha$, $r_0 $ and $U(r)$.

Note that if the physical low energy parameters, say $\alpha$, $r_0$
and so on, and the potential $U$ are known, the result is
unique~\footnote{In practice this situation may be too optimistic,
since low energy parameters may be directly deduced from the data,
which are analyzed with a given model. In
Ref.~\cite{Valderrama:2005ku} we provide a thorough determination
of the low energy parameters for NN interaction in all partial waves
for the high quality potentials of Ref.~\cite{Stoks:1994wp}.}. Another
important point is that if we have a singular potential we cannot
start directly at the origin, but at a given small radius $ R$. An
advantage of our method that will become clear below is that we can make
$R$ much smaller than any other scale in the problem. Even if we do
this numerically, this is an effective way of eliminating the
regularization. Moreover, if we take the limit of large $R$, much
larger than the range of the potential $a$ , we are effectively having
a constant variable phase given by the short distance theory. It thus
makes sense to compare the full result including the potential with
that of the low energy expansion.

\section{Higher partial waves and coupled channels} 
\label{sec:higher_pw}

\def\u{{\bf u}} \def\U{{\bf U}} \def\S{{\bf S}} \def\h{{\bf h}}
\def\L{{\bf L}} \def\E{{\bf 1}} \def\j{{\bf j}} \def\y{{\bf y}}
\def\K{{\bf K}} 
\def\f{{\bf f}} 
\def\k{{\bf k}}
\def\uin{{\bf u}_{\rm in}}
\def\uout{{\bf u}_{\rm out}}

The generalization of the present ideas to coupled channels is in
principle straightforward and runs parallel to what was done in
Sect.~\ref{sec:rg} with some modification which we outline in the
following. The coupled channel Schr\"odinger equation for the relative
motion reads
\begin{eqnarray}
-\u '' (r) + \left[ \U (r) + \frac{{\bf l}^2}{r^2} \right] \u (r) 
= \k^2 \u (r) \, , 
\label{eq:sch_cp} 
\end{eqnarray} 
where $\U (r)$ is the coupled channel matrix potential, $\u(r)$ is the
reduced matrix wave function in the initial and final state and
\begin{eqnarray} 
{\bf l}^2 &=& {\rm diag} ( l_1 (l_1+1), \dots, l_N (l_N +1) ) \, ,
\nonumber \\ \\ \k^2 &=& {\rm diag} ( 2 \mu_1 (E-E_1), \dots, 2\mu_N
(E-E_N) ) \, , \nonumber \\  
\label{eq:k_coupled}
\end{eqnarray} 
are the angular momentum and the CM momentum in the coupled channel
space respectively. $E_i$ is the threshold energy and $\mu_i $ the
reduced mass in the i-th channel. We assume for $\u(r)$ the boundary condition,
\begin{eqnarray}
\u ' (R)  - \L_k (R) \u (R) = 0 \, ,
\label{eq:bc_coupled}
\end{eqnarray}   
where $\L_k(R)$ is a real hermitean matrix in coupled channel space,
which in our framework encodes the {\it unknown} physics at distances
$r$ below the boundary radius $R$. In addition, we assume the
asymptotic normalization condition for scattering states
\begin{eqnarray}
\u (r)  \to \uin (r) - \uout (r) \S \, ,
\label{eq:asym}
\end{eqnarray} 
with $\S$ the standard coupled channel unitary S-matrix, 
$\S^\dagger \S ={\bf 1} $, from which the scattering amplitude 
${\bf f} = {\bf 1} + 2 i {\bf k }^{1/2} \S {\bf k}^{1/2}$ can be obtained. 
The out-going and in-going wave functions can be defined as
\begin{eqnarray}
\uin(r) &=& \hat \h^{(-)} (r) \, ,\\
\uout(r) &=& \hat \h^{(+)} (r) \, ,
\end{eqnarray}
where $\h^{(+)}$ and $\h^{(-)}$ are given by
\begin{eqnarray}
\h^{(+)} (r) &=& {\rm diag} ( \hat h^+_{l_1} ( k_1 r) , \dots , \hat
h^+_{l_N} (k_N r) ) \, , \\ 
\h^{(-)} (r) &=& {\rm diag} ( \hat h^-_{l_1} ( k_1 r) ,
\dots , \hat h^-_{l_N} (k_N r) ) \, ,
\end{eqnarray} 
with $ \hat h^{\pm}_l ( x) $ the reduced Haenkel functions of order
$l$, $ \hat h_l^{\pm} (x) = x H_{l+1/2}^{\pm} (x) $ ($ \hat h_0^{\pm} =
e^{ \pm i x}$), which satisfy the Schr\"odinger's equation 
for a free particle.  

To determine how should the matrix boundary condition depend on the
boundary radius in order to achieve the {\it same} S-matrix, we
proceed similarly as in Sect.~\ref{sec:rg}. Making an infinitesimal
displacement of the radius, $R \to R + \Delta R$, and taking into
account the total derivative of the wave function with respect to the
boundary radius
\begin{equation}
\frac{\partial \u (r,R)}{\partial R}  = \u_R (r,R) \, ,
\end{equation}
then, the derivative of the boundary condition is given by
\begin{eqnarray}
\u'' (R,R) + \u_R ' (R,R) - {\bf L} ' _k (R) \u(R,R) - &&
\nonumber \\
{\bf L}_k (R) \u' (R,R) - {\bf L}_k (R) \u_R (R,R) &=& 0 \, .
\label{eq:bc_def_coupled} 
\end{eqnarray}
Deriving also Schr\"odinger's equation with respect to the inner
radius $R$ 
\begin{equation}
-\u_R '' (r,R) + \U(r) \u_R (r,R) = k^2 \u_R (r,R)
\label{eq:sch_def_coupled}
\end{equation}
and the asymptotic wave function, Eq.~(\ref{eq:asym})
\begin{eqnarray}
\u_R (r,R) &\to&  -  \uout (r) \frac{ d \S}{dR} \, ,\\ 
\u' (r,R)  & \to & \uin'(r) -  \uout' (r) \S \, ,\\  
\u_R' (r,R) & \to & \uout' (r) \frac{ d \S}{dR} \, .  
\end{eqnarray}
Thus, using Lagrange's identity we get 
\begin{eqnarray}
0 &=& \u(r,R)^\dagger \u_R (r,R)'' - \u''(r,R)^\dagger \u_R (r,R) \, ,
\nonumber \\ &=&
\left( \u(r,R)^\dagger \u_R (r,R)' - \u'(r,R)^\dagger \u_R (r,R)
\right)' \, .\nonumber \\
\label{eq:bc_asy_coupled}
\end{eqnarray}
Integrating between $R$ and $\infty$ and using the boundary condition,
Eq.~(\ref{eq:bc_coupled}) and Eq.~(\ref{eq:bc_asy_coupled}) and the identity
$\uin^\dagger \uout$ as well as the Wronskian $ \uin^\dagger \uout' -
\uin'^\dagger \uout = 2 {\rm i} {\bf k} $ we finally get
\begin{eqnarray}
2 {\rm i} \S^\dagger \k \frac{ d \S }{dR} = 
\u(R,R)^\dagger &\Big[& \k^2  -\U(R) -\frac{{\bf l}^2}{r^2} + \nonumber \\ && 
{\bf L}' (R) + {\bf L} (R)^2 \Big]\,\u(R,R)\, . \nonumber\\
\label{eq:del2_coupled}
\end{eqnarray}
This equation tell us how the S-matrix changes as the inner boundary
radius is changed. If we require the S-matrix {\it not} to be
dependent on the particular choice of $R$ we get
\begin{equation}
{\bf L} (R)^2 + {\bf L }' (R) = \U(R) + \frac{{\bf l}^2}{r^2} - {\bf k}^2 \, ,
\label{eq:renorm_coupled} 
\end{equation}
which is the coupled channel generalization of Eq.~(\ref{eq:renorm}),
which likewise accounts for the coupled channel momentum ${\bf k} $
dependent evolution of the boundary condition. The evolution of the
low energy parameters can be translated into the corresponding
evolution of the short distance boundary condition as a function of
the boundary radius. Defining the dimensionless quantity
\begin{eqnarray}
{\bf \Xi}_\k (R) = R\,{\bf L}_\k (R) = 
R\,{\bf u}'_\k (R) {\bf u}_\k (R)^{-1}\, , 
\end{eqnarray}   
and using Eq.~(\ref{eq:renorm_coupled}) we get
\begin{eqnarray}
R\,\frac{d {\bf \Xi}_\k}{d R} (R) = {\bf \Xi}_\k ( {\bf 1} - {\bf \Xi}_\k ) + 
{\bf U}(R) R^2 + {\bf l}^2 - \k^2 R^2 \, , \nonumber\\
\label{eq:bc_ren_ck}
\end{eqnarray}   
a result already derived in our previous
work~\cite{PavonValderrama:2004nb} by different means. Assuming for
simplicity the degenerate case, $ k_i = k $ and expanding into powers
of the common momentum $k$ one gets,
\begin{eqnarray}
{\bf \Xi}_k (R) =  {\bf \Xi}_0 (R) + (k R)^2\,{\bf \Xi}_2 (R)  + \dots 
\end{eqnarray} 
so the RG flow at zero energy becomes 
\begin{eqnarray}
R\,\frac{d {\bf \Xi}_0}{d R}  = 
{\bf {\Xi}_0} ( {\bf 1} - {\bf {\Xi}_0} ) + {\bf U}(R) R^2
+ {\bf l}^2  \, . 
\label{eq:bc_ren_c0}
\end{eqnarray}   
A thorough study for the general multichannel case is beyond the scope
of the present work. Nevertheless, for some special cases some general
conclusions may be drawn. The most appealing regime has to do with the
possible appearance of chaos. If that would be so, the whole
renormalization program would not be implementable in practice, since
an absolute knowledge of the short distance conditions would be
needed.  It is well known, that the minimal order for a dynamical
autonomous system to develop chaotic solutions is three since for two
independent variables the Poincar\'e-Bendixon theorem~\cite{Coddington} 
guarantees integrability. Actually, if we introduce the variable 
$t=- \log (R/a)$, as a new variable with $R$ as a dependent variable, 
the simplest case for potential scattering would correspond to two-coupled 
channels. In Ref.~\cite{PavonValderrama:2004nb} we have discussed the case
corresponding to OPE NN potential in the $^3S_1-^3D_1$ coupled
channel, and infrared fixed points have been determined.

To analyze the short distance behaviour, let us assume the potential
to behave as an inverse singular power 
\begin{eqnarray}
{\bf U} (R) R^2  = -\frac{\bf a^n}{R^n} \, ,
\end{eqnarray} 
where ${\bf a} $ is a matrix with length dimension. We can diagonalize
the potential by a {\it global} transformation, say ${\bf G} $, so
that if we have a set of diagonal ${\bf G} {\bf C}_0 {\bf G}^{-1}$ at
some initial value the solution will always be diagonal. This means
that the system is integrable and hence chaos is precluded.

Similarly to the findings of Sect.~\ref{sec:vf} (see also
Ref.~\cite{PavonValderrama:2004nb}) the coupled channel boundary
condition, Eq.~(\ref{eq:bc_def_coupled}), for the {\it outer} boundary
values problem, Eq.~(\ref{eq:sch_cp}) and Eq.~(\ref{eq:asym}), can
be interpreted in simple physical terms of a complementary {\it inner}
problem where the potential ${\bf U}(r)$ acts in the interval $ R \le
r < \infty$. If we switch off the potential above a given boundary
radius $R$ we have, at the boundary
\begin{eqnarray} 
 {\bf L}_k (R) &=& \u' (R) \u^{-1} (R) \nonumber \\ &=& \left[ \uin'
(R) - \uout' (R) \S (R) \right] \nonumber \\ &\times& \left[ \uin (R)
- \uout (R) \S (R) \right]^{-1} \, ,
\end{eqnarray}  
where ${\S (R)}$ is the S-matrix associated to the potential ${\bf
U} (r) $ acting in the region $ 0 < r \le R$, which inherits the
dependence on the chosen boundary radius $R$. It is straightforward
to obtain the equation for the variable S-matrix,
\begin{eqnarray}
2 {\rm i} k \frac{ d \S (R)}{dR} &=& \left[ \S(R) \hat \h^{(+)} (R) -
\hat \h^{(-)} (R) \right] \U(R) \nonumber \\ & \times& \left[ \hat
\h^{(-)} (R) - \hat \h^{(+)} (R) \S (R) \right] \, .
\label{eq:vs}
\end{eqnarray} 
Further consequences of this equation, in particular its low energy
limit can be looked up in Ref.~\cite{Valderrama:2005ku}. 

\section{Summary and Conclusions}
\label{sec:conclusions} 

In the present work we have analyzed the role of boundary conditions
in potential two-body scattering from a renormalization point of view.
We remind that in quantum mechanical problems the most general form of
interactions can be accommodated not only in terms of a given
potential, but also as a nontrivial mixed boundary condition. The
discussion has been carried out in configuration space, because there
the disentanglement of the short and long distance physics is
naturally formulated using both a suitable boundary condition at the
origin and a local potential respectively. The suitability of the
boundary condition is governed by a simple renormalization group
equation, and depends explicitly on the choice of the potential, the
scattering energy as well as some renormalization conditions. In
potential scattering the most appropriate conditions at low energies
are the threshold parameters, such as the scattering length, the
effective range and so on.  The generalization of such an equation for
the case of coupled channels is straightforward and has also been
analyzed in some detail.

The resulting equations can be applied to many cases of interest, 
like the study of both infrared and ultraviolet fixed points. 
We find stable and unstable infrared fixed points corresponding to 
the limit of small and large scattering lengths respectively, 
in agreement with previous authors~\cite{Birse:1998dk,Barford:2002je}. 
The same kind of fixed points is also found for regular potentials and
repulsive singular potentials in the ultraviolet regime.
In contrast, for the case of attractive singular power law potentials,
we also describe ultraviolet limit cycles and attractors 
with a computable fractal dimension.
Actually, these exotic renormalization group solutions are genuinely 
non-perturbative short distance phenomena, and have a close relationship 
with the cut-off dependence of observables and hence with power counting.
Once the RG trajectory of the boundary condition enters the attractor,
renormalized observables show a very smooth dependence with respect to
the cut-off, as can be seen in 
Refs.~\cite{PavonValderrama:2005gu,Valderrama:2005wv,PavonValderrama:2005uj}. 
Moreover, these results imply fine tuning short-distance conditions as 
already found in our previous works~\cite{PavonValderrama:2003np,
PavonValderrama:2004nb}.
Obviously, the appearance of an RG trajectory attractor and fractality 
immediately suggests the search for possible short distance chaotic 
behaviour of the renormalization group equations.  
Let us note that the presence of chaos at short distances would demand not 
only fine tuning but absolute tuning of renormalization constants in the 
ultraviolet limit. 
Although our equations look very similar to those appearing in dynamical 
systems with chaotic behaviour, for the obvious candidate of coupled channel 
zero energy s-wave scattering with a singular power law potential 
we have not seen any traces of a chaotic pattern.

Furthermore, the ultraviolet renormalization group behaviour of singular 
potentials provides interesting insights into the power counting rules for
non-perturbative effective field theories.
Boundary conditions allow a clear estimate of the cut-off behaviour of
scattering observables when certain renormalization conditions are imposed.
Since one expects that an effective theory leads to a low energy description 
which becomes increasingly insensitive to detailed short distance dynamics,
then the cut-off dependence of the observables can be used to construct a
power counting for the unknown short range physics, i.e. the counterterms.
One consequence is that counterterms are no longer completely independent of
the long range potential, as happened in the original Weinberg's counting.
Thus, a counterterm must be included in channels with 
an attractive singular potential, in agreement with previous 
works~\cite{Beane:2000wh,Nogga:2005hy,Valderrama:2005wv}.
The inclusion of extra counterterms in these channels do not change
the power law suppression in the cut-off of short range physics.
However, the proportionality coefficient does depend on the
renormalization conditions. 

On the contrary in channels with a repulsive singular potential no
counterterm can be added if the cut-off is to be
removed~\cite{Valderrama:2005wv}.  This result, which seems very
counterintuitive from a naive point of view, is indeed very natural
when one takes into account that a long range repulsive potential acts
as a potential barrier for the unknown short distance physics, thus
completely screening their contribution to low energy physics.

A very appealing perspective of the renormalization group method is
provided by the definition of a complementary scattering problem,
which can be mapped into a variable phase
equation~\cite{Calogero:1965} with nontrivial initial
(short-distance) conditions and used extensively in our previous
works~\cite{PavonValderrama:2003np,PavonValderrama:2004nb,
PavonValderrama:2004td}. 
Actually, using the long distance fixed points we find a kind of short-distance
modified effective range expansion both for large and small scattering
lengths, which in some cases is amenable to a perturbative
discussion. Moreover, our method allows to handle the case of
singular potentials at the origin. This complements the long distance
modified effective range expansion proposed long-ago~\cite{Haeringen}
and also deduced more recently by renormalization group arguments in
Ref.~\cite{Barford:2002je}, where by construction genuinely singular
potentials at the origin where excluded. 

One of the most rewarding aspects of the previous and present
investigations has to do with the practical elimination of short
distance cut-offs in the NN scattering problem which long range pion
exchange contributions become singular at the origin. Since Effective
Field Theory ideas were proposed to study the NN problem, a lot a
progress has been made, but the existence of finite cut-offs has
clouded the key renormalization issues.  The renormalization program
becomes necessary to make truly model independent calculations based
on both the long distance physics explicitly governed by pion
exchanges and hence sensitive to chiral symmetry, and the unknown
short distance dynamics. The present approach suggests a way from a
renormalization group viewpoint to effectively remove these cut-offs,
and suggests that the non-perturbative counterterms to achieve scale
independence can indeed have a quite unexpected behaviour, as we have
discussed and illustrated in detail for the $^1S_0$ channel with the
singular TPE Potential. Actually, our results are more
general and can be applied in other contexts where the unknown short
distance physics plays a significant role.

\begin{acknowledgments}

One of us (E.R.A.) thanks Rafael Ortega (Matem\'atica Aplicada) for
suggesting the method to look for absence of chaos in the coupled
channel case and Juan Campos for discussions at very early stages of
this work. M.P.V. thanks U. van Kolck for discussions on singular potentials
and W. Broniowski and P. Bo{\.z}ek for their kind hospitality in Krakow
where part of this work has been done.

This work is supported in part by funds provided by the
Spanish DGI and FEDER funds with grant
no. FIS2005-00810, Junta de Andaluc{\'\i}a grants no.  FQM225-05, EU
Integrated Infrastructure Initiative Hadron Physics Project contract
no. RII3-CT-2004-506078

\end{acknowledgments}

\appendix 

\section{Solutions to the Schr\"odinger Equation for Singular Potentials}
\label{app:singular}

Singular Potentials are commonplace on Effective Theories. 
Since the effective interaction is developed as a long range (or low energy) 
expansion, each new order gives rise to potentials which are more singular
at short range scales. This can be seen in the NN chiral potential, 
which at $Q^{\nu}$ order displays a $1 / r^{3+\nu}$ singularity for 
distances below the pion Compton wave length~\footnote{The degree of 
singularity can be milder in some cases, as in the $^1S_0$ singlet channel
at $Q^{0}$ order, in which the potential behaves as $1 / r$.}.
Consequently the study of the solutions of the Schr\"odinger equation for
singular potentials is of main interest for this work, in which we treat the
renormalization of the effective interactions in coordinate space.

We will consider the case of a potential which shows a power law divergence 
near the origin, i.e. $U(r) \to \pm a^{n-2} / r^{n}$. For it, we can write
the Schr\"odinger equation as
\begin{eqnarray}
-u''(r) + \left[\,U(r) + \frac{l(l+1)}{r^2}\,\right]\,u = k^2\,u \, .
\end{eqnarray}
For short enough distances the singular potential dominates and the centrifugal
barrier becomes irrelevant; the conditions of applicability of the WKB 
approximation hold for $r \ll a\,{(n/2)}^{2/(n+2)}$, so we can use it 
to estimate the behaviour of the wave functions near the origin.
For an attractive singular potential we have
\begin{eqnarray}
u(r) \to C\,{\left(\frac{r}{a}\right)}^{n/4}\,
\sin{\left[ \frac{1}{n/2-1}\,{\left( \frac{a}{r} \right)}^{n/2-1} + 
\varphi \right]} \, ,
\end{eqnarray}
where $\varphi$ is the semiclassical phase, which is a free parameter that
must be fixed from some given long distance information. 
It should be noted that there is no unique regular solution at the origin,
as happened with regular potentials; both linearly independent solutions
are regular at the origin.
On the contrary, for a repulsive singular potential we have
\begin{eqnarray}
u(r) &\to& 
C_{+}\,{\left(\frac{r}{a}\right)}^{n/4}\,
\exp{\left[ + \frac{1}{n/2-1}\,{\left( \frac{a}{r} \right)}^{n/2-1} \right]}
+ \nonumber\\
&& C_{-}\,{\left(\frac{r}{a}\right)}^{n/4}\,
\exp{\left[ - \frac{1}{n/2-1}\,{\left( \frac{a}{r} \right)}^{n/2-1} \right]}\,,
\end{eqnarray}
which consists in a linear combination of an irregular (the positive sign
exponential) and regular solution (the negative sign exp.). 
Physical solutions demand that $C_{+} = 0$, thus precluding the possibility
of fixing any observable via counterterms 
(see end of Sect.~\ref{sub:error_estimates}).

It is also interesting to consider the momentum expansion of the wave function
\begin{eqnarray}
u_k(r) = u_0(r) + k^2\,u_2(r) + k^4\,u_4(r) + \dots 
\end{eqnarray}
which is used all along this work, specially when considering the momentum
expansion of the boundary condition. The terms in these expansions obey the
following differential equations
\begin{eqnarray}
-u_0'' + \left[U(r) + \frac{l(l+1)}{r^2}\right]\,u_0 &=& 0 \, ,\\
-u_2'' + \left[U(r) + \frac{l(l+1)}{r^2}\right]\,u_2 &=& u_0 \, ,\\
-u_4'' + \left[U(r) + \frac{l(l+1)}{r^2}\right]\,u_4 &=& u_2 \, ,
\end{eqnarray}
and so on. 
Once a certain solution for $u_0(r)$ is given, the higher order terms can be 
solved via Green functions
\begin{eqnarray}
u_2(r) &=& \int_{0}^{\infty}\,dr' G_0(r,r') \,u_0(r') \, ,\\
u_4(r) &=& \int_{0}^{\infty}\,dr' G_0(r,r') \,u_2(r') \, ,
\end{eqnarray}
where the Green function can be defined in terms of two linear independent
solutions $u_{0,a}$ and $u_{0,b}$ of the zero energy Schr\"odinger equation
\begin{eqnarray}
G_0(r,r') &=& 
u_{0,a}(r) \, u_{0,b}(r')\,\theta (r'-r) + \nonumber\\
&& u_{0,a}(r') \, u_{0,b}(r)\,\theta (r-r') \, ,
\end{eqnarray}
where $u_{0,a}$ and $u_{0,b}$ fulfill the relation
\begin{eqnarray}
u_{0,a}'(r) u_{0,b}(r) - u_{0,a}(r) u_{0,b}'(r) = 1 \, ,
\end{eqnarray}
corresponding to unity Wronskian normalization. 
From this representation of $u_2$, $u_4$, and so on, in terms of Green function
it is very easy to obtain their power law behaviour near the origin. Since any
solution of the zero energy Schr\"odinger equation for a singular potential
behaves as $r^{n/4}$, times some trigonometric or exponential function 
depending whether the potential is attractive or repulsive, we see that
$u_2$ behaves as $r^{n/2+1} \times r^{n/4}$, $u_4$ as $r^{n+2} \times r^{n/4}$,
and so on, each power of $k^2$ adding an $r^{n/2+1}$ to the wave function.
This means that there is a strong short distance suppression in the expansion
of the wave function in powers of momenta.

\section{Evaluation of the Truncation of the Boundary Condition for 
Attractive Singular Potentials}
\label{app:truncation}

In this appendix we want to discuss in further detail the error
estimates when truncations are made in the boundary condition,
Sect.~\ref{sub:error_estimates}, thus breaking exact RG invariance.
The idea behind these truncations is to fix some physical observables
of the system, and then make predictions for the other observables.
This violates exact RG invariance, which would imply a complete
knowledge of the phase shifts of the system at all energies, something
which is not possible in practical computations, and which would also
lack predictive power.  Nevertheless, as we saw in
Sect.~\ref{sub:error_estimates}, although exact RG invariance is
broken for an arbitrary cut-off $R$, under some circumstances we
recover it in the $R \to 0$ limit.

In all these cases, higher order RG equations are not fulfilled. Thus, if we
only fix $\alpha_0$ for example, we can rewrite the expansion of the boundary
condition as
\begin{eqnarray}
L_k(R) &=& L_0(R) + k^2\,\Delta_2(R) + \dots
\end{eqnarray}
meaning by this notation that $L_0$ fulfills its RG equation, while $\Delta_2$
and higher order terms don't. If we fix both $\alpha_0$ and $r_0$, we could
write accordingly
\begin{eqnarray}
L_k(R) &=& L_0(R) + k^2\,L_2(R) + k^4\,\Delta_4(R) + \dots
\end{eqnarray}
in which now both $L_0$ and $L_2$ fulfill their respective RG equations,
while the higher orders don't. In many occasion, the higher order terms
can be neglected, and then the boundary condition reduces to
\begin{eqnarray}
L_k(R) &=& L_0(R) \, ,\\
L_k(R) &=& L_0(R) + k^2\,L_2(R) \, ,
\end{eqnarray}
which in turn implies that $\Delta_2 / L_0 \to 0$ or $\Delta_4 / L_2 \to 0$
when the cut-off $R$ goes to zero, thus simplifying the computations.

For computing $L_0$, $L_2$, and higher order terms in the expansion of the
boundary condition it is useful to define the next momentum expansion of 
the wave function
\begin{eqnarray}
\frac{u_k(r,R)}{k} &=& u_0(r,R) + k^2\,u_2(r,R) + \nonumber\\
&& k^4\,u_4(r,R) + \dots
\label{eq:uk_exp}
\end{eqnarray}
in which we have normalized the wave functions according 
to the asymptotic normalization
\begin{eqnarray}
u_k(r,R) & \to & \sin{(k\,r + \delta(k,R))} \, ,\\
\nonumber\\
u_0(r,R) &\to & r - \alpha_0 \, , \\
u_2(r,R) &\to & \frac{r}{6}\,(3\,\alpha_0\,(r - r_0) - r^2) \, ,\\
u_4(r,R) &\to & \frac{r}{120}\,\Big( r^4 - 5\,\alpha_0\,r^3 + \nonumber\\
&& 10\,\alpha_0\,r_0 r^2  - 120\,\alpha_0\,v_2 \Big) \, ,
\end{eqnarray}
and so on.  Note that here $\delta(k,R)$ is the phase shift dependence
on the short distance cut-off (not to be confused with the variable
phase). This normalization for the wave functions is consistent
with the one used in the derivation of Eq.~(\ref{eq:del1}). The terms
$u_0$, $u_2$, etc, obey the following equations
\begin{eqnarray}
- u_0'' + U(r)\,u_0 &=& 0 \, ,\\
- u_2'' + U(r)\,u_2 &=& u_0 \, ,\\
- u_4'' + U(r)\,u_4 &=& u_2 \, ,
\end{eqnarray}
which can be trivially deduced by applying the reduced Schr\"odinger equation
to the expansion in powers of momenta of the wave function, 
Eq.~(\ref{eq:uk_exp}).

Now let's study the case in which we fix the scattering length of the system,
thus fulfilling Eq.~(\ref{eq:xi_0'}), while Eq.~(\ref{eq:xi_2'}) and higher 
order ones are not. In such a case we can write the expansion of the boundary
condition as
\begin{eqnarray}
L_k(R) = L_0(R) + k^2\,\Delta_2(R) + \dots
\end{eqnarray}
in which the relation of the terms in this expansion and the ones of the
expansion of the wave function is given by
\begin{eqnarray}
L_0(R) &=& \frac{u_0'}{u_0} \\
\Delta_2(R) &=& \frac{u_2' u_0 - u_2 u_0'}{u_0^2} \, .
\end{eqnarray}
From the behaviour of the zero energy wave function for power-law singular
potential, it is trivial to check that for small cut-offs
\begin{eqnarray}
L_0(R) \to \frac{1}{R^{n/2}}\,
\cot\left[\frac{2}{n-2}\,(\frac{a}{R})^{n/2-1} + \varphi_0 \right] \, . 
\label{eq:L0_singular}
\end{eqnarray} 
On the other side, the behaviour of $\Delta_2$ can be easily evaluated if we
take into account the following Lagrange identity
\begin{eqnarray}
(u_2' u_0 - u_2 u_0')' = u_0(r)^2 \, ,
\label{eq:u2u0_lagrange}
\end{eqnarray}
where we have dropped the dependence of $u_0$ on the cut-off, since this wave
function is cut-off independent. From this we find that
\begin{eqnarray}
\Delta_2(R) = \frac{1}{u_0(R)^2}\,\int_{0}^{R}\,dr\,u_0(r)^2 \, ,
\end{eqnarray}
which for small cut-offs approximately behaves as
\begin{eqnarray}
\Delta_2(R) \to R\,f\left[\frac{2}{n-2}\,(\frac{a}{R})^{n/2-1} + 
\varphi_0\right] \, ,
\label{eq:D2_singular}
\end{eqnarray}
with $f(x)$ some unspecified trigonometric function. It can be easily shown 
that $\Delta_2$ obeys the differential equation
\begin{eqnarray}
\Delta_2'(R) + 2\,L_0(R)\,\Delta_2(R) - 1 = 0 \, ,
\end{eqnarray}
in contrast with the renormalization group equation
\begin{eqnarray}
L_2'(R) + 2\,L_0(R)\,L_2(R) + 1 = 0 \, .
\end{eqnarray}
As we can see, $k^2\,\Delta_2 \ll L_0$, which means that we can drop the 
$\Delta_2$ term, without jeopardizing the smooth cut-off dependence of 
the phase shifts for small cut-offs. 
In fact, it is curious to see that dropping $\Delta_2$ improves the convergence.
When we take $L_k(R) = L_0(R)$, we arrive at the following
cut-off dependence for the phase shift
\begin{eqnarray}
\frac{d\,\delta}{d R} = - k^3 {\left( \frac{u_k(R,R)}{k} \right)}^2 
\sim k^3 R^{n/2} \, ,
\end{eqnarray}
while when considering the contribution for the $\Delta_2$ term, i.e. we take
$L_k(R) = L_0(R) + k^2\,\Delta_2(R)$, we arrive at the following result
\begin{eqnarray}
\frac{d\,\delta}{d R} = - 2\,k^3 {\left( \frac{u_k(R,R)}{k} \right)}^2
\sim k^3 R^{n/2} \, ,
\end{eqnarray}
which is qualitatively the same dependence, although the cut-off dependence
is doubled. In short, what we can see is that, when fixing the scattering 
length, $\Delta_2(R)$ is negligible in comparison with $L_0(R)$, 
and can therefore be removed.

Unfortunately this changes when considering the theory in which one fixes both
the scattering length and the effective range. In it, the boundary condition
can be expanded as
\begin{eqnarray}
L_k(R) = L_0(R) + k^2\,L_2(R) + k^4\,\Delta_4(R) + \dots
\end{eqnarray}
and, as we will show below, the $\Delta_4$ contribution cannot be ignored.
The behaviour of $L_0(R)$ is the one given by Eq.~(\ref{eq:L0_singular}), i.e.
exactly the same as in the previous case. On the contrary, the behaviour of
$L_2(R)$ differs significantly from that of $\Delta_2(R)$, given by
Eq.~(\ref{eq:D2_singular}). The reason lies in the fact that we are 
fixing the effective range. This can be deduced from the behaviour
of the $u_2$ wave function, which obeys the equation
\begin{eqnarray}
-u_2'' + U(r)\,u_2 = u_0 \, .
\label{eq:u2_G}
\end{eqnarray}
The solutions of the above equation can be separated into an homogeneous and
inhomogeneous piece
\begin{eqnarray}
u_2 = u_{2,H} + u_{2,I} \, ,
\end{eqnarray}
which in turn means that $L_2(R)$ can be written as the contribution of
these two pieces
\begin{eqnarray}
L_2 &=& L_{2,H} + L_{2,I} \, ,
\end{eqnarray}
where
\begin{eqnarray}
L_{2,H} &=& \frac{u_{2,H}' u_0 - u_{2,H} u_0'}{u_0^2} \, ,\\
L_{2,I} &=& \frac{u_{2,I}' u_0 - u_{2,I} u_0'}{u_0^2} \, .
\end{eqnarray}
The inhomogeneous piece of $u_2$ comes from the nontrivial contribution 
stemming from the $u_0$ wave function and can be computed via Green's functions.
Its short distance behaviour is much smoother than that of $u_0$, 
$u_{2,I}$ scales as $r^{3 n / 4 + 1}$ times an oscillating function $f(x)$
with $x = \frac{2}{n-2}\,(\frac{a}{R})^{n/2-1}$, while 
$u_0$ just scales as $r^{n / 4}$. 
From these behaviours it is trivial to see that
\begin{eqnarray}
L_{2,I}(R) = \Delta_2(R) \sim R \, ,
\end{eqnarray}
in which we wanted to make clear the identification between the inhomogeneous
piece of $L_2$ and what we previously called $\Delta_2$.

The problem with the inhomogeneous piece of $u_2$ is that its contribution
to the effective range is fixed, and is given by the following formula
\begin{eqnarray}
r_{0,I} = \frac{2}{\alpha_0^2}\,\int_{0}^{\infty}\,dr\,
\left[ (r - \alpha_0)^2 - u_0^2(r) \right] \, ,
\end{eqnarray}
as can be trivially deduced from Eq.~(\ref{eq:u2u0_lagrange}), and in which 
we have taken the cut-off to zero for simplicity. For modifying the total
effective range we need to add one contribution from the asymptotic behaviour
at large distances of the homogeneous wave function in such a way that
\begin{eqnarray}
u_{2,H} \to - \frac{1}{2}\,\alpha_0\,r_{0,H}\,r \, .
\end{eqnarray}
Then, the complete effective range of the wave function 
$u_2 = u_{2,H} + u_{2,I}$ will be $r_0 = r_{0,H} + r_{0,I}$, 
thus fixing the effective range to the desired value.
But the homogeneous contribution to $u_2$ is much more singular than 
the inhomogeneous one. The short distance behaviour of $u_{2,H}$ is given by
\begin{eqnarray}
u_{2,H}(R) \to R^{n/4}\,
\sin\left[\frac{2}{n-2}\,(\frac{a}{R})^{n/2-1} + \varphi_2) \right] \, ,
\label{eq:u2H_short}
\end{eqnarray}
with $\varphi_2$ a semiclassical phase. 
From the previous behaviour for $u_{2,H}$, Eq.~(\ref{eq:u2H_short}), one can
see that $L_{2,H}$ (and therefore $L_2$) will behave in a very similar way 
to $L_0$, i.e. a $1/R^{n/2}$ singularity times some oscillating function.
But one can get a much better evaluation of $L_{2,H}$ by considering the
following Lagrange identity for $u_0$ and $u_{2,H}$
\begin{eqnarray}
(u_{2,H}' u_0 - u_{2,H} u_0')' = 0 \, ,
\label{eq:u2Hu0_lagrange}
\end{eqnarray}
in contrast with $u_{2,I}$ which obeys the Lagrange identity 
Eq.~(\ref{eq:u2u0_lagrange}). This means that $u_{2,H}' u_0 - u_{2,H} u_0'$,
which appears in the numerator of $L_{2,H}$, is a constant value.
Evaluating at large distances, we obtain
\begin{eqnarray}
u_{2,H}' u_0 - u_{2,H} u_0' = \frac{r_{0,H}}{2}\,\alpha_0^2 \, ,
\end{eqnarray}
from which trivially follows
\begin{eqnarray}
L_{2,H}(R) = {\left( \frac{\alpha_0}{u_0(R)} \right)}^2\,\frac{r_{0,H}}{2} \, .
\end{eqnarray}

It is curious to see the equations which both contributions to $L_2$ follow
\begin{eqnarray}
L_{2,H}' + 2\,L_0(R)\,L_{2,H}(R) + 2 &=& 0 \, ,\\
L_{2,I}' + 2\,L_0(R)\,L_{2,I}(R) - 1 &=& 0 \, ,
\end{eqnarray}
from which taking into account that $L_2 = L_{2,H} + L_{2,I}$ one recovers
the renormalization group equation.

Finally, to  evaluate $\Delta_4$ we need its expression in terms of wave 
functions
\begin{eqnarray}
\Delta_4 (R) = \frac{u_4'\,u_0 - u_4\,u_0'}{u_0^2} + \frac{u_2}{u_0}\,L_2(R)
\, ,
\end{eqnarray}
where the $u_4$ contribution stems solely from the inhomogeneous piece 
(since we are not fixing $v_2$). By considering the appropriate Lagrange's 
identity
\begin{eqnarray}
(u_{4}' u_0 - u_{4} u_0')' = u_0\,u_2 \, ,
\label{eq:u4u0_lagrange}
\end{eqnarray}
it is trivial to deduce the entire evaluation of $\Delta_4(R)$
\begin{eqnarray}
\Delta_4 (R) = \frac{1}{u_0^2}\,\int_{0}^{R}\,dr\,u_0\,u_2 + 
\frac{u_2}{u_0}\,L_2(R) \,.
\end{eqnarray}
Due to the $u_2 L_2 / u_0$ contribution, we see that $\Delta_4$ is as singular
as $L_2$, and therefore as $L_0$, so its presence cannot be neglected in the
momentum expansion of the boundary condition. The same happens with 
$\Delta_6$, $\Delta_8$, and so on, but for the purposes of this appendix
is enough to consider only $\Delta_4$.

With all this we are prepared to evaluate the cut-off dependence of the phase
shifts, which is given by
\begin{eqnarray}
\frac{d\,\delta}{d R} &=& -k^5\,
\left[ L_2^2 + \Delta_4' + 2\,L_0\,\Delta_4 \right]\,u_0^2(R) \nonumber \\
&& + {\cal O}(k^7) \, .
\end{eqnarray}
After evaluating $\Delta_4$ inside the brackets, we are left with the much
simpler expression
\begin{eqnarray}
\frac{d\,\delta}{d R} &=& -k^5\,u_2(R)\,u_0(R) + {\cal O}(k^7) \nonumber \\
&& \sim - k^5\,R^{n/2} \, ,
\end{eqnarray}
which is just the same cut-off dependence as in the previous case, but 
somewhat diminished by the appearance of an extra $k^2$ factor.

Of course this is only the first step in showing that the phase shift is
well behaved. The ${\cal O}(k^7)$ term needs the explicit consideration of 
$\Delta_6$, the ${\cal O}(k^9)$ term of $\Delta_8$, and so on. 
Although it is straightforward to show that including the appropriate 
expansion of the boundary condition, the phase shift has a convergent
behaviour for $R \to 0$, it is very inconvenient for doing practical
computations in which we want an specific expression for the boundary
condition at the boundary radius $R$. 

The solution to these problems is relatively straightforward if we realize
that the contribution to $\Delta_4$, $\Delta_6$, etc, which cannot be ignored
comes from the homogeneous contributions to the wave functions, i.e. from
$u_0$ and $u_2$ (when we fix $\alpha_0$ and $r_0$).
For the general case in which we fix $\alpha_0$, $r_0$, $v_2$, ... , $v_n$,
the correct way to {\it truncate} the boundary condition is 
\begin{eqnarray}
L_k(R) = 
\frac{u_0' + k^2\,u_2' + \dots + k^{2 n}\,u_{2 n}'}
{u_0 + k^2\,u_2 + \dots + k^{2 n}\,u_{2 n}} + k^{2 n + 2}\,\Delta \, ,
\end{eqnarray}
where $\Delta$ contains the inhomogeneous higher order pieces. It is easy
to see that $\Delta(R) \sim R$, so it is negligible with respect to the main
piece, which behaves as $\sim 1/R^{n/2}$. The main piece generates all the
nontrivial terms needed for having a well-behaved $R \to 0$ limit on 
the phase shift. In fact, this probes the complete uniqueness of the 
result when fixing, for example, either $\alpha_0$ and $r_0$ in the
$k\,\cot{\delta}$ expansion, or $\alpha_0$ and $\beta_0$ in the 
$\tan{\delta} / k$ expansion.

\section{Soft and hard boundary conditions} 
\label{sec:soft} 

In this appendix we analyze how can we define contact interactions
through boundary conditions and the relation between what we call soft
and hard boundary conditions. Naively, the scattering amplitude
corresponding to a contact interaction is given by
\begin{eqnarray} 
f(k)=\frac1{-1/\alpha - {\rm i} k}  \, . 
\label{eq:f_contact}
\end{eqnarray} 
This amplitude {\it always} has a pole at $ k = i / \alpha $ which
corresponds to a virtual state for $\alpha < 0 $ and to a bound state
for $ \alpha > 0 $~\footnote{There are situations where this pole is
absent. For instance a square well repulsive potential}. The limit of
trivial scattering $ \alpha \to 0 $ corresponds to send the pole,
either positive or negative, to infinity. Or rather, to take $\delta =
n \pi $. By continuity of $\delta (k) $ in $ \alpha $ we should have a
difference of $\pi $ if $\alpha \to 0^+$ or if $\alpha \to 0^-$, thus
the limit of trivial scattering $\alpha \to 0$ is not uniquely
defined. We can obtain the amplitude (\ref{eq:f_contact}) by solving the
free Schr\"odinger equation using the ``soft'' boundary condition at
the origin
\begin{eqnarray}
\alpha u'(0) + u(0) =0 \, .
\label{eq:f_soft}
\end{eqnarray} 
If we take the limit $\alpha \to 0$ we get the ``hard'' condition $
u(0)=0$. This gives a trivial phase-shift $\delta=0$. Note that this
limit is indeed singular since it does not distinguish between $
\alpha \to 0^+ $ and $ \alpha \to 0^-$. So, how can we have
nontrivial scattering with no pole? As discussed in
Ref.~\cite{Albeverio} there is no way of doing this with a finite
range potential in the limit of vanishing range.

The proper way is to use a ``hard'' condition at the point $r=\alpha$
for $\alpha > 0$,
\begin{eqnarray}
u(\alpha)=0  \, ,
\label{eq:f_hard}
\end{eqnarray} 
yielding 
\begin{eqnarray} 
f(k)=\frac1{k \cot(-\alpha k) - {\rm i} k}  \qquad  0 < \alpha < \infty   \, .
\end{eqnarray} 
This amplitude does not have any pole in the complex plane.  Thus, one
starts from $ \alpha = \infty$ to $\alpha=0^+ $ using
Eq.~(\ref{eq:f_hard}) which corresponds to a repulsive core and then
goes on for $ \alpha=0^-$ with Eq.~(\ref{eq:f_hard}).

For the purely short distance theory, the generalization to coupled
channel scattering is almost trivial. If we assume $s-$wave
scattering, and using the boundary condition at the origin
\begin{eqnarray}
{\bf \alpha} \u'(0) + \u(0) =0 \, , 
\end{eqnarray} 
with ${\bf \alpha}$ the s-wave scattering length matrix, we get for
the coupled channel amplitude
\begin{eqnarray}
{\bf f}^{-1} =  -{\bf \alpha}^{-1} + {\rm i} \k \, ,  
\end{eqnarray} 
with $\k$ given by Eq.~(\ref{eq:k_coupled}). The simplicity of the
derivation contrasts with the cumbersome treatment of the
Lippmann-Schwinger equation with a sharp three-momentum cut-off
presented in Ref.~\cite{Cohen:2004kf}.

\section{Why a boundary condition instead of a potential as a
regulator in coordinate space ?}
\label{sec:app_why} 

In this appendix we elaborate further on the suitability of the BC as
compared to the use of square well or delta shell regulators. Our main
concern has to do with choosing a regularization method where a
truncated low energy expansion of the amplitude corresponds {\it
exactly } with a truncated low energy expansion of the regulator. One
of the advantages of a boundary condition as a short distance
regulator, as opposed to a short range potential, is related to the
non appearance of spurious high order effects in a low energy
expansion of the amplitude~\footnote{Actually, one of the advantages
of using Dimensional regularization (DR) in the MS scheme was that
there was a one-to-one connection between the bare potential and the
renormalized amplitude. Unfortunately, DR has never been applied to
the non-perturbative regularization of singular potentials at the
origin.}. As we will see here this requires infinitely many
counter-terms in an energy dependent potential.  Of course, any of
these methods when considered to all orders ought to yield the same
result; the difference has to do with truncating the regularized
potential to a given order. As we will show the BC is the only method
of the three analyzed where truncation is consistent order by
order. To understand the situation let us consider the three
regularizations in coordinate space: Square well
regularization~\cite{Beane:2000wh}, delta shell
regularization~\cite{Barford:2002je} and boundary condition
regularization~\cite{PavonValderrama:2003np}.
\begin{itemize}
\item Square well regulator.  
In this case we get 
\begin{eqnarray} 
U_k (r) = \left( U_0 + k^2 U_2 + \dots \right) \theta (a -r) \, .
\label{eq:sw_k}
\end{eqnarray} 
The phase-shift reads 
\begin{eqnarray} 
k \cot \delta = \frac{\sqrt{k^2+U_k} \cot (k a ) \cot( \sqrt{k^2+U_k} a) +
k } {k \cot(k a ) - \sqrt{k^2+U_k} \cot( \sqrt{k^2+U_k} a) } \, .
\end{eqnarray} 
Thus, the scattering length is given by  
\begin{eqnarray} 
\alpha = a \left( 1-\frac{\tan ( \sqrt{U_0} a) }{\sqrt{U_0} a}
\right) 
\label{eq:alpha_well} 
\end{eqnarray} 
and the effective range 
\begin{eqnarray} 
r_0 &=& 2 a \left( 1 - \frac{1}{\alpha U_0 a } - \frac{a^2}{3
\alpha^2} \right) \nonumber \\ &+& U_2 \frac{a-\alpha}{\alpha^2 U_0} \left( 1 -
\sqrt{U_0} a + \alpha a U_0 - U_0^2 a^2 \right)
\end{eqnarray} 
and so on for $v_2$, etc. Obviously, we can always fit any set of low
energy parameters by adding sufficient energy dependent terms to the
potential $U_0$, $U_2$ etc. any value of the low energy constants we
want. The mapping, however, is not one to one; we have infinitely many
solutions for $U_0$ Eq.~(\ref{eq:alpha_well}) for a given value of
$\alpha$ at a given scale $a$. This multi-valuation propagates to
higher order low energy parameters. As a consequence, if we only
specify a set of low energy parameters, the remaining higher order
ones are multivalued, and hence the phase-shift is not uniquely
predicted. To make the ambiguity more explicit let us take the limit $
a \to 0 $ and consider the LO truncated potential, $U_0$.  In this
limit we obtain from inversion of Eq.~(\ref{eq:alpha_well}),
\begin{eqnarray}
\sqrt{U_0} a = \left( n + \frac12 \right) \pi \left( 1 +
\frac{a}\alpha + \dots \right) \, ,
\end{eqnarray}  
with $n$ an {\it arbitrary} integer number. Using this asymptotic
solution we get for the effective range,  
\begin{eqnarray}
r_0 = 2 a\left( 1- \frac{a}\alpha \frac{1}{(n+1/2)^2 \pi^2} + \dots
\right) \, ,
\end{eqnarray} 
which depends manifestly on the arbitrary value of $n$. The $n$
dependence might be cancelled by including a NLO term, $U_2$. More
generally, for any finite $a$ one should include infinitely many terms
in the energy expansion of the square well potential,
Eq.~(\ref{eq:sw_k}) to get rid of these multi-valuation in the pure
short range theory. 

\item Delta Shell regulator.  In this case we get
\begin{eqnarray}
U(r) = \left( U_0 + k^2 U_2 + \dots \right) \delta (a -r) \, . 
\end{eqnarray}
Then one gets for the scattering length 
\begin{eqnarray}
\alpha = \frac{a^3 U_0}{1+U_0 a^2 } \, ,
\end{eqnarray}
whereas the effective range reads, 
\begin{eqnarray}
r_0 = \frac{4 a }3 \left(1 -\frac{a}{2 \alpha} \right)  + 2 a U_2
\left( 1 - \frac{a}{\alpha} \right)^2  \, .
\end{eqnarray}
In this case there are no multiple solutions, but similarly to
the square well regularization we get spurious terms at higher orders 

\item Boundary condition regulator. In this case we have 
$$
\frac{u_k' (a) } {u_k (a) } = k \cot( k a + \delta) = \xi_0 + (ka)^2
\xi_2 + \dots \,\, .
$$ In this case the translation between the boundary condition and an
effective range expansion is most straightforward; each energy
contribution to the BC provides a new term in the effective range. If
we take the limit $a \to 0 $ we get a one-to-one mapping. Thus, the BC
method provides a compatible hierarchy of equations in a low momentum
expansion.

\end{itemize}

\section{Renormalization Group Equation for Potential Regularization} 
\label{sec:potential_rg} 

The standard way of visualizing renormalization is by means of
potentials. In this section we use our renormalization group ideas
similar to those presented in Sect.~\ref{sec:rg} to determine the RG
evolution of these short distance regulators. 

\subsection{Square well Potential Regularization}

In Refs.~\cite{Beane:2000wh} a short range energy dependent square
well has been employed to regulate the short distance behaviour. That
means taking the family of potentials
\begin{eqnarray} 
U_k (r,R) = U_k (R) \theta ( R-r) + U (r) \theta (r-R) \, .
\end{eqnarray} 
The regular solution for $r < R$ is (assuming $ U_k(R) < 0 $ for
definiteness) 
\begin{eqnarray} 
u(r) = A \sin \left( \sqrt{k^2-U_k(R)} r\right)   \qquad r < R \, ,
\end{eqnarray} 
where $A$ has to be determined by matching to the long range piece of
the wave function. Then 
\begin{widetext} 
\begin{equation}
-\frac1k \frac{d \delta }{dR} = \left\{ \sin^2 \left(
\sqrt{k^2-U_k(R)} R\right) \left( U_k ( R) - U ( R)\right)+ U_k' (R)
\left[ \frac{R}2-\frac{\sin \left(2 \sqrt{k^2-U_k (R)} R \right)}{ 4 
\sqrt{k^2-U_k (R)} } \right] \right\} \frac{A^2}{k^2} \,. 
\end{equation}
\end{widetext}
If we demand the phase shift to be independent on the short-distance
regulator we get the renormalization group equation,  
\begin{equation}
R U_k' (R) = \left( U_k (R) - U ( R)\right) \frac{ 2 \sin^2 (\phi)}
{1-\phi \cos{\phi} \sin{\phi} } \, ,
\end{equation}
where we have defined the dimensionless combination, 
\begin{eqnarray}
\phi= \sqrt{k^2-U_k(R)} R \, .
\end{eqnarray} 
A similar equation has also been found in Ref.~\cite{Eiras:2001hu} by
different means. The low energy fixed points are given by
\begin{equation}
\sin (\phi) = 0 \qquad 1 = \phi \sin ( \phi) \cos ( \phi) \, .
\end{equation}
The first equation has the analytical solution $\phi_n = n \pi$. The
other equation has also infinitely many solutions $\phi_m$.  All these
fixed points are stable, so we have infinitely many branches, in
agreement with the observation of Ref.~{\cite{Eiras:2001hu}.

\subsection{Delta shell Potential}

In Ref.~\cite{Barford:2002je} a delta shell potential regularization
for the short range potential has been introduced, and a RG equation
in momentum space has been obtained by cutting off the high energy
components. According to our point of view, we should cut-off the long
range potential at short distances. Then, we have
\begin{eqnarray}
U(r) &=&  U_k (R ) \delta (r-R ) + U (r) \theta(r-R) \, .
\end{eqnarray} 
Let us denote by $u_L$ and $v_L(r)$ the wave functions regular and
singular at the origin respectively associated to the long range
potential alone, $U (r)$, and fulfilling 
\begin{eqnarray}
u_L \to \sin (kr + \delta_L) \, ,\\
v_L \to \cos (kr + \delta_L) \, .
\end{eqnarray} 
If we take the $ u(r)=A u_L(r) + B v_L (r) $ for $r > R $, then we have
\begin{eqnarray}
\frac{u' (R)}{u (R)} - k \cot(kR) = U_S (R) \, .
\label{eq:disc} 
\end{eqnarray}  
To obtain the RG equation let us compute the change of
the phase shift,
\begin{eqnarray} 
\Delta \delta = -\frac1k \Delta\left[ U_S (R) u (R)^2 \right] +\frac1k
U_L (R) \Delta R u(R)^2 \, .
\end{eqnarray} 
Using the condition \ref{eq:disc} we get 
\begin{eqnarray}
U_S '(R) = U_L(R) -2  \left[ U_S (R) + k \cot(kR) \right] U_S(R) \, .
\end{eqnarray} 


\begin{thebibliography}{59}
\expandafter\ifx\csname natexlab\endcsname\relax\def\natexlab#1{#1}\fi
\expandafter\ifx\csname bibnamefont\endcsname\relax
  \def\bibnamefont#1{#1}\fi
\expandafter\ifx\csname bibfnamefont\endcsname\relax
  \def\bibfnamefont#1{#1}\fi
\expandafter\ifx\csname citenamefont\endcsname\relax
  \def\citenamefont#1{#1}\fi
\expandafter\ifx\csname url\endcsname\relax
  \def\url#1{\texttt{#1}}\fi
\expandafter\ifx\csname urlprefix\endcsname\relax\def\urlprefix{URL }\fi
\providecommand{\bibinfo}[2]{#2}
\providecommand{\eprint}[2][]{\url{#2}}

\bibitem[{\citenamefont{Wilson and Kogut}(1974)}]{Wilson:1973jj}
\bibinfo{author}{\bibfnamefont{K.~G.} \bibnamefont{Wilson}} \bibnamefont{and}
  \bibinfo{author}{\bibfnamefont{J.~B.} \bibnamefont{Kogut}},
  \bibinfo{journal}{Phys. Rept.} \textbf{\bibinfo{volume}{12}},
  \bibinfo{pages}{75} (\bibinfo{year}{1974}).

\bibitem[{\citenamefont{Weinberg}(1990)}]{Weinberg:1990rz}
\bibinfo{author}{\bibfnamefont{S.}~\bibnamefont{Weinberg}},
  \bibinfo{journal}{Phys. Lett.} \textbf{\bibinfo{volume}{B251}},
  \bibinfo{pages}{288} (\bibinfo{year}{1990}).

\bibitem[{\citenamefont{Weinberg}(1991)}]{Weinberg:1991um}
\bibinfo{author}{\bibfnamefont{S.}~\bibnamefont{Weinberg}},
  \bibinfo{journal}{Nucl. Phys.} \textbf{\bibinfo{volume}{B363}},
  \bibinfo{pages}{3} (\bibinfo{year}{1991}).

\bibitem[{\citenamefont{Bedaque and van Kolck}(2002)}]{Bedaque:2002mn}
\bibinfo{author}{\bibfnamefont{P.~F.} \bibnamefont{Bedaque}} \bibnamefont{and}
  \bibinfo{author}{\bibfnamefont{U.}~\bibnamefont{van Kolck}},
  \bibinfo{journal}{Ann. Rev. Nucl. Part. Sci.} \textbf{\bibinfo{volume}{52}},
  \bibinfo{pages}{339} (\bibinfo{year}{2002}), \eprint{nucl-th/0203055}.

\bibitem[{\citenamefont{van Kolck}(1999{\natexlab{a}})}]{vanKolck:1998bw}
\bibinfo{author}{\bibfnamefont{U.}~\bibnamefont{van Kolck}},
  \bibinfo{journal}{Nucl. Phys.} \textbf{\bibinfo{volume}{A645}},
  \bibinfo{pages}{273} (\bibinfo{year}{1999}{\natexlab{a}}),
  \eprint{nucl-th/9808007}.

\bibitem[{\citenamefont{van Kolck}(1999{\natexlab{b}})}]{vanKolck:1999mw}
\bibinfo{author}{\bibfnamefont{U.}~\bibnamefont{van Kolck}},
  \bibinfo{journal}{Prog. Part. Nucl. Phys.} \textbf{\bibinfo{volume}{43}},
  \bibinfo{pages}{337} (\bibinfo{year}{1999}{\natexlab{b}}),
  \eprint{nucl-th/9902015}.

\bibitem[{\citenamefont{Epelbaum}(2006)}]{Epelbaum:2005pn}
\bibinfo{author}{\bibfnamefont{E.}~\bibnamefont{Epelbaum}},
  \bibinfo{journal}{Prog. Part. Nucl. Phys.} \textbf{\bibinfo{volume}{57}},
  \bibinfo{pages}{654} (\bibinfo{year}{2006}), \eprint{nucl-th/0509032}.

\bibitem[{\citenamefont{Hammer et~al.}(2006)\citenamefont{Hammer,
  Kalantar-Nayestanaki, and Phillips}}]{Hammer:2006qj}
\bibinfo{author}{\bibfnamefont{H.~W.} \bibnamefont{Hammer}},
  \bibinfo{author}{\bibfnamefont{N.}~\bibnamefont{Kalantar-Nayestanaki}},
  \bibnamefont{and} \bibinfo{author}{\bibfnamefont{D.~R.}
  \bibnamefont{Phillips}} (\bibinfo{year}{2006}), \eprint{nucl-th/0611084}.

\bibitem[{\citenamefont{Lomon and Feshbach}(1967)}]{LF67}
\bibinfo{author}{\bibfnamefont{E.}~\bibnamefont{Lomon}} \bibnamefont{and}
  \bibinfo{author}{\bibfnamefont{H.}~\bibnamefont{Feshbach}},
  \bibinfo{journal}{Rev. Mod. Phys.} \textbf{\bibinfo{volume}{39}},
  \bibinfo{pages}{611} (\bibinfo{year}{1967}).

\bibitem[{\citenamefont{Stoks et~al.}(1993)\citenamefont{Stoks, Kompl,
  Rentmeester, and de~Swart}}]{Stoks:1993tb}
\bibinfo{author}{\bibfnamefont{V.~G.~J.} \bibnamefont{Stoks}},
  \bibinfo{author}{\bibfnamefont{R.~A.~M.} \bibnamefont{Kompl}},
  \bibinfo{author}{\bibfnamefont{M.~C.~M.} \bibnamefont{Rentmeester}},
  \bibnamefont{and} \bibinfo{author}{\bibfnamefont{J.~J.}
  \bibnamefont{de~Swart}}, \bibinfo{journal}{Phys. Rev.}
  \textbf{\bibinfo{volume}{C48}}, \bibinfo{pages}{792} (\bibinfo{year}{1993}).

\bibitem[{\citenamefont{Rentmeester et~al.}(1999)\citenamefont{Rentmeester,
  Timmermans, Friar, and de~Swart}}]{Rentmeester:1999vw}
\bibinfo{author}{\bibfnamefont{M.~C.~M.} \bibnamefont{Rentmeester}},
  \bibinfo{author}{\bibfnamefont{R.~G.~E.} \bibnamefont{Timmermans}},
  \bibinfo{author}{\bibfnamefont{J.~L.} \bibnamefont{Friar}}, \bibnamefont{and}
  \bibinfo{author}{\bibfnamefont{J.~J.} \bibnamefont{de~Swart}},
  \bibinfo{journal}{Phys. Rev. Lett.} \textbf{\bibinfo{volume}{82}},
  \bibinfo{pages}{4992} (\bibinfo{year}{1999}), \eprint{nucl-th/9901054}.

\bibitem[{\citenamefont{Beane and Savage}(2001)}]{Beane:2000fi}
\bibinfo{author}{\bibfnamefont{S.~R.} \bibnamefont{Beane}} \bibnamefont{and}
  \bibinfo{author}{\bibfnamefont{M.~J.} \bibnamefont{Savage}},
  \bibinfo{journal}{Nucl. Phys.} \textbf{\bibinfo{volume}{A694}},
  \bibinfo{pages}{511} (\bibinfo{year}{2001}), \eprint{nucl-th/0011067}.

\bibitem[{\citenamefont{Calogero}(1967)}]{Calogero:1965}
\bibinfo{author}{\bibfnamefont{F.}~\bibnamefont{Calogero}},
  \emph{\bibinfo{title}{Variable Phase Approach to Potential Scattering}}
  (\bibinfo{publisher}{Academic Press, New York}, \bibinfo{year}{1967}).

\bibitem[{\citenamefont{Pavon~Valderrama and
  Ruiz~Arriola}(2004{\natexlab{a}})}]{PavonValderrama:2003np}
\bibinfo{author}{\bibfnamefont{M.}~\bibnamefont{Pavon~Valderrama}}
  \bibnamefont{and}
  \bibinfo{author}{\bibfnamefont{E.}~\bibnamefont{Ruiz~Arriola}},
  \bibinfo{journal}{Phys. Lett.} \textbf{\bibinfo{volume}{B580}},
  \bibinfo{pages}{149} (\bibinfo{year}{2004}{\natexlab{a}}),
  \eprint{nucl-th/0306069}.

\bibitem[{\citenamefont{Pavon~Valderrama and
  Ruiz~Arriola}(2004{\natexlab{b}})}]{PavonValderrama:2004nb}
\bibinfo{author}{\bibfnamefont{M.}~\bibnamefont{Pavon~Valderrama}}
  \bibnamefont{and}
  \bibinfo{author}{\bibfnamefont{E.}~\bibnamefont{Ruiz~Arriola}},
  \bibinfo{journal}{Phys. Rev.} \textbf{\bibinfo{volume}{C70}},
  \bibinfo{pages}{044006} (\bibinfo{year}{2004}{\natexlab{b}}),
  \eprint{nucl-th/0405057}.

\bibitem[{\citenamefont{Birse et~al.}(1999)\citenamefont{Birse, McGovern, and
  Richardson}}]{Birse:1998dk}
\bibinfo{author}{\bibfnamefont{M.~C.} \bibnamefont{Birse}},
  \bibinfo{author}{\bibfnamefont{J.~A.} \bibnamefont{McGovern}},
  \bibnamefont{and} \bibinfo{author}{\bibfnamefont{K.~G.}
  \bibnamefont{Richardson}}, \bibinfo{journal}{Phys. Lett.}
  \textbf{\bibinfo{volume}{B464}}, \bibinfo{pages}{169} (\bibinfo{year}{1999}),
  \eprint{hep-ph/9807302}.

\bibitem[{\citenamefont{Barford and Birse}(2003)}]{Barford:2002je}
\bibinfo{author}{\bibfnamefont{T.}~\bibnamefont{Barford}} \bibnamefont{and}
  \bibinfo{author}{\bibfnamefont{M.~C.} \bibnamefont{Birse}},
  \bibinfo{journal}{Phys. Rev.} \textbf{\bibinfo{volume}{C67}},
  \bibinfo{pages}{064006} (\bibinfo{year}{2003}), \eprint{hep-ph/0206146}.

\bibitem[{\citenamefont{van Haeringen and Kok}(1980)}]{Haeringen}
\bibinfo{author}{\bibfnamefont{H.}~\bibnamefont{van Haeringen}}
  \bibnamefont{and} \bibinfo{author}{\bibfnamefont{L.}~\bibnamefont{Kok}},
  \bibinfo{journal}{Phys. Rev.} \textbf{\bibinfo{volume}{A26}},
  \bibinfo{pages}{1218} (\bibinfo{year}{1980}).

\bibitem[{\citenamefont{Steele and Furnstahl}(1999)}]{Steele:1998zc}
\bibinfo{author}{\bibfnamefont{J.~V.} \bibnamefont{Steele}} \bibnamefont{and}
  \bibinfo{author}{\bibfnamefont{R.~J.} \bibnamefont{Furnstahl}},
  \bibinfo{journal}{Nucl. Phys.} \textbf{\bibinfo{volume}{A645}},
  \bibinfo{pages}{439} (\bibinfo{year}{1999}), \eprint{nucl-th/9808022}.

\bibitem[{\citenamefont{Birse and McGovern}(2004)}]{Birse:2003nz}
\bibinfo{author}{\bibfnamefont{M.~C.} \bibnamefont{Birse}} \bibnamefont{and}
  \bibinfo{author}{\bibfnamefont{J.~A.} \bibnamefont{McGovern}},
  \bibinfo{journal}{Phys. Rev.} \textbf{\bibinfo{volume}{C70}},
  \bibinfo{pages}{054002} (\bibinfo{year}{2004}), \eprint{nucl-th/0307050}.

\bibitem[{\citenamefont{Birse}(2006)}]{Birse:2005um}
\bibinfo{author}{\bibfnamefont{M.~C.} \bibnamefont{Birse}},
  \bibinfo{journal}{Phys. Rev.} \textbf{\bibinfo{volume}{C74}},
  \bibinfo{pages}{014003} (\bibinfo{year}{2006}), \eprint{nucl-th/0507077}.

\bibitem[{\citenamefont{Pavon~Valderrama and
  Ruiz~Arriola}(2005)}]{PavonValderrama:2005gu}
\bibinfo{author}{\bibfnamefont{M.}~\bibnamefont{Pavon~Valderrama}}
  \bibnamefont{and}
  \bibinfo{author}{\bibfnamefont{E.}~\bibnamefont{Ruiz~Arriola}},
  \bibinfo{journal}{Phys. Rev.} \textbf{\bibinfo{volume}{C72}},
  \bibinfo{pages}{054002} (\bibinfo{year}{2005}), \eprint{nucl-th/0504067}.

\bibitem[{\citenamefont{Nogga et~al.}(2005)\citenamefont{Nogga, Timmermans, and
  van Kolck}}]{Nogga:2005hy}
\bibinfo{author}{\bibfnamefont{A.}~\bibnamefont{Nogga}},
  \bibinfo{author}{\bibfnamefont{R.~G.~E.} \bibnamefont{Timmermans}},
  \bibnamefont{and} \bibinfo{author}{\bibfnamefont{U.}~\bibnamefont{van
  Kolck}}, \bibinfo{journal}{Phys. Rev.} \textbf{\bibinfo{volume}{C72}},
  \bibinfo{pages}{054006} (\bibinfo{year}{2005}), \eprint{nucl-th/0506005}.

\bibitem[{\citenamefont{Epelbaum and Meissner}(2006)}]{Epelbaum:2006pt}
\bibinfo{author}{\bibfnamefont{E.}~\bibnamefont{Epelbaum}} \bibnamefont{and}
  \bibinfo{author}{\bibfnamefont{U.}~\bibnamefont{Meissner}}
  (\bibinfo{year}{2006}), \eprint{arXiv:nucl-th/0609037}.

\bibitem[{\citenamefont{Valderrama and Arriola}(2006)}]{Valderrama:2005wv}
\bibinfo{author}{\bibfnamefont{M.~P.} \bibnamefont{Valderrama}}
  \bibnamefont{and} \bibinfo{author}{\bibfnamefont{E.~R.}
  \bibnamefont{Arriola}}, \bibinfo{journal}{Phys. Rev.}
  \textbf{\bibinfo{volume}{C74}}, \bibinfo{pages}{054001}
  (\bibinfo{year}{2006}), \eprint{nucl-th/0506047}.

\bibitem[{\citenamefont{Pavon~Valderrama and
  Ruiz~Arriola}(2006)}]{PavonValderrama:2005uj}
\bibinfo{author}{\bibfnamefont{M.}~\bibnamefont{Pavon~Valderrama}}
  \bibnamefont{and}
  \bibinfo{author}{\bibfnamefont{E.}~\bibnamefont{Ruiz~Arriola}},
  \bibinfo{journal}{Phys. Rev.} \textbf{\bibinfo{volume}{C74}},
  \bibinfo{pages}{064004} (\bibinfo{year}{2006}), \eprint{nucl-th/0507075}.

\bibitem[{\citenamefont{Bogner et~al.}(2001)\citenamefont{Bogner, Schwenk, Kuo,
  and Brown}}]{Bogner:2001jn}
\bibinfo{author}{\bibfnamefont{S.~K.} \bibnamefont{Bogner}},
  \bibinfo{author}{\bibfnamefont{A.}~\bibnamefont{Schwenk}},
  \bibinfo{author}{\bibfnamefont{T.~T.~S.} \bibnamefont{Kuo}},
  \bibnamefont{and} \bibinfo{author}{\bibfnamefont{G.~E.} \bibnamefont{Brown}}
  (\bibinfo{year}{2001}), \eprint{nucl-th/0111042}.

\bibitem[{\citenamefont{Bogner et~al.}(2003{\natexlab{a}})\citenamefont{Bogner,
  Kuo, Schwenk, Entem, and Machleidt}}]{Bogner:2001gq}
\bibinfo{author}{\bibfnamefont{S.~K.} \bibnamefont{Bogner}},
  \bibinfo{author}{\bibfnamefont{T.~T.~S.} \bibnamefont{Kuo}},
  \bibinfo{author}{\bibfnamefont{A.}~\bibnamefont{Schwenk}},
  \bibinfo{author}{\bibfnamefont{D.~R.} \bibnamefont{Entem}}, \bibnamefont{and}
  \bibinfo{author}{\bibfnamefont{R.}~\bibnamefont{Machleidt}},
  \bibinfo{journal}{Phys. Lett.} \textbf{\bibinfo{volume}{B576}},
  \bibinfo{pages}{265} (\bibinfo{year}{2003}{\natexlab{a}}),
  \eprint{nucl-th/0108041}.

\bibitem[{\citenamefont{Holt et~al.}(2004)\citenamefont{Holt, Kuo, Brown, and
  Bogner}}]{Holt:2003rj}
\bibinfo{author}{\bibfnamefont{J.~D.} \bibnamefont{Holt}},
  \bibinfo{author}{\bibfnamefont{T.~T.~S.} \bibnamefont{Kuo}},
  \bibinfo{author}{\bibfnamefont{G.~E.} \bibnamefont{Brown}}, \bibnamefont{and}
  \bibinfo{author}{\bibfnamefont{S.~K.} \bibnamefont{Bogner}},
  \bibinfo{journal}{Nucl. Phys.} \textbf{\bibinfo{volume}{A733}},
  \bibinfo{pages}{153} (\bibinfo{year}{2004}), \eprint{nucl-th/0308036}.

\bibitem[{\citenamefont{Bogner et~al.}(2003{\natexlab{b}})\citenamefont{Bogner,
  Kuo, and Schwenk}}]{Bogner:2003wn}
\bibinfo{author}{\bibfnamefont{S.~K.} \bibnamefont{Bogner}},
  \bibinfo{author}{\bibfnamefont{T.~T.~S.} \bibnamefont{Kuo}},
  \bibnamefont{and} \bibinfo{author}{\bibfnamefont{A.}~\bibnamefont{Schwenk}},
  \bibinfo{journal}{Phys. Rept.} \textbf{\bibinfo{volume}{386}},
  \bibinfo{pages}{1} (\bibinfo{year}{2003}{\natexlab{b}}),
  \eprint{nucl-th/0305035}.

\bibitem{Harada:2005tw}
K.~Harada, K.~Inoue and H.~Kubo,
Phys.\ Lett.\  B {\bf 636}, 305 (2006)
[arXiv:nucl-th/0511020].

\bibitem[{\citenamefont{Kaiser et~al.}(1997)\citenamefont{Kaiser, Brockmann,
  and Weise}}]{Kaiser:1997mw}
\bibinfo{author}{\bibfnamefont{N.}~\bibnamefont{Kaiser}},
  \bibinfo{author}{\bibfnamefont{R.}~\bibnamefont{Brockmann}},
  \bibnamefont{and} \bibinfo{author}{\bibfnamefont{W.}~\bibnamefont{Weise}},
  \bibinfo{journal}{Nucl. Phys.} \textbf{\bibinfo{volume}{A625}},
  \bibinfo{pages}{758} (\bibinfo{year}{1997}), \eprint{nucl-th/9706045}.

\bibitem[{\citenamefont{Frederico et~al.}(1999)\citenamefont{Frederico,
  Timoteo, and Tomio}}]{Frederico:1999ps}
\bibinfo{author}{\bibfnamefont{T.}~\bibnamefont{Frederico}},
  \bibinfo{author}{\bibfnamefont{V.~S.} \bibnamefont{Timoteo}},
  \bibnamefont{and} \bibinfo{author}{\bibfnamefont{L.}~\bibnamefont{Tomio}},
  \bibinfo{journal}{Nucl. Phys.} \textbf{\bibinfo{volume}{A653}},
  \bibinfo{pages}{209} (\bibinfo{year}{1999}), \eprint{nucl-th/9902052}.

\bibitem[{\citenamefont{Kaplan}(1997)}]{Kaplan:1996nv}
\bibinfo{author}{\bibfnamefont{D.~B.} \bibnamefont{Kaplan}},
  \bibinfo{journal}{Nucl. Phys.} \textbf{\bibinfo{volume}{B494}},
  \bibinfo{pages}{471} (\bibinfo{year}{1997}), \eprint{nucl-th/9610052}.

\bibitem[{\citenamefont{Nieves}(2003)}]{Nieves:2003uu}
\bibinfo{author}{\bibfnamefont{J.}~\bibnamefont{Nieves}},
  \bibinfo{journal}{Phys. Lett.} \textbf{\bibinfo{volume}{B568}},
  \bibinfo{pages}{109} (\bibinfo{year}{2003}), \eprint{nucl-th/0301080}.

\bibitem[{\citenamefont{Phillips et~al.}(2000)\citenamefont{Phillips, Afnan,
  and Henry-Edwards}}]{Phillips:1999bf}
\bibinfo{author}{\bibfnamefont{D.~R.} \bibnamefont{Phillips}},
  \bibinfo{author}{\bibfnamefont{I.~R.} \bibnamefont{Afnan}}, \bibnamefont{and}
  \bibinfo{author}{\bibfnamefont{A.~G.} \bibnamefont{Henry-Edwards}},
  \bibinfo{journal}{Phys. Rev.} \textbf{\bibinfo{volume}{C61}},
  \bibinfo{pages}{044002} (\bibinfo{year}{2000}), \eprint{nucl-th/9910063}.

\bibitem[{\citenamefont{Case}(1950)}]{Case:1950}
\bibinfo{author}{\bibfnamefont{K.}~\bibnamefont{Case}}, \bibinfo{journal}{Phys.
  Rev.} \textbf{\bibinfo{volume}{80}}, \bibinfo{pages}{797}
  (\bibinfo{year}{1950}).

\bibitem[{\citenamefont{Beane et~al.}(2002)\citenamefont{Beane, Bedaque,
  Savage, and van Kolck}}]{Beane:2001bc}
\bibinfo{author}{\bibfnamefont{S.~R.} \bibnamefont{Beane}},
  \bibinfo{author}{\bibfnamefont{P.~F.} \bibnamefont{Bedaque}},
  \bibinfo{author}{\bibfnamefont{M.~J.} \bibnamefont{Savage}},
  \bibnamefont{and} \bibinfo{author}{\bibfnamefont{U.}~\bibnamefont{van
  Kolck}}, \bibinfo{journal}{Nucl. Phys.} \textbf{\bibinfo{volume}{A700}},
  \bibinfo{pages}{377} (\bibinfo{year}{2002}), \eprint{nucl-th/0104030}.

\bibitem[{\citenamefont{Beane et~al.}(2001)}]{Beane:2000wh}
\bibinfo{author}{\bibfnamefont{S.~R.} \bibnamefont{Beane}}
  \bibnamefont{et~al.}, \bibinfo{journal}{Phys. Rev.}
  \textbf{\bibinfo{volume}{A64}}, \bibinfo{pages}{042103}
  (\bibinfo{year}{2001}), \eprint{quant-ph/0010073}.

\bibitem[{\citenamefont{Epelbaum et~al.}(2000)\citenamefont{Epelbaum, Gloeckle,
  and Meissner}}]{Epelbaum:1999dj}
\bibinfo{author}{\bibfnamefont{E.}~\bibnamefont{Epelbaum}},
  \bibinfo{author}{\bibfnamefont{W.}~\bibnamefont{Gloeckle}}, \bibnamefont{and}
  \bibinfo{author}{\bibfnamefont{U.-G.} \bibnamefont{Meissner}},
  \bibinfo{journal}{Nucl. Phys.} \textbf{\bibinfo{volume}{A671}},
  \bibinfo{pages}{295} (\bibinfo{year}{2000}), \eprint{nucl-th/9910064}.

\bibitem[{\citenamefont{Eiras and Soto}(2003)}]{Eiras:2001hu}
\bibinfo{author}{\bibfnamefont{D.}~\bibnamefont{Eiras}} \bibnamefont{and}
  \bibinfo{author}{\bibfnamefont{J.}~\bibnamefont{Soto}},
  \bibinfo{journal}{Eur. Phys. J.} \textbf{\bibinfo{volume}{A17}},
  \bibinfo{pages}{89} (\bibinfo{year}{2003}), \eprint{nucl-th/0107009}.

\bibitem[{\citenamefont{Ince}(1956)}]{Ince:1926}
\bibinfo{author}{\bibfnamefont{E.}~\bibnamefont{Ince}},
  \emph{\bibinfo{title}{Ordinary Differential Equations}}
  (\bibinfo{publisher}{Dover Publications, New York}, \bibinfo{year}{1956}).

\bibitem[{\citenamefont{Campos}(1997)}]{Campos:1997}
\bibinfo{author}{\bibfnamefont{J.}~\bibnamefont{Campos}},
  \bibinfo{journal}{Bull. London Math. Soc.} \textbf{\bibinfo{volume}{29}},
  \bibinfo{pages}{205} (\bibinfo{year}{1997}).

\bibitem[{\citenamefont{Braaten and Hammer}(2006)}]{Braaten:2004rn}
\bibinfo{author}{\bibfnamefont{E.}~\bibnamefont{Braaten}} \bibnamefont{and}
  \bibinfo{author}{\bibfnamefont{H.~W.} \bibnamefont{Hammer}},
  \bibinfo{journal}{Phys. Rept.} \textbf{\bibinfo{volume}{428}},
  \bibinfo{pages}{259} (\bibinfo{year}{2006}), \eprint{cond-mat/0410417}.

\bibitem[{\citenamefont{Leclair et~al.}(2003)\citenamefont{Leclair, Roman, and
  Sierra}}]{Leclair:2003xj}
\bibinfo{author}{\bibfnamefont{A.}~\bibnamefont{Leclair}},
  \bibinfo{author}{\bibfnamefont{J.~M.} \bibnamefont{Roman}}, \bibnamefont{and}
  \bibinfo{author}{\bibfnamefont{G.}~\bibnamefont{Sierra}},
  \bibinfo{journal}{Nucl. Phys.} \textbf{\bibinfo{volume}{B675}},
  \bibinfo{pages}{584} (\bibinfo{year}{2003}), \eprint{hep-th/0301042}.

\bibitem[{\citenamefont{Braaten and Phillips}(2004)}]{Braaten:2004pg}
\bibinfo{author}{\bibfnamefont{E.}~\bibnamefont{Braaten}} \bibnamefont{and}
  \bibinfo{author}{\bibfnamefont{D.}~\bibnamefont{Phillips}}
  (\bibinfo{year}{2004}), \eprint{hep-th/0403168}.

\bibitem[{\citenamefont{Hammer and Swingle}(2006)}]{Hammer:2005sa}
\bibinfo{author}{\bibfnamefont{H.~W.} \bibnamefont{Hammer}} \bibnamefont{and}
  \bibinfo{author}{\bibfnamefont{B.~G.} \bibnamefont{Swingle}},
  \bibinfo{journal}{Annals Phys.} \textbf{\bibinfo{volume}{321}},
  \bibinfo{pages}{306} (\bibinfo{year}{2006}), \eprint{quant-ph/0503074}.

\bibitem[{\citenamefont{Epelbaum
  et~al.}(2004{\natexlab{a}})\citenamefont{Epelbaum, Gloeckle, and
  Meissner}}]{Epelbaum:2003gr}
\bibinfo{author}{\bibfnamefont{E.}~\bibnamefont{Epelbaum}},
  \bibinfo{author}{\bibfnamefont{W.}~\bibnamefont{Gloeckle}}, \bibnamefont{and}
  \bibinfo{author}{\bibfnamefont{U.-G.} \bibnamefont{Meissner}},
  \bibinfo{journal}{Eur. Phys. J.} \textbf{\bibinfo{volume}{A19}},
  \bibinfo{pages}{125} (\bibinfo{year}{2004}{\natexlab{a}}),
  \eprint{nucl-th/0304037}.

\bibitem[{\citenamefont{Epelbaum
  et~al.}(2004{\natexlab{b}})\citenamefont{Epelbaum, Gloeckle, and
  Meissner}}]{Epelbaum:2003xx}
\bibinfo{author}{\bibfnamefont{E.}~\bibnamefont{Epelbaum}},
  \bibinfo{author}{\bibfnamefont{W.}~\bibnamefont{Gloeckle}}, \bibnamefont{and}
  \bibinfo{author}{\bibfnamefont{U.-G.} \bibnamefont{Meissner}},
  \bibinfo{journal}{Eur. Phys. J.} \textbf{\bibinfo{volume}{A19}},
  \bibinfo{pages}{401} (\bibinfo{year}{2004}{\natexlab{b}}),
  \eprint{nucl-th/0308010}.

\bibitem[{\citenamefont{Epelbaum et~al.}(2005)\citenamefont{Epelbaum, Glockle,
  and Meissner}}]{Epelbaum:2004fk}
\bibinfo{author}{\bibfnamefont{E.}~\bibnamefont{Epelbaum}},
  \bibinfo{author}{\bibfnamefont{W.}~\bibnamefont{Glockle}}, \bibnamefont{and}
  \bibinfo{author}{\bibfnamefont{U.-G.} \bibnamefont{Meissner}},
  \bibinfo{journal}{Nucl. Phys.} \textbf{\bibinfo{volume}{A747}},
  \bibinfo{pages}{362} (\bibinfo{year}{2005}), \eprint{nucl-th/0405048}.

\bibitem[{\citenamefont{Phillips and Cohen}(1997)}]{Phillips:1996ae}
\bibinfo{author}{\bibfnamefont{D.~R.} \bibnamefont{Phillips}} \bibnamefont{and}
  \bibinfo{author}{\bibfnamefont{T.~D.} \bibnamefont{Cohen}},
  \bibinfo{journal}{Phys. Lett.} \textbf{\bibinfo{volume}{B390}},
  \bibinfo{pages}{7} (\bibinfo{year}{1997}), \eprint{nucl-th/9607048}.

\bibitem[{\citenamefont{Entem et~al.}(2007)\citenamefont{Entem,
  Pavon~Valderrama, and Ruiz~Arriola}}]{EPR07}
\bibinfo{author}{\bibfnamefont{D.}~\bibnamefont{Entem}},
  \bibinfo{author}{\bibfnamefont{M.}~\bibnamefont{Pavon~Valderrama}},
  \bibnamefont{and}
  \bibinfo{author}{\bibfnamefont{E.}~\bibnamefont{Ruiz~Arriola}},
  \bibinfo{journal}{Preprint (unpublished)}  (\bibinfo{year}{2007}).

\bibitem[{\citenamefont{Richardson}(1999)}]{Richardson:1999hj}
\bibinfo{author}{\bibfnamefont{K.~G.} \bibnamefont{Richardson}}
  (\bibinfo{year}{1999}), \eprint{hep-ph/0008118}.

\bibitem[{\citenamefont{Valderrama and Arriola}(2005)}]{Valderrama:2005ku}
\bibinfo{author}{\bibfnamefont{M.~P.} \bibnamefont{Valderrama}}
  \bibnamefont{and} \bibinfo{author}{\bibfnamefont{E.~R.}
  \bibnamefont{Arriola}}, \bibinfo{journal}{Phys. Rev.}
  \textbf{\bibinfo{volume}{C72}}, \bibinfo{pages}{044007}
  (\bibinfo{year}{2005}).

\bibitem[{\citenamefont{Stoks et~al.}(1994)\citenamefont{Stoks, Klomp,
  Terheggen, and de~Swart}}]{Stoks:1994wp}
\bibinfo{author}{\bibfnamefont{V.~G.~J.} \bibnamefont{Stoks}},
  \bibinfo{author}{\bibfnamefont{R.~A.~M.} \bibnamefont{Klomp}},
  \bibinfo{author}{\bibfnamefont{C.~P.~F.} \bibnamefont{Terheggen}},
  \bibnamefont{and} \bibinfo{author}{\bibfnamefont{J.~J.}
  \bibnamefont{de~Swart}}, \bibinfo{journal}{Phys. Rev.}
  \textbf{\bibinfo{volume}{C49}}, \bibinfo{pages}{2950} (\bibinfo{year}{1994}),
  \eprint{nucl-th/9406039}.

\bibitem[{\citenamefont{Coddington and Levinson}(1955)}]{Coddington}
\bibinfo{author}{\bibfnamefont{E.}~\bibnamefont{Coddington}} \bibnamefont{and}
  \bibinfo{author}{\bibfnamefont{N.}~\bibnamefont{Levinson}},
  \emph{\bibinfo{title}{Theory of ordinary differential equations}}
  (\bibinfo{publisher}{Mc. Graw-Hill}, \bibinfo{year}{1955}).

\bibitem[{\citenamefont{Pavon~Valderrama and
  Ruiz~Arriola}(2004{\natexlab{c}})}]{PavonValderrama:2004td}
\bibinfo{author}{\bibfnamefont{M.}~\bibnamefont{Pavon~Valderrama}}
  \bibnamefont{and}
  \bibinfo{author}{\bibfnamefont{E.}~\bibnamefont{Ruiz~Arriola}}
  (\bibinfo{year}{2004}{\natexlab{c}}), \eprint{nucl-th/0410020}.

\bibitem[{\citenamefont{Albeverio et~al.}(1988)\citenamefont{Albeverio,
  Gesztesy, R., and Holden}}]{Albeverio}
\bibinfo{author}{\bibfnamefont{S.}~\bibnamefont{Albeverio}},
  \bibinfo{author}{\bibfnamefont{F.}~\bibnamefont{Gesztesy}},
  \bibinfo{author}{\bibfnamefont{H.-K.} \bibnamefont{R.}}, \bibnamefont{and}
  \bibinfo{author}{\bibfnamefont{H.}~\bibnamefont{Holden}},
  \emph{\bibinfo{title}{Solvable Models in Quantum Mechanics}}
  (\bibinfo{publisher}{Texts and Monographs in Physics; Springer},
  \bibinfo{year}{1988}).

\bibitem[{\citenamefont{Cohen et~al.}(2004)\citenamefont{Cohen, Gelman, and van
  Kolck}}]{Cohen:2004kf}
\bibinfo{author}{\bibfnamefont{T.~D.} \bibnamefont{Cohen}},
  \bibinfo{author}{\bibfnamefont{B.~A.} \bibnamefont{Gelman}},
  \bibnamefont{and} \bibinfo{author}{\bibfnamefont{U.}~\bibnamefont{van
  Kolck}}, \bibinfo{journal}{Phys. Lett.} \textbf{\bibinfo{volume}{B588}},
  \bibinfo{pages}{57} (\bibinfo{year}{2004}), \eprint{nucl-th/0402054}.

\end{thebibliography}

\end{document}